\documentclass[twoside,11pt]{article}

\usepackage{blindtext}

\usepackage{jmlr2e}

\usepackage{xspace}
\usepackage[english]{babel}
\usepackage{amsmath,amsfonts,amssymb,bm,mathptmx,dsfont}
\usepackage[table,dvipsnames]{xcolor}
\usepackage{tikz}
\usepackage[normalem]{ulem}

\newtheorem{assumption}[theorem]{Assumption}

\usepackage[nameinlink]{cleveref}

\crefname{figure}{Figure}{Figures}
\crefname{section}{Section}{Sections}
\crefname{proposition}{Proposition}{Propositions}
\crefname{appendix}{Appendix}{Appendices}
\crefname{assumption}{Assumption}{Assumptions}
\crefname{lemma}{Lemma}{Lemmata}
\crefname{definition}{Definition}{Definitions}
\crefname{remark}{Remark}{Remarks}
\crefname{algorithm}{Algorithm}{Algorithms}
\crefname{table}{Table}{Tables}

\newcommand \titA {Relaxed Gaussian Process Interpolation}
\newcommand \titB {a Goal-Oriented Approach to Bayesian Optimization}

\newcommand \kwA {Gaussian processes}
\newcommand \kwB {Bayesian optimization}
\newcommand \kwC {Expected improvement}
\newcommand \kwD {Goal-oriented modeling}
\newcommand \kwE {Reproducing kernel Hilbert spaces}

\renewcommand \mathbb  {\mathds}
\newcommand   \UU      {\mathds{U}}
\newcommand   \XX      {\mathds{X}}
\newcommand   \RR      {\mathds{R}}
\newcommand   \NN      {\mathds{N}}
\newcommand   \one     {\mathds{1}}
\newcommand   \Lset    {L}
\newcommand   \HH      {\mathcal{H}}
\newcommand   \Pcal    {\mathcal{P}}
\newcommand   \Ncal    {\mathcal{N}}
\newcommand   \tr      {^{\mathsf{T}}}
\newcommand   \EE      {\mathsf{E}}
\newcommand   \PP      {\mathsf{P}}
\newcommand   \ddiff   {\mathrm{d}}
\newcommand   \du      {\ddiff u}
\newcommand   \CRPS    {\ensuremath{\mathrm{CRPS}}}
\newcommand   \tCRPS   {\ensuremath{\mathrm{tCRPS}}}
\newcommand   \EI      {\mathrm{EI}}
\newcommand   \xn      {\underline{x}_n}

\newcommand   \z       {\underline{z}}
\newcommand   \zn      {\underline{z}_n}
\newcommand   \Zn      {\underline{Z}_n}
\newcommand   \xnzero  {\underline{x}_{n}^0}
\newcommand   \znzero  {\underline{z}_{n}^0}
\newcommand   \xnB  {\underline{x}_{n}^B}
\newcommand   \znB  {\underline{z}_{n}^B}
\newcommand   \R       {R}
\newcommand   \GP      {\mathop{\mathrm{GP}}}
\newcommand   \reGP    {\mbox{reGP}\xspace}

\providecommand{\abs}[1]{\lvert#1\rvert}
\providecommand{\ns}[1]{\lVert#1\rVert}

\renewcommand{\hat}{\widehat}

\DeclareMathOperator*{\argmax}{arg\,max}
\DeclareMathOperator*{\argmin}{arg\,min}

\usepackage{algorithm}
\usepackage{algorithmic}

\usepackage{lastpage}
\jmlrheading{26}{2025}{1-\pageref{LastPage}}{7/22}{8/25}{22-0828}{Sébastien J. Petit and Julien Bect and Emmanuel Vazquez}

\hypersetup{ hidelinks }

\ShortHeadings{Relaxed Gaussian Process Interpolation}{Petit and Bect and Vazquez}
\firstpageno{1}

\begin{document}

\title{\titA:\\ \titB}

\author{\name Sébastien J. Petit \email sebastien.petit@lne.fr \\
       \addr Laboratoire National de Métrologie et d’Essais,\\
       78197, Trappes Cedex, France\thanks{Part of this work was funded by the
       French Agence Nationale de la Recherche et de la Technologie (ANRT) under a CIFRE grant, and was carried out
       while S.~J. Petit was affiliated with Safran Aircraft Engines (Moissy-Cramayel, France) and L2S.}
       \AND
       \name Julien Bect \email julien.bect@centralesupelec.fr \\
       \addr Université Paris-Saclay,\\
       CNRS, CentraleSupélec,\\
       Laboratoire des signaux et systèmes,\\
       91190, Gif-sur-Yvette, France
       \AND
       \name Emmanuel Vazquez \email emmanuel.vazquez@centralesupelec.fr \\
       \addr Université Paris-Saclay,\\
       CNRS, CentraleSupélec,\\
       Laboratoire des signaux et systèmes,\\
       91190, Gif-sur-Yvette, France}       
       
\editor{Ryan Adams}

\maketitle

\begin{abstract}
  This work presents a new procedure for obtaining predictive distributions
  in the context of Gaussian process (GP) modeling, with a relaxation of
  the interpolation constraints outside ranges of interest:
  the mean of the predictive distribution no longer necessarily
  interpolates the observed values when they are outside ranges of
  interest, but is simply constrained to remain outside.
  This method called relaxed Gaussian process (\reGP) interpolation
  provides better predictive distributions in ranges of
  interest, especially in cases where a stationarity assumption for the
  GP model is not appropriate. It can be viewed as a goal-oriented
  method and becomes particularly interesting in Bayesian optimization,
  for example, for the minimization of an objective function, where good
  predictive distributions for low function values are important.  When
  the expected improvement criterion and \reGP are used for sequentially
  choosing evaluation points, the convergence of the resulting
  optimization algorithm is theoretically guaranteed (provided that the
  function to be optimized lies in the reproducing kernel Hilbert space
  attached to the known covariance of the underlying Gaussian process).
  Experiments indicate that using \reGP instead of stationary GP models
  in Bayesian optimization is beneficial.
\end{abstract}

\begin{keywords}
  \kwA; \kwB; \kwC; \kwD; \kwE
\end{keywords}

\section{Introduction}

\subsection{Context and Motivation}
Gaussian process (GP) interpolation and regression \citep[see,
e.g.,][]{stein1999:_interpolation_of_spatial_data,
  rasmussen06:_gauss_proces_machin_learn} is {a classical} method
for predicting an unknown function from data. It has found applications
in active learning techniques, and notably
in Bayesian optimization, a popular derivative-free global optimization
technique for functions whose evaluations are time-consuming.

A GP model is defined by a {mean function and a covariance function}, which are
generally selected from data within parametric families.  The most
popular models assume stationarity and rely on standard covariance
functions such as the Matérn covariance. The assumption of stationarity
yields models with relatively low-dimensional parameters. However, such a
hypothesis can sometimes result in poor models when the function to be
predicted has different scales of variation or different local
regularities across the domain.

This is the case for instance in the motivating example given by
\cite{gramacy2008:_tgp}, or in the even simpler toy minimization problem shown in
\cref{fig:rb_function}.  The objective function in this example, which we
shall call the Steep function, is smooth with an obvious global minimum
around the point $x = 8$. However, the variations around the minimum are
overshadowed by some steep variations on the left.
\cref{fig:rb_model} shows a stationary GP fit with $n = 8$ points,
where the parameters of the covariance function have been selected using
maximum likelihood.  Observe that the
confidence bands are too large and that the conditional mean varies too
much in the neighborhood of the global minimum, consistently with the
stationary GP model that reflects the prior that our function oscillates
around a mean value with a constant scale of variations. In this case,
even if GP interpolation is consistent \citep{vazquez2010:_pointwise_consistency},
stationarity seems an unsatisfactory assumption for the Steep function. One
expects Bayesian optimization techniques to be somehow inefficient on
this problem with such a stationary model, whose posterior distributions
are too pessimistic in the region of the minimum.

Nevertheless, the Steep function has the characteristics of an easy
optimization problem: it has only two local minima, with the global minimum lying in a
valley of significant volume. Consequently, a Bayesian optimization
technique could be competitive if it relied on a model giving good
predictions in regions where the function takes low values. In this
work, we propose to explore goal-oriented GP modeling, where we want
predictive models in regions of interest, even if it means being less
predictive elsewhere.

\begin{figure}
  \hspace{-0.5cm}
  \begin{tikzpicture}
    \draw (0,0) node[below] {\scalebox{0.79}{\input{figures/gplike_full.pgf}}};
    \draw (7.8cm,0) node[below] {\scalebox{0.79}{\input{figures/gplike_zoom.pgf}}};
  \end{tikzpicture}
    \caption{%
      Left: the Steep function. Right: same illustration with a
      restrained range on the $y$-axis.
      The variations on the left tend to overshadow the global minimum on the right.
    }
  \label{fig:rb_function}
\end{figure}

\begin{figure}
  \hspace{-0.5cm}
  \begin{tikzpicture}
    \draw (0,0) node[below] {\scalebox{0.79}{\input{figures/model_gplike_full.pgf}}};
    \draw (7.8cm,0) node[below] {\scalebox{0.79}{\input{figures/model_gplike_zoom.pgf}}};
  \end{tikzpicture}
  \caption{%
    Left: GP fit on the Steep function. Right: same illustration with a
    restrained range on the $y$-axis. The squares represent the
    data. The red line represents the posterior mean $\mu_n$ given by
    the model and the gray envelopes represent the associated
    uncertainties.  }
  \label{fig:rb_model}
\end{figure}

\subsection{Related Works}\label{sec:review}

\subsubsection{Local Models}
\label{sec:review:local}

Going beyond the stationary hypothesis has been an active direction of
research. %
Local models are one popular solution, which still uses stationary
Gaussian processes as a core building block.

A first class of local models is obtained by considering partitions of
the input domain with different GP models on each subset. Partitions can
be built by splitting the domain along the coordinate axes. This is the
case of the treed Gaussian process models proposed by
\cite{gramacy2008:_tgp}, which combines a fully Bayesian framework and
the use of RJ-MCMC techniques for the inference, or the trust-region
method by
\cite{eriksson_et_al_2019:_scalable_local_bayesian_optimization}.
\cite{park2018:_patchwork_kriging} also propose partition-based local
models built by splitting the domain along principal component
directions. In~such techniques, there are parameters
related to, e.g., the way the partitions evolve with the data, the
size of the partitions, or how local Gaussian processes interact with
each other.

A second class of local models is obtained by spatially weighting one or
several GP models.  Many schemes have been proposed, including methods
based on partition of unity
\citep{nott_and_dunsmuir_2002:_estimation_of_nonstationary_spatial_covariance_structure},
weightings of covariance functions
\citep{pronzato_and_rendas_2017:_bayesian_local_kriging,
  rivoirard_2010:_continuity_kriging_moving_neigh}, and convolution
techniques \citep[see,
e.g.,][]{higdon_1998:_a_process_convolution_approach_to_modeling_temperatures,
  gibbs1998:_bayesian,
  higdon_2002:_space_time_modeling_using_process_convolutions,
  hoef_et_al_2004:_flexible_spatial_models_using_ma_and_fft,
  stein2005:_nonstationary}. Let us also mention data-driven aggregation
techniques: composite Gaussian process models \citep{ba2012:_cgp}, and
mixture of experts techniques \citep[see, e.g.,][]{
  tresp2001:_mixtures_of_experts, rasmussen2002:_infinite_moe,
  meeds2006:_alternative_moe, yuan2009:_variational_moe,
  yang2011:_efficient_em, yuksel2012:_survey_moe}. In the latter
framework, the weights are called gating functions and the estimation of
the parameters and the inference are usually performed using EM, MCMC,
or variational techniques. Weighting methods generally have parameters
specifying weighting functions, with an increased need to watch for
overfitting phenomena.

\subsubsection{Transformation and Composition of Models}
\label{sec:review:compo}

Transformation and composition of models is another popular
alternative to stationary GP modeling, which also uses stationary
Gaussian processes as a core building block.

A first technique for composition of models consists in using a
parametric transformation of a~GP \citep{rychlik1997:_transformed_gp,
  snelson2004:_warped}.

Another route is to transform the input domain, using for instance a
parametric density \citep{xiong2007:_input_transforms}, or other
parametric transformations involving possible dimension reduction \citep{marmin2018:_warped}.
\cite{bodin2020:_modulating} proposed a framework that uses additional
input variables, serving as nuisance parameters, 
to smooth out some badly behaved data. The practitioner has to specify a
prior over the variance of the nuisance parameter and inference is based
on MCMC.

\cite{lazaro2012:_bwgp} takes the step of choosing a GP prior on the
output transform and resorts to variational inference techniques.
This type of idea can be viewed as an ancestor of deep Gaussian
processes \cite[see, e.g.,][]{damianou2013:_deep, dunlop2018:_deep_gp,
  hebbal2021:_dgp_bo, jakkala2021:_deep_gp, bachoc2021:_dgp}, which
stack layers of linear combinations of GPs. The practitioner has to
specify a network structure among other parameters and resort to
variational inference.

\cite{picheny2019:_obo} proposed another approach {where
predictions are} made only from pairwise comparisons between data points,
relying on the variational framework of ordinal GP regression proposed
by \cite{chu2005:_ogpr} for the inference.

\subsubsection{{Goal-oriented approximate inference techniques}}
\label{sec:review:goai}

Another related area of research is goal-oriented approximate
inference, where the task to be performed with the predictive
model is taken into account in the approximations that are used in the
inference (a.k.a. training) step. %

For instance, in the context of scalable Bayesian optimization,
\cite{mcintire_2016:_sparse_gp_bo} and
\cite{moss_2023:_high_throughput_bo} have proposed goal-oriented
criteria to choose inducing points. %
Another example is \cite{yang_2021:_sparse_spectrum_gp_bo}, who
propose a goal-oriented criterion to select the frequencies of a
sparse spectrum approximation. %
Loss-calibrated approximate Bayesian computation
\citep{lacoste_julien_2011:_abc_loss_calibrated_bayes,
  morais_2022:_abc_loss_calibrated_bayes_ep}, which takes into account
the decision-making task when approximating the posterior
distribution, also falls in this category.

These works focus on fine-tuning approximations of classical GP
models, by taking into account the task to be performed. %
This is in contrast with the method introduced in this article, which
aims to \emph{replace} the classical GP model with a new,
goal-oriented GP-based predictive distribution, which no longer stems
from conditioning a GP prior. %
(The development of a massively scalable version of the proposed
method is beyond the scope of this article, but could be based on
approximate goal-directed inference techniques.)

\subsection{Contributions and Outline}

The main contribution of this article is a method called \emph{relaxed
  Gaussian processes} (\reGP) for building goal-oriented GP-based
models targeting regions of interest specified through function
values. %
The objective is to obtain global models that exhibit good
predictive distributions on a range of interest---in the case of a
minimization problem, the range of interest would be the values below
a threshold---while possibly being less predictive outside the range
of interest. %
This is achieved by relaxing interpolation constraints
outside this range. %
Such a model is presented in \cref{fig:rb_slack}: compared to
the situation in \cref{fig:rb_model}, the model is more
predictive in the region where the Steep function takes low values,
with expected benefits for the efficiency of Bayesian optimization.

\begin{figure}[h!]
  \hspace{-0.5cm}
  \begin{tikzpicture}
    \draw (0,0) node[below] {\scalebox{0.79}{\input{figures/slack_gplike.pgf}}};
    \draw (7.8cm,0) node[below] {\scalebox{0.79}{\input{figures/slack_gplike_transfo.pgf}}};
  \end{tikzpicture}
  \caption{%
    Left: prediction of the Steep function with the proposed methodology
    (black line: relaxation threshold $t$;  blue points:
    relaxed observations).
    Right: $\mu_n$ versus $f$ (with more observations for illustration purposes).
    The model interpolates the data below $t$. The blue points are relaxed observations.
    }
  \label{fig:rb_slack}
\end{figure}

This article provides {two other} main contributions. %
{On the one hand, } we give theoretical and empirical results justifying the
method and its use for Bayesian optimization. %
{On the other hand,} to assess the predictivity of \reGP\ {models}, we adopt the formalism
of scoring rules \citep{gneiting:2007:scoring} and propose the use of
a goal-oriented scoring rule that we call \emph{truncated continuous
  ranked probability score} (tCRPS), which is designed to assess the
predictivity of a model in a range of interest. %
The tCRPS is used to construct a leave-one-out (LOO) goodness-of-fit
criterion  to automatically select the range of function
values (outside the range of interest) where the interpolation
constraints are relaxed. %

An important feature of the proposed approach is that, with respect to
classical GP modeling, it only requires the choice of one additional
key parameter---namely, the range of interest where good predictions
are expected. %
We propose several heuristics to set this range of interest in
Bayesian optimization and level-set estimation problems.

The organization of this article is as follows.  \cref{sec:basics} briefly recalls
the formalism of Gaussian processes and Bayesian optimization.
\cref{sec:slack} presents \reGP and its theoretical
properties. %
The tCRPS and its use for selecting {the relaxation range} are then presented in
\cref{sec:threshold_choice}.  \cref{sec:rgpi_bo} presents
a \reGP-based Bayesian optimization algorithm called EGO-R, together
with the convergence analysis of this algorithm and
a numerical benchmark.
Finally, \cref{sec:concl} presents our
conclusions and perspectives for future work.

An open source implementation of the reGP method and the
  numerical experiments is available online at
  \url{https://github.com/relaxedGP/regp_paper_experiments}.

\section{Background and Notations}\label{sec:basics}

\subsection{Gaussian Process Modeling}\label{sec:gp}

Consider a real-valued function $f: \XX \to \RR$, where
$\XX \subseteq \RR^d$, and suppose we want to infer $f$ at a given
$x \in \XX$ from evaluations of~$f$ on a finite set of points
$\xn = (x_1, \ldots, x_n)\in\XX^{n}$, $n\geq 1$.  A standard Bayesian
approach to this problem consists in using a GP model
$\xi \sim \GP \left( \mu, \, k \right)$ as a prior about~$f$,
where $\mu: \XX \to \RR$ is a mean function and
$k: \XX \times \XX \to \RR$ is a covariance function, which is supposed
to be strictly positive-definite in this article.

The posterior distribution of $\xi$ given
$\Zn = \left( \xi(x_1), \dots, \xi(x_n) \right)\tr$ is still a Gaussian
process, whose mean and covariance functions are given by the standard
kriging equations \citep{matheron71}. More precisely:
\begin{equation}\label{eq:posterior-distribution}
\xi \, | \, \Zn \sim \GP \left( \mu_n, \, k_n \right),
\end{equation}
with
\begin{equation}
  \label{eq:posterior-mean}
  \mu_n(x) = \mu(x) + k \left(x , \, \xn \right) K_n^{-1} \left( \Zn
    - \mu(\xn) \right)
\end{equation}    
and
\begin{equation*}
  k_n(x, \, y) = k(x, \, y) - k \left(x , \, \xn \right) K_n^{-1} k \left(y, \, \xn \right)\tr\,,
\end{equation*}
and
where $\mu(\xn) = \left( \mu(x_1), \dots, \mu(x_n) \right)\tr$, 
$k \left(x , \, \xn \right) = \left( k (x, \, x_1), \dots, k (x, \, x_n)  \right)$,
and $K_n$ is the $n \times n$ matrix with entries $k (x_i, \, x_j)$.
We shall also use the notation $\sigma_n^2(x) =
k_n(x, \, x)$ for the posterior variance, a.k.a. the kriging variance, a.k.a.
the squared power function, so that
$\xi(x) \, | \, \Zn \sim \Ncal \left( \mu_n(x), \, \sigma_n^2(x)
\right)$.

The functions $\mu$ and $k$ control the posterior
distribution~\eqref{eq:posterior-distribution} and must be chosen
carefully.  The standard practice is to select them from data within
a parametric family
$\{ \left(\mu_{\theta}, k_{\theta}\right), \ \theta \in \Theta \}$.
A~common approach is to suppose stationarity for the GP, which means
choosing a constant mean function $\mu \equiv c \in \RR$ and a
stationary covariance function $k(x, y) = {\tau^2} r(x - y)$,
where~$r:\RR^d \to \RR$ is a stationary correlation function. 

A correlation function often recommended in the literature
\citep{stein1999:_interpolation_of_spatial_data} is the
(geometrically anisotropic) Mat\'ern correlation function
\begin{equation}
  \label{eq:matern-cov}
  r(h) = \frac{2^{1 - \nu}}{\Gamma(\nu)}
  \left( \sqrt{2 \nu} \lVert h \rVert_{\rho} \right)^{\nu} \mathcal{K}_{\nu} \left( \sqrt{2 \nu} \lVert h \rVert_{\rho}
  \right)\,,  \quad 
  \lVert h \lVert_{\rho}^{2}= \sum_{j = 1}^d \frac{h_{\left[ j \right]}^2}{\rho_j^2}\,,  
\end{equation}
for $h = (h_{[1]},\,\ldots,\,h_{[d]})\in\RR^{d}$,
and where $\Gamma$ is the Gamma function and $\mathcal{K}_{\nu}$ is the
modified Bessel function of the second kind. The covariance parameters to be
selected in this case are
$({\tau^2},\, \rho_1,\,\ldots,\, \rho_d,\, \nu) \in \left( 0, \infty
\right)^{d+2}$ with ${\tau^2}$ the process variance, $\rho_i$ the range
parameter along the \hbox{$i$-th} dimension, and $\nu$ a regularity parameter
controlling the smoothness of the process. Two other standard covariance
functions can be recovered for specific values of $\nu$: the exponential
covariance function for $\nu = 1/2$ and the squared-exponential
covariance function for $\nu \to \infty$.

A variety of techniques for selecting the parameter $\theta$ have
been proposed in the literature, but we can safely say that maximum
likelihood estimation is the most popular and can be recommended in the case of interpolation
\citep{petit2021:_gaussian_process_model_selection}. It simply consists
in minimizing the negative log-likelihood
\begin{equation}
  \label{eq:likelihood}
  \mathcal{L} \left(\theta; \, \Zn \right) = - \log \left( p \left( \Zn \, \lvert \, \theta \right) \right) 
  \propto \log \left( \det \left( K_n \right) \right) + (\Zn - \mu(\xn))\tr K_n^{-1} (\Zn - \mu(\xn)) + \text{constant},
\end{equation}
where $p$ stands for the probability density of $\Zn$. Other
methods for selecting the parameters include the restricted maximum
likelihood method and leave-one-out strategies
\citep[see, e.g.,][]{stein1999:_interpolation_of_spatial_data,
rasmussen06:_gauss_proces_machin_learn}.

\subsection{Bayesian Optimization}\label{sec:bo}

The framework of GPs is well suited to the problem of sequential design
of experiments, or active learning. In particular, for minimizing a
 real-valued function $f$ defined on a compact domain $\XX$, the Bayesian approach consists in
{sequentially choosing} evaluation points ${x_1,\,x_2},\,\ldots \in \XX$ using a GP model
$\xi$ for $f$, which makes it to possible to build a
sampling criterion
that represents an expected information gain on the minimum of $f$
when an
evaluation is made at a new point. One of the most popular sampling
criteria (also called acquisition function) is the \emph{Expected Improvement} (EI)
\citep{mockus1978:_bayesian_extremum, jones1998:_ego}, which can be
expressed as
\begin{equation}\label{eq:ei}
\rho_n(x) = \EE \left( \left( m_n - \xi(x) \right)_{+} \, \lvert \, \Zn \right),
\end{equation}
where
$m_n = \min(\xi(x_1), \dots, \xi(x_n))$.  The EI criterion corresponds
to the expectation of the excursion of $\xi$ below the minimum given $n$
observations, and can be written in closed form:
\begin{proposition}{\citep{jones1998:_ego,
      vazquez_and_bect2010:_convergence_ei}}\label[proposition]{prop:ei}
  The EI criterion may be written as
  $\rho_n(x) = \gamma \left( m_n - \mu_n(x), \, \sigma_n^2(x) \right)$ with
  \begin{equation*}
    \gamma: (z, \, s) \in \RR \times \RR_+ \mapsto \left\{
      \begin{array}{ll}
        \sqrt{s} \phi \left( \frac{z}{\sqrt{s}} \right) + z \Phi \left( \frac{z}{\sqrt{s}} \right) & \ \mathrm{if} \ s > 0, \\
        \max(z, 0) & \ \mathrm{if} \ s = 0,
      \end{array}
    \right.
  \end{equation*}
  where $\phi$ and $\Phi$ stand for the probability density and
  cumulative distribution functions of the standard Gaussian
  distribution. %
  Moreover, the function $\gamma$ is continuous, satisfies
  $\gamma(z, \, s) > 0$ if $s > 0$ and is non-decreasing with respect
  to $z$ and $s$ on $\RR \times \RR_+$.
\end{proposition}
When the EI criterion is used for optimization, that is, when the
sequence of evaluation points~${(x_n)_{n>0} }$ of $f$ is chosen using the
rule
\begin{equation*}
  {x_{n+1}} = \argmax_{x\in\XX} \rho_{n}(x)\,,
\end{equation*}
the resulting algorithm is generally called the Efficient Global
Optimization (EGO) algorithm, as proposed by \cite{jones1998:_ego}. %
The EGO algorithm has known convergence properties
\citep{vazquez_and_bect2010:_convergence_ei, bull2011:_convergence}.

A variety of other sampling criteria for the minimization problem can
be found in the literature \citep[see,
e.g.,][]{frazier2008:_knowledge_gradients, villemonteix2009:_iago,
  srinivas2010:_ucb, vazquez2014:_ei_square}, but we shall focus on
the EI {criterion} in this article. %
(\cref{app:bench_ucb10} also considers the Upper
Confidence Bound (UCB) criterion by \citealp{srinivas2010:_ucb}, and
\cref{app:levelset} deals with the
task of estimating excursion sets using the straddle heuristic by
\citealp{bryan2005activelearning}.)%

\subsection{Reproducing Kernel Hilbert Spaces}\label{sec:rkhs}

Reproducing kernel Hilbert spaces \citep[RKHS, see,
e.g.,][]{aronszajn1950:_rkhs, berlinet2011:_reproducing} are Hilbert
spaces of functions commonly used in the field of approximation theory
\citep[see, e.g.,][]{wahba1990:_splines_models, wendland04:_scatt}.  A
Hilbert space $\HH(\XX)$ of real-valued functions on $\XX$ with an inner
product $(\cdot\,,\cdot)_{\HH(\XX)}$ is called an RKHS if it has a
reproducing kernel, that is, a function $k\colon \XX \times \XX \to \RR$
such that $k(x, \, \cdot) \in \HH(\XX)$, and
\begin{equation}\label{eq:reproducing_prop}
\left( f, \, k(x, \, \cdot) \right)_{\HH(\XX)} = f(x)
\end{equation}
(the reproducing property), for all $x \in \XX$ and $f \in \HH(\XX)$.
Furthermore, given a (strictly) positive definite covariance
function~$k$, there exists a unique RKHS admitting $k$ as reproducing
kernel.

Given locations $\xn = ( x_1, \dots, x_n ) \in \XX^n$, and
corresponding values $\zn\in\RR^n$, suppose we want to find a function
$h\in \HH(\XX)$ with minimal norm, such that
$h(\xn) = \left(h(x_1), \dots, h(x_n) \right)\tr = \zn$.  %
Then the solution is given by the following classical result, which
can be derived from the generalized representer theorem
\citep{scholkopf2001generalized} and appears in early work on optimal
interpolation in RKHS \citep{kw1970:_correspondance}. %
\begin{proposition}{(Minimum-norm interpolant)}\label[proposition]{prop:min_norm_blup}
The problem
\begin{equation}\label{eq:min_norm_extension}
  \left\{\; \begin{aligned}
    \text{minimize}   & \quad \ns{h}_{\HH(\XX)}\\
    \text{subject to} & \quad h \in \HH(\XX)\\
                      & \quad h(\xn) = \zn
  \end{aligned} \right.
\end{equation}
has a unique solution given by $s_{\zn} = k(\cdot, \, \xn) K_n^{-1} \zn$.
\end{proposition}
Observe that the solution $s_{\zn}$ is equal to the
posterior mean~\eqref{eq:posterior-mean} when $\mu = 0$.

Moreover, for any $f \in \HH(\XX)$ and $x\in\XX$, the reproducing
property~\eqref{eq:reproducing_prop} yields the upper bound
\begin{equation}\label{eq:power_function}
  \left| f(x) - s_{\zn}(x) \right| \leq \sigma_n(x) \, \ns{f}_{\HH(\XX)},
\end{equation}
with $\sigma_n(x) = \sqrt{k_n \left( x, \, x \right)}$ and~$\zn = f(\xn)$
\citep[see, e.g.,][p.~13]{scheuerer_et_al2013:_interpolation_det_stoch}.
Note that
$\sigma_n(x)$ is the worst-case error at $x$ for the interpolation of
functions in the unit ball of $\HH(\XX)$.

\section{Relaxed Gaussian Process Interpolation}\label{sec:slack}

\subsection{Relaxed Interpolation}\label{sec:relaxed_interpolation}

The example in the introduction (see
\crefrange{fig:rb_function}{fig:rb_slack}) suggests that, in
order to gain accuracy over a range of values of interest, it can be
beneficial to relax interpolation constraints outside this range. More
precisely, the probabilistic model in \cref{fig:rb_slack}
interpolates data lying below a selected threshold~$t$, and when data
are above $t$, the model only keeps the information that the data
exceeds~$t$.

In the following, we consider the general setting where relaxation is
carried out on a set {of function values} of the form $\R = \bigcup_{j = 1}^J R_j$, where
$R_1,\, \ldots,\, R_J \subset \RR$ are disjoint closed intervals with
non-zero lengths. (The set $\R= \left[t, +\infty\right)$ was used in
the example of \cref{fig:rb_slack}.)

As above, we shall write $\xn = (x_1, \dots, x_n) \in \XX^n$ for a
sequence of locations with corresponding function values
$\zn = (z_1, \dots, z_n)\tr \in \RR^n$.  Then, we introduce the set
$C_{\R,\,n} = C_1 \times \cdots \times C_n \subset \RR^n$ of relaxed
constraints, where
\begin{equation}\label{eq:goal_oriented_constraints}
  \left\{
    \begin{array}{llll}
      C_i = R_j & \mathrm{if} \ z_i \in R_j \ \mathrm{for \ some} \ j,  \\
      C_i = \{ z_i \} & \mathrm{otherwise}. \\
    \end{array}
  \right.
\end{equation}
Let also
$$\HH_{\R,\,n} = \{h \in \HH(\XX) \ \lvert \ h(\xn) \in C_{\R,\,n} \}$$
be the set of relaxed-interpolating functions. The~following
proposition gives the definition of the minimum-norm relaxed
predictor.
\begin{proposition}\label[proposition]{prop:finite_constrained_problem}
  The problem
  \begin{equation}\label{eq:min_norm_slack}
    \left\{\; \begin{aligned}
        \text{minimize}   & \quad \ns{h}_{\HH(\XX)}\\
        \text{subject to} & \quad h \in \HH_{\R,\,n}\\
      \end{aligned}
    \right.
  \end{equation}
  has a unique solution given by $s_{\zn^{\star}}$, where
  $\zn^{\star}$ is the unique solution of the quadratic problem
  \begin{equation}\label{eq:finite_constrained_problem}
    \argmin_{\underline{z} \in C_{\R,\,n}} \ \underline{z}\tr K_n^{-1} \underline{z}.
  \end{equation}
\end{proposition}

An extension of this relaxation scheme to the case of noisy
observations, where the function no longer exactly interpolates the
data, is given in \cref{prop:relaxed_kernel_ridge}.

\subsection{Relaxed Gaussian Process Interpolation}\label{sec:gp_r}

The main advantage of Gaussian processes is the possibility to obtain
not only point predictions but also predictive distributions.
However, \cref{prop:finite_constrained_problem} only
defines a function approximation. We~now turn relaxed interpolation
into a probabilistic model providing predictive distributions whose
mean is not constrained to interpolate data on a given {relaxation} range~$R$. The
following proposition makes a step in this direction.

\begin{proposition}\label[proposition]{prop:joint_posterior}
  Let $\xi \sim \GP(0, \, k)$, $\xn = \left(x_1, \dots, x_{n}\right) \in \XX^n$,
  $\zn \in \RR^n$ and $\underline{x}_{m}^{\prime} = \left(x_1^{\prime}, \dots, x_{m}^{\prime}\right) \in \XX^m$
  be a set of locations of interest where predictions should be made.
  Write $\Zn = \left(\xi(x_1), \dots, \xi(x_n)\right)\tr$
  and $\underline{Z}_m^{\prime} = \left(\xi(x_{1}^{\prime}), \dots, \xi(x_{m}^{\prime}) \right)\tr$. %
  Then the mode of the probability density function
  \begin{equation}\label{eq:joint_posterior}  
  p \left( \underline{Z}_m^{\prime}, \, \Zn \, \lvert \, \Zn \in C_{\R,\,n} \right)
  \end{equation}  
  is given by $\bigl(s_{\zn^{\star}}(\underline{x}_{m}^{\prime}),~\zn^{\star}\bigr)$.
\end{proposition}

In other words, the relaxed interpolation solution of
\cref{prop:finite_constrained_problem} corresponds to the
maximum a posteriori (MAP) estimate under the predictive
model~\eqref{eq:joint_posterior}. %
Conditioning on events of the form
$\Zn \in C_{\R,\,n}$ has been used in Bayesian statistics as a
principled way to handle outliers and model misspecification; see,
e.g., \citet{lewis2021:_bayesian} and references therein.
Such conditional distributions also appear in other settings: in
constrained Gaussian process modeling, where constraints encode expert
knowledge \citep{da2012:_constrained, maatouk2017:_gaussian,
  lopez2018:_finite}, and in classification, ordinal regression, and
preference learning problems, where observations impose inequality
constraints \citep{benavoli2021unified}.

However, the predictive distribution~\eqref{eq:joint_posterior} is non-Gaussian
since the support of $\Zn$ is truncated. In particular, no closed-form
expression is available for any of its moments, and sampling requires advanced
techniques (e.g., variational, MCMC; see \cref{sec:concl} for a discussion).
Motivated by this observation, we propose instead to build a goal-oriented
probabilistic model using the following definition.
\begin{definition}[Relaxed-GP predictive distribution; fixed~$\mu$ and~$k$]\label[definition]{def:fixed-rgpi}
  Given $\xn \in \XX^n$, $\zn\in\RR^{n}$, and a relaxation {range}~$\R$ (finite union of closed intervals),
  the \emph{relaxed-GP} (\reGP) predictive distribution with fixed mean
  function~$\mu$ and covariance function~$k$ is defined as the
  (Gaussian) conditional distribution of $\xi\sim \GP(\mu,\, k)$ given
  $\Zn = \zn^{\star}$, where
  \begin{equation}
  \label{eq:fixedrgpi_problem}
    \zn^{\star} = \argmin_{\underline{z} \in C_{\R,\,n}}\; %
    \left( \underline{z} - \mu(\xn) \right)\tr K_n^{-1} \left( \underline{z} - \mu(\xn) \right),
  \end{equation}
  with $C_{\R,\,n}$ defined by~\eqref{eq:goal_oriented_constraints}.
\end{definition}

Observe that~\eqref{eq:fixedrgpi_problem} reduces
to~\eqref{eq:finite_constrained_problem} when $\mu = 0$. %
Consequently, the mean of the distribution is the
predictor~$s_{\zn^{\star}}$ from
\cref{prop:finite_constrained_problem} in this particular
case. %
Moreover, the \reGP predictive distribution can be seen as an
approximation of~\eqref{eq:joint_posterior}, where
$p \left( \Zn \, \lvert \, \Zn \in C_{\R,\,n} \right)$ has been
replaced by its mode. %
As discussed earlier, the main advantage of the \reGP predictive
distribution compared to~\eqref{eq:joint_posterior} is its reasonable
computational {cost} since it is a GP. %
Therefore, it makes it possible to use adaptive strategies for the
choice of $\R$, as {demonstrated} in \cref{sec:threshold_choice}. %
Moreover, it also has appealing theoretical approximation properties,
as discussed in \cref{sec:theory_base}.

As discussed in \cref{sec:gp}, the standard practice is to
select the mean and the covariance functions within a parametric
family
$\{ \left(\mu_{\theta}, k_{\theta}\right), \ \theta \in \Theta \}$.
{
Leveraging the connection between~\eqref{eq:fixedrgpi_problem}
and~\eqref{eq:likelihood}, we propose}
to perform the parameter selection and the
relaxation jointly.  This is formalized by the following definition of
relaxed Gaussian process interpolation.
\begin{definition}[Relaxed-GP predictive distribution; estimated parameters]\label[definition]{def:rgpi}
  Given $\xn \in \XX^n$, $\zn\in\RR^{n}$, a relaxation {range}~$\R$ (finite union of closed intervals),
  and parametric families~$\left( \mu_\theta \right)$
  and~$\left( k_\theta \right)$ as in \cref{sec:gp}, the
  \emph{relaxed-GP} (\reGP) predictive distribution with estimated
  parameters is the (Gaussian) conditional distribution of
  $\xi \sim \GP(\mu_{\theta},\, k_{\theta})$ given
  $\Zn = \zn^{\star}$, where $\zn^{\star}$ and
  $\theta = \hat{\theta}_n$ are obtained jointly by minimizing the
  negative log-likelihood:
  \begin{equation}
    \label{eq:rgpi_problem}
    \left( \hat{\theta}_n, \ \zn^{\star} \right) \;=\; \argmin_{\theta \in \Theta, \ \underline{z} \in C_{\R,\,n}} \mathcal{L} \left( \theta; \, \underline{z}  \right)\,,
  \end{equation}
  with $C_{\R,\,n}$ defined by~\eqref{eq:goal_oriented_constraints}.
\end{definition}

An extension of this definition to the case of noisy
observations is presented in \cref{app:noise}.

The mean function and the covariance function usually vary continuously with~$\theta$
on an open subset of~$\RR^q$.
In this case, it can be shown that the
mapping~$\theta \mapsto \min_{\underline{z} \in C_{\R,\,n}} \mathcal{L} \left( \theta; \, \underline{z}  \right)$
is continuous~\citep[using, e.g., Theorem~2.2 from][]{best1995continuity}.
Consequently, the minimum of~\eqref{eq:rgpi_problem} is reached
when~$\theta$ is restricted to a compact subset~$\Theta \subset \RR^q$.

\begin{remark}[On minimizing~\eqref{eq:rgpi_problem} jointly]
  Let $\underline{Z}_{n, \, 1}$ be
  the values within the {relaxation} range $\R$, and $\underline{Z}_{n, \,
    0}$ the values in $\R^{c} = \RR\setminus \R$ that are not relaxed. The negative
  log-likelihood can be written as
\begin{equation*}
\mathcal{L} \left( \theta ; \, \Zn \right) =
- \ln \left( p \left( \underline{Z}_{n, \, 0} \, \lvert \, \theta \right) \right)
- \ln \left( p \left( \underline{Z}_{n, \, 1}  \, \lvert \, \theta, \, \underline{Z}_{n, \, 0} \right) \right),
\end{equation*}
where the first term is a goodness-of-fit criterion based on the values
in $\R^{c}$, and where the second term can mainly be viewed as an imputation term,
which ``reshapes'' the values in $\R$ with the information from~$\underline{Z}_{n, \, 0}$.
(Note also that~$\theta$ appears in the second term. When this
term is minimized with respect to~$\underline{Z}_{n, \, 1}$, it becomes a
parameter selection term that promotes the~$\theta$s compatible
with the excursions in~$C_{\R,\,n}$.)
\end{remark}

\begin{figure}[tbp]
  \centering
  \includegraphics[width=12cm]{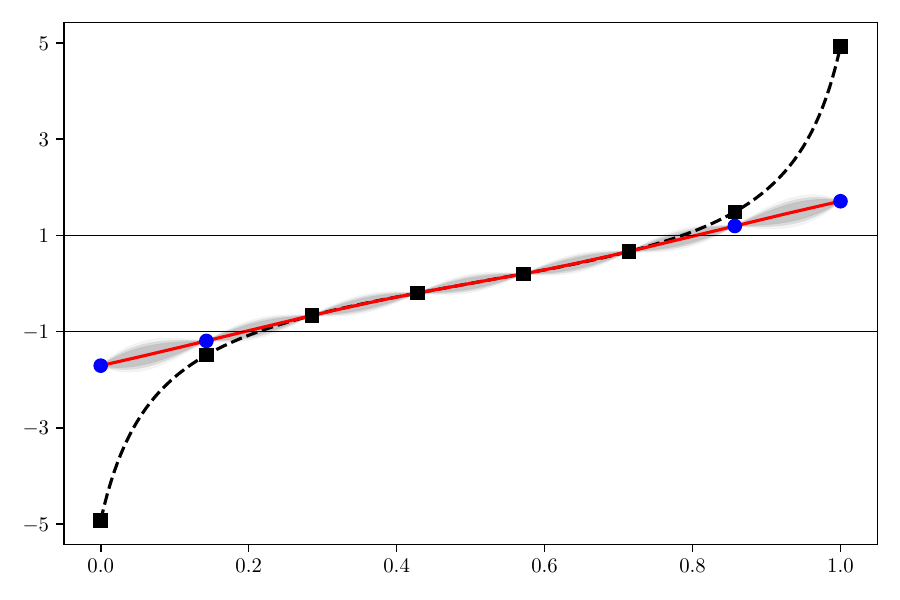}
  \caption{%
    An example of \reGP predictive distribution with
    $\R = \left( - \infty, -1 \right] \cup \left[1, + \infty \right)$
    on a function~$f$ represented in dashed black lines. %
    The solid black lines represent the relaxation thresholds. %
    The problem~\eqref{eq:rgpi_problem} was solved only in
    $\underline{z}$ as the parameters of the (constant) mean and
    ($\nu=5/2$ Matérn) covariance functions were held fixed for
    illustration purposes.}
  \label{fig:gpr_nicer}
\end{figure}

For illustration, we provide an example of a \reGP predictive distribution in
\cref{fig:gpr_nicer}, with an union of two intervals for the relaxation {range}~$R$.

\begin{remark}[Numerical details]
\label[remark]{rem:numerics}
  Minimizing~\eqref{eq:rgpi_problem} with respect to $\underline{z} \,\,$ 
  falls under the scope of quadratic programming \cite[see,
  e.g.,][]{nocedal2006:_quadratic} and could be solved efficiently using
  dedicated algorithms. This suggests that specific algorithms could be
  developed for the problem. In this work, we simply use a
  standard SLSQP solver \citep[see][]{schittkowski1982nonlinear}
  using the gradient
  of~\eqref{eq:rgpi_problem}.
  The SciPy implementation was used with default parameters.
  Good performances were obtained by the following parameter
  initialization procedure:
  first, initialize the GP parameter~$\theta$ using common rules while
  excluding the most extreme observations, and then optimize the relaxed
  observations~$\underline{z}$ using quadratic programming.
\end{remark}

\subsection{Convergence Analysis of \reGP}\label{sec:theory_base}

In this section, we provide theoretical results concerning the convergence
of the method proposed. This section can be skipped on first reading.

\subsubsection{Known Convergence Results about Interpolation in RKHS}\label{sec:theory_base_known}

Recall that the fractional-order Sobolev space
$W_2^{\beta} ( \RR^d )$, with regularity $\beta \geq 0$, is the
space of functions on $\RR^{d}$ defined by
$$
W_2^{\beta} ( \RR^d ) = \Bigl\lbrace h \in \Lset^2 ( \RR^d ), \ \ \
\ns{h}_{W_2^{\beta} ( \RR^d )}^2 =
\int_{\RR^d} \left( 1 + \ns{\omega}^2 \right)^{\beta}\, \lvert \hat{h}(\omega) \lvert^2 \ddiff \omega  < +\infty
\Bigr\rbrace,
$$%
where $\hat{h} \in \Lset^2 ( \RR^d )$ is the Fourier transform of $h \in \Lset^2 ( \RR^d )$. %
For~$\beta > d/2$, the usual identification with a set of continuous
functions given by the Sobolev embedding theorem will be used in this article.

For a given $\XX \subset \RR^d$, define the Sobolev
spaces $W_2^{\beta} ( \XX ) = \bigl\{h_{\lvert \XX}, \, h \in W_2^{\beta} ( \RR^d ) \bigr\}$ endowed
with the norm
\begin{equation}\label{eq:restriction_sobolev_norm}
\ns{h}_{W_2^{\beta} ( \XX )} = \inf_{g \in W_2^{\beta} ( \RR^d ), \, g_{\lvert \XX} = h} \ns{g}_{W_2^{\beta} ( \RR^d )}.
\end{equation}

The following assumption about~$\XX$ will sometimes be used in this section.
\begin{assumption}\label[assumption]{assump:domain}
  The domain is non-empty, compact, connected, has locally Lipschitz
  boundary \citep[see, e.g.,][Section 4.9]{adams2003:_sobolev}, and is
  equal to the closure of its interior.
\end{assumption}
\cref{assump:domain} ensures that the previous definition
coincides with other commons definitions, and makes it possible to use
well-known results from the field of scattered data approximation, by
preventing the existence of cusps. %
Many common domains---such as hyperrectangles or balls, for
instance---{satisfy} \cref{assump:domain}.

A strictly positive-definite reproducing kernel $k:\XX\times\XX\to\RR$
is said to have regularity $\alpha > 0$ if the associated RKHS
$\HH(\XX)$ coincides with $W_2^{\alpha + d/2}(\XX)$ as a function
space, with equivalent norms. %
As such, the Matérn stationary kernels~\eqref{eq:matern-cov} have
correlation functions~$r$ whose Fourier transform satisfies \citep[see,
e.g.,][Theorem 6.13]{wendland04:_scatt}
$$
C_1 \left( 1 + \ns{\cdot}^2 \right)^{- \nu - d/2} 
\leq
\hat{r}
\leq
C_2
\left( 1 + \ns{\cdot}^2 \right)^{- \nu - d/2}
$$ 
for some $C_2 \geq C_1 > 0$, and have therefore Sobolev regularity $\alpha=\nu$
on $\RR^d$ \citep[see, e.g.,][Corollary 10.13]{wendland04:_scatt}
and consequently also on $\XX$, using~\eqref{eq:restriction_sobolev_norm}
and \cref{lem:extension}.
Other examples are given by \citet{wendland04:_scatt}, for instance.

We now recall a classical convergence result about interpolation in
RKHS with evaluation points in a bounded domain.  Consider a kernel
$k: \XX \times \XX \to \RR$, and let $\left(x_n\right)_{n \geq 1} \in \XX^{\NN}$ be
a sequence  of distinct points. %
The following property (a minor reformulation of Theorem 4.1 of
\citealp{arcangeli2007:_error_bounds}) gives error bounds that depend on
the Sobolev regularity of $k$ and the so-called fill distance of $\xn
\in \XX^n$, defined by
\begin{equation*}
h_{n} = \sup_{x \in \XX} \min_{1 \leq i \leq n} \ns{x - x_i}.                   
\end{equation*}
\begin{proposition}\label[proposition]{prop:power_function_bounds}
Let $k$ be a reproducing kernel with regularity $\alpha > 0$.
If~$\XX$ satisfies \cref{assump:domain}, then
\begin{equation}\label{eq:power_function_bounds}
\sup_{x \in \XX} \sigma_n(x) \lesssim h_{n}^{\alpha},\quad n\geq 1\,,
\end{equation}
where $\lesssim$ denotes inequality up to a constant, that does not depend
on $\left(x_n\right)_{n \geq 1}$.
\end{proposition}
Using~\eqref{eq:power_function} and
\cref{prop:power_function_bounds}, this yields the following
uniform bound.
\begin{corollary}  
  Let $k$ be a reproducing kernel with regularity $\alpha > 0$,
  $\HH(\XX)$ the RKHS generated by $k$, and let $f \in \HH(\XX)$. As
  above, let $s_{\zn}$ be the solution of~\eqref{eq:min_norm_extension}
for $\zn = \left(f(x_1), \dots, f(x_n)\right)\tr$, $n \geq 1$. 
If~$\XX$ satisfies \cref{assump:domain}, then
\begin{equation}\label{eq:standard_linf_bound}
\ns{f - s_{\zn}}_{\Lset^{\infty}(\XX)}
\lesssim h_{n}^{\alpha} \ns{f}_{\HH(\XX)}\,.
\end{equation}
\end{corollary}

\subsubsection{Convergence of Relaxed Interpolation with Fixed Covariance Function}\label{sec:convergence_regp}

Let $k\colon \XX \times \XX \to \RR$ be a continuous strictly positive-definite reproducing kernel.
In this section, we consider the zero-mean \reGP predictive distribution
obtained from $\xi\sim \GP(0, \, k)$, with relaxed interpolation
constraints on a union  $\R = \bigcup_{j = 1}^q R_j$ of disjoints
closed intervals~$R_j$ with non-zero length.
Let $\HH(\XX)$ be the RKHS attached to $k$, $f \in \HH(\XX)$,
and consider a sequence $\left(x_n\right)_{n \geq 1} \in \XX^{\NN}$ of distinct points.
Furthermore, define the regions
$\XX_j = \{x \in \XX, \ f(x) \in R_j \}$ for $1 \leq j \leq q$ and
$\XX_0 = \XX \setminus \bigcup_{j \geq 1} \XX_j$. We give results about
the limit of the sequence of \reGP predictive distributions that indicate
an improved fit in $\XX_0$.

Let $ s_{\R,\,n} = s_{\zn^{\star}}$ be
the relaxed predictor from
\cref{prop:finite_constrained_problem} based
on~$\left(x_1, \dots, x_n\right)$ and~$\left(f(x_1), \dots, f(x_n)\right)\tr$,
$n \geq 1$. %
The following proposition establishes the limit behavior of~$( s_{\R,\,n})_{n \geq 1}$.
\begin{proposition}\label[proposition]{prop:constrained_problem}
Let $\UU \subset \XX$ and let $\HH_{\R,\,\UU}$ denote the set
of functions $h \in \HH(\XX)$ such that, for all $x \in \UU$,
    \begin{equation}\label{eq:continuous_constraints_convex}
     \left\{
      \begin{array}{llll}
           h(x) \in R_j & \mathrm{if} \ f(x) \in R_j \ \mathrm{for \ some} \ j,  \\
           h(x) = f(x) & \mathrm{otherwise}. \\
    \end{array}
    \right.
  \end{equation}
Then the problem
\begin{equation}\label{eq:continuous_extension}
\min_{h \in \HH_{\R,\,\UU}} \ns{h}_{\HH(\XX)} 
\end{equation}
has a unique solution denoted by $s_{\R, \UU}$. Moreover, when $\UU$ is the closure of~$\{ x_n \}$,
\begin{equation*}
  s_{\R, n} \xrightarrow{\HH(\XX)} s_{\R,\UU}\,.
\end{equation*}
\end{proposition}

In particular, when $\{ x_n \}$ is dense in $\XX$, then $\UU = \XX$ and
$(s_{\R,\,n})_{n \geq 1}$ converges to $s_{\R,\,\XX}$, which is the minimal-norm element of the set
$\HH_{\R,\, \XX}$.

The next proposition tells us that the interpolation error on~$\XX_0$ can be
bounded by a term that depends on the norm of $s_{\R,\,\XX}$.
\begin{proposition}\label[proposition]{prop:convergence_bounds_model_cuvette}
  For any $x \in \XX_0$ and $n \geq 1$,
  \begin{equation}\label{eq:power_function_slack_cuvette}
    \abs{f(x) - s_{\R,\,n}(x)  }
    \leq 2 \sigma_{n, \, 0}(x) \ns{s_{\R,\,\XX}}_{\HH(\XX)},
  \end{equation}
  where $\sigma_{n, \, 0}$ is the power function obtained using only points in
  $\XX_0$ for predictions {and with the convention that~$\sigma_{n, \, 0}^2(x) = k( x, \, x)$
  if there are no points}.
\end{proposition}
This yields the following error bounds when the design is dense.
\begin{proposition}\label[proposition]{prop:h_bounds}
  Suppose that~$\XX$ is compact, $\{x_n\}$ is dense, and that $k$ has regularity $\alpha > 0$.
  Let $B \subset \XX_0$ {satisfy} \cref{assump:domain}. Then, for all $n \geq 1$,
    \begin{equation}\label{eq:rate_cuvette}
      \ns{f - s_{R,\,n}}_{\Lset^{\infty}(B)}  
      \lesssim h_{n}^{\alpha} \ns{s_{\R,\,\XX}}_{\HH(\XX)}.
    \end{equation}
    
  Let $d(y, \, A)$ be the distance of $y \in \RR$ to $A \subset
    \RR$. For $j \geq 1$, $x \in \XX_j$, and for all $n \geq 1$:
    \begin{equation*}
      d(s_{\R,\,n}(x), \, R_j)  
      \lesssim h_{n}^{\alpha} \ns{s_{\R,\,\XX}}_{\HH(\XX)} \quad \mathrm{if} \ \alpha < 1,
    \end{equation*}
    \begin{equation*}
      d(s_{\R,\,n}(x), \, R_j)  
      \lesssim \left( \abs{\ln(h_n)} + 1 \right)^{1/2} \, h_{n}\, \ns{s_{\R,\,\XX}}_{\HH(\XX)} \quad \mathrm{if} \ \alpha = 1,
    \end{equation*}
    and
    \begin{equation*}
      d(s_{\R,\,n}(x), \, R_j)  
      \lesssim h_{n} \ns{s_{\R,\,\XX}}_{\HH(\XX)} \quad \mathrm{if} \ \alpha > 1\,,
    \end{equation*}
  where~$\lesssim$ denotes inequality up to a constant, that does not depend
  on $f$, $n$, $x$ or $(x_n)$.
\end{proposition}

Finally, we investigate the following question: how large {can the norm of~$f$ be}
compared to that of the approximation~$\ns{s_{\R,\,\XX}}_{\HH(\XX)}$?
\begin{proposition}\label[proposition]{prop:finite_smooth_norm_reduction}
  Suppose that $k$ has regularity $\alpha > 0$ and that there exists
  some $j \geq 1$ such that $\XX_j$ has a non-empty interior. We have
  \begin{equation}\label{eq:infty_norm_sobolev_convex}
    \sup_{h \in \HH_{\R,\, \XX}} \ns{h}_{\HH(\XX)} = +\infty,
  \end{equation}
  with $\HH_{\R,\, \XX}$ given by~\eqref{eq:continuous_constraints_convex} for $f \in \HH(\XX)$.
\end{proposition}
This result shows that the norm reduction obtained by approximating $f$
with relaxed interpolation constraints can therefore be arbitrarily high in the
finite-smoothness case.  A stronger version of
\cref{prop:finite_smooth_norm_reduction} for the special
case where $\R = [t, +\infty)$ can be derived, and shows 
that
$$
\sup_{h \in \HH_{\R,\, \XX}} \ns{h}_{\Lset^{\infty}(\XX)} = +\infty\,.
$$

Overall, no matter the element of $\HH_{\R,\, \XX}$ at hand,
\reGP converges to a function~$s_{\R,\,\XX}$ which:
coincides with $f$ on $\XX_0$,
satisfies $f(x) \in R_j \Leftrightarrow s_{\R,\,\XX}(x) \in R_j$ for all $x \in \XX$,
and is ``nicer'' than~$f$ in the sense of~$\ns{\cdot}_{\HH(\XX)}$.
Furthermore, \reGP yields error bounds carrying the norm
of $s_{\R,\,\XX}$, which can be arbitrarily smaller than the norm of~$f$
in the case of a finite-smoothness covariance function.

 \begin{remark}\label[remark]{remark:regp_uq}
 Note that $\sigma_n \leq \sigma_{n, \, 0}$ due to the projection residuals interpretation.
 Empirical and theoretical results about the \emph{screening effect} \citep[see,
 e.g.,][]{stein2010:_screening, bao2020:_screening}, suggests
 that~$\sigma_n \simeq \sigma_{n, \, 0}$, if $k$ has smoothness~$\alpha > 0$.
 In this case, observe that---no matter the element
 of~$\HH_{\R,\, \XX}$ at hand---the bound~\eqref{eq:power_function_slack_cuvette}
 is larger by only a small factor compared to~\eqref{eq:power_function}
 with~$f = s_{\R,\,\XX}$.
 (However, to the best of our knowledge, no result exists concerning the screening effect
 for arbitrary designs.)
 \end{remark}

\begin{remark}
\eqref{eq:infty_norm_sobolev_convex}
does not hold in general
for infinitely smooth covariance
functions. For instance, \citet[][Corollary
3.9]{steinwart:_rkhs_gaussian}
show that~$\HH_{\R,\, \XX} = \{ f \}$
if the interior of $\XX_0$ is not empty
and~$k$ is the squared-exponential covariance
function, i.e., \eqref{eq:matern-cov}, with $\nu \to \infty$.
\end{remark}

\cref{app:more_convergence} offers insights into the convergence
of relaxed interpolation for functions that lie outside the RKHS
attached to the covariance. Additionally, we present results on the
behavior of the approximation when the covariance parameters are not
fixed.

\section{Choice of the Relaxation {Range}}\label{sec:threshold_choice}

\subsection{Towards Goal-Oriented Cross-Validation}\label{sec:cv}

The framework of \reGP makes it possible to predict a function $f$
from point evaluations of $f$. Suppose we are specifically interested
in obtaining good predictive distributions in a range
  $Q\subset\RR$ of function values,
and accept degraded predictions outside this range. %
To achieve this goal, the idea of \reGP is to relax interpolation
constraints. %
Naturally, it makes sense to relax interpolation constraints outside
the {range of interest~$Q$}, but it {can} happen that relaxing {all the} interpolation
constraints {outside~$Q$} does not improve predictive distributions on~$Q$. %
Therefore, the question arises as to how to automatically select a
range $R$ in $\RR \setminus Q$, on which interpolation constraints
should be relaxed.

In the following, we put $R^{(0)} = \RR \setminus Q$, and we view the
relaxation {range}~$R$ as a parameter of the \reGP model, which has to be
chosen in $R^{(0)}$ along with the parameters $\theta$ of the
underlying GP $\xi$. A first idea for the selection of $R$ is to rely
on the standard leave-one-out cross-validation approach to select the
parameters of a GP \citep{dubrule1983:_cv_kriging,
  rasmussen06:_gauss_proces_machin_learn,
  zhang10:_krigin_cross_valid}. Using the formalism of~\emph{scoring
  rules} \citep[see, e.g.,][]{gneiting:2007:scoring,
  petit2021:_gaussian_process_model_selection}, selecting parameters
by a leave-one-out approach amounts to minimizing a selection
criterion written as
\begin{equation}\label{eq:s_cv}
  J_n(R) = \frac{1}{n} \sum_{i = 1}^n S \left(P_{R,\, n,\, -i},\, f(x_i) \right),
\end{equation}
where $P_{R,\, n,\, -i}$ is the \reGP predictive distribution with data
$\z_{n,-i} = (z_1,\,\ldots, z_{i-1},\, z_{i+1},\, \ldots, z_n)$ and
relaxation {range}~$R$. %
The function $S$ in~\eqref{eq:s_cv} is a scoring rule, that is, a function
$S: \Pcal \times \RR \to \RR\cup\left\{-\infty,
  +\infty\right\}$, acting on a class~$\Pcal$ of probability
distributions on~$\RR$, such that $S(P,\,z)$ assigns a loss for choosing
a predictive distribution $P\in\Pcal$, while observing $z\in\RR$.
Scoring rules make it possible to quantify the quality of probabilistic
predictions.

Since the user is not specifically interested in good predictive
distributions in $R^{(0)}$, validating the model on $R^{(0)}$ should
not be a primary focus.  However, simply restricting the
sum~\eqref{eq:s_cv} by removing indices~$i$ such that
$f(x_i) \in R^{(0)}$ would make it impossible to assess if the model
is good at predicting that $f(x) \in \R^{(0)}$ for a given
$x \in \XX$.  For instance, in the case of minimization, with
$Q = (-\infty, \, t^{(0)})$ and $R^{(0)} = [\,t^{(0)}, + \infty)$, it is
important to identify the regions corresponding to $f$ being above
$t^{(0)}$, even if we are not interested in accurate predictions above
$t^{(0)}$, because we expect that an optimization algorithm should
avoid the exploration of these regions.

In the next section, we propose instead to keep the whole
leave-one-out sum~\eqref{eq:s_cv}, but to choose a scoring rule~$S$
that serves our goal-oriented approach.

\subsection{Truncated Continuous Ranked Probability Score}\label{sec:tcrps}

An appealing class of scoring rules for goal-oriented predictive
distributions is the class of weighted scoring rules for binary
predictors \citep{gneiting:2007:scoring, matheson1976:_scoring}, which
may be written as
\begin{equation}\label{eq:w_binary}
  S\left( P,\, z \right) = \int_{- \infty}^{+\infty} s(F_P(u), \one_{z \leq u})\, \mu(\du)\,,
\end{equation}
where
$s\colon \left[0, \, 1\right] \times \{0, \, 1\} \to \RR \cup \{ -
\infty, + \infty \}$ is a scoring rule for binary predictors,
$F_P$ is the cumulative distribution function of~$P$, and
$\mu$ is a Borel measure on $\RR$. %
A well-known instance of \eqref{eq:w_binary} is the continuous ranked
probability score \citep{gneiting2005:_crps} written as
$$
S^\CRPS(P, \,z) = \int_{-\infty}^{+\infty} \left(F_P(u)- \one_{z \leq u} \right)^2 \du\,,
$$
which is obtained by choosing the Brier score for $s$ and the Lebesgue
measure for $\mu$.

For the case where we are specifically interested in obtaining good
predictive distributions in a range of interest~$Q \subset \RR$, we
propose to use the following scoring rule, which we call
\emph{truncated continuous ranked probability score} (tCRPS):
\begin{equation}\label{eq:tcrps}
  S_Q^{\tCRPS}(P,\,z)
  \;=\; \int_Q \left(F_P(u)- \one_{z \leq u} \right)^2 \du.
\end{equation}
This scoring rule, proposed by \cite{lerch:2013:comparison} in a
different context, reduces to~$S^\CRPS$ when $Q = \RR$. %
It can be seen as a special case of the weighted CRPS
\citep{matheson1976:_scoring, gneiting:2007:scoring,
  gneiting:2011:comparing}, in which the indicator function~$\one_Q$
plays the role of the weight function---in other words, the
measure~$\mu$ in~\eqref{eq:w_binary} has density~$\one_Q$ with respect
to Lebesgue's measure.

Consider for instance the case
$Q = \left( -\infty,\, t^{(0)} \right)$: %
\begin{equation*}
  S_Q^{\tCRPS}(P,\,z)
  \;=\; \int_{-\infty}^{t^{(0)}} \left(F_P(u)- \one_{z \leq u} \right)^2 \du.
\end{equation*}
The upper endpoint~$t^{(0)}$ of the range will be referred to as the
validation threshold. %
Note that, in this case, $S_Q^{\tCRPS}(P,\, z)$~does not depend on the
specific value of~$z$ when $z$ is above the validation threshold. %
This scoring rule is thus well
suited to the problem of measuring the performance of a predictive
distribution in such a way as to fully assess the goodness-of-fit of the
distribution when the true value is below a threshold, and only ask that
the support of the predictive distribution is concentrated above the threshold when the
true value is above the threshold.

We provide in \cref{app:tcrps_expr} some properties of the
scoring rule~\eqref{eq:tcrps} and closed-form expressions for the case
where~$Q$ is an interval (or a finite union of intervals) and $P$~is
Gaussian. %
To the best of our knowledge, these expressions are new.

\subsection{Choosing the Relaxation {Range} using the tCRPS Scoring Rule}
\label{sec:choos-relax-set}

Given a range of interest~$Q$, the tCRPS scoring rule makes it
possible to derive a goal-oriented leave-one-out selection criterion {for the relaxation range~$R$},
{which} we call the LOO-tCRPS criterion:
\begin{equation}\label{eq:loo_tcrps}
  J_n \left( R \right) = \frac{1}{n} \sum_{i = 1}^n S_Q^\tCRPS
  \left( P_{R,\,n,\, -i} ,\, f(x_i) \right)\,.
\end{equation}

Using~\eqref{eq:loo_tcrps}, we suggest the following procedure to
select a \reGP model. First, choose a sequence of nested candidate
relaxation {ranges}
$R^{(0)} \supset R^{(1)} \supset \cdots \supset
R^{(G-1)}=\emptyset$. The next step is the computation of
$J_n( R^{(g)})$, $g=0,\,\ldots,\, {G - 1}$, which involves the predictive
distributions $P_{R^{(g)},\,n,\, -i}$.

In principle, \eqref{eq:rgpi_problem} should be solved again each time
a data point $(x_i,\,z_i)$ is removed, to obtain a
pair~$(\hat{\theta}_{n, -i}^{(g)},\, \z_{n, -i}^{(g)})$ and then the
corresponding \reGP distribution~$P_{R^{(g)},\,n,\, -i}$. %
To alleviate computational cost, a simple idea is to rely on the fast
leave-one-out formulas \citep{dubrule1983:_cv_kriging} for Gaussian
processes: for each set~$R^{(g)}$, solve~\eqref{eq:rgpi_problem} to
obtain $\hat{\theta}_n^{(g)}$ and
$\zn^{(g)}=(z_1^{(g)},\,\ldots,\, z_{n}^{(g)})\tr$, and then compute
the conditional distributions
$\xi(x_i) \mid \{\xi(x_j) = z_{j}^{(g)}, j \neq i\}$, where
$\xi \sim \GP(\mu,\, k)$, and where $\mu$ and $k$ have
parameter~$\hat{\theta}_n^{(g)}$, using the fast leave-one-out
formulas. By doing so, we neglect the difference between
$\hat{\theta}_{n, -i}^{(g)}$ and $\hat{\theta}_n^{(g)}$ and the
difference between $\z_{n, -i}^{(g)}$ and the vector
$(z_1^{(g)},\,\ldots, \, z_{i-1}^{(g)},\,z_{i+1}^{(g)}, \,\ldots,\,
z_{n}^{(g)})\tr$.
The procedure ends
by choosing the relaxation {range}~$R^{(g)}$ that achieves the best
LOO-tCRPS value.

\cref{fig:gpr_t} illustrates the selection of the relaxation {range}
used in \cref{fig:rb_slack}. %
\cref{algo:regp} summarizes the general \reGP procedure using LOO-tCRPS
for selecting relaxation range in a given
list of candidates for relaxation range. \cref{sec:bench_optim_methodo} and \cref{app:levelset}
give specific implementations for Bayesian optimization 
and the problem of estimation of an excursion set.

\begin{figure}
  \centering
  \scalebox{0.9}{\input{figures/slack_gplike_t_star.pgf}}
  \caption{%
    Illustration of the choice of a relaxation {range}. The range of
    interest~$Q$ is determined by the threshold $t^{(0)}$. The
    relaxation {range}~$R$ corresponding to the region above $t$ has been obtained by the procedure described in
    \cref{sec:choos-relax-set}.
    }
  \label{fig:gpr_t}
\end{figure}

\begin{algorithm}
   \caption{%
     \reGP with automatic selection of the relaxation range.}
   \label{algo:regp}
\begin{algorithmic}
   \STATE {\bfseries Input:} Data $(\xn, \, \zn)$; a range of interest $Q$;
   and a list $\RR \setminus Q = R^{(0)} \supset \dots \supset R^{(G - 1)} = \emptyset$
   of relaxation range candidates.
   \FOR{$g=0$ {\bfseries to} $G - 1$}
   \STATE Obtain $\hat{\theta}_n^{(g)}$ and $\zn^{(g)}$ by solving~\eqref{eq:rgpi_problem} with $R^{(g)}$
   \STATE Compute~$J_n (R^{(g)})$ with~$Q$, $\hat{\theta}_n^{(g)}$, and $\zn^{(g)}$ using~\eqref{eq:loo_tcrps}
   \ENDFOR   
   \STATE {\bfseries Output:}  The pair $\hat{\theta}_n^{(g)}, \,
   \zn^{(g)}$ that minimizes~\eqref{eq:loo_tcrps}. 
\end{algorithmic}
\end{algorithm}

\subsection{An Example for the Estimation of an Excursion Set}\label{sec:tcrps_example}

We illustrate the method on the problem of estimating an excursion
set~$\{x \in \XX, \, f(x) \leq 0 \}$. We~consider the G10 optimization
problem used by \cite{regis2014:_constrained}, and focus on the
constraint $c_6 \leq 0$. %
Finding solutions satisfying the $c_6 \leq 0$ constraint using a GP
model is difficult, probably because the values of $c_6$ are very
bi-modal, as illustrated in \cref{fig:g10_type_problem}. %
However, \cite{feliot2017:_jogo} found that the difficulty could be
overcome by performing an ad-hoc monotonic transformation
$z\mapsto z^\alpha$, with $\alpha = 7$, on the constraint.

\begin{figure}
    \centering
    \resizebox{0.50\textwidth}{!}{{\input{figures/g10mod.pgf}}}%
    \resizebox{0.50\textwidth}{!}{{\input{figures/g10modmod.pgf}}}
    \vspace{-0.5cm}
    \label{fig:g10mod}
    \caption{%
      Left: Histogram of the values of the function $c_6$ from the G10
      problem.  Right: Same illustration but for the function
      $c_6^\alpha$, with $\alpha=7$.  The histograms are obtained from
      the values of the functions on a space-filling design of size
      $n = 100$.  On the left, the values are very separated and
      concentrated on two modes, yielding a function close to a
      piecewise constant function. After transformation, the
      phenomenon is mitigated.  }
  \label{fig:g10_type_problem}
\end{figure}

The estimation of an excursion set $\left\{ f \leq 0 \right\}$ involves
capturing precisely the behavior of $f$ around zero. Thus, we define a
range of interest $Q = ( - t^{(0)}, \, t^{(0)} )$ centered on zero, with $t^{(0)}$
sufficiently small (note that there may be no data in $Q$).
Then, we consider 
relaxation range candidates
$\R^{(g)} = (-\infty, -t^{(g)}\,] \cup [\, t^{(g)}, +\infty)$ with a
sequence of thresholds $t^{(0)} < \cdots < t^{(G - 1)}=+\infty$, and we
select $t^{(g)}$ by minimizing the LOO-tCRPS as described in the
previous section.

\cref{app:levelset} presents numerical experiments
on active learning of the set~$\{x \in \XX, \, c_6(x) \leq 0 \}$
with \reGP.
\cref{fig:g10_type_problem_slack} illustrates the relaxation
obtained using a value of~$t^{(0)}$ such that~$Q$ contains~$25\%$
of the observations of a design of size~$n = 300$ in the experiments
presented in \cref{app:levelset}.
Here, the LOO-tCRPS chooses
$t^{(g)}= t^{(0)}$.
Observe that the
transformation after relaxation resembles the
transformation $z \mapsto z^\alpha$ proposed by
\cite{feliot2017:_jogo}.
{Applying} the \reGP framework on the
transformed function $c_6^\alpha$ (details omitted for brevity), we
find that the LOO-tCRPS chooses a large~$t^{(g)}$ such that the
interpolation constraints are relaxed for only a few observations.

\begin{figure}
    \centering
    \scalebox{1.1}{\input{figures/g10mod_slack.pgf}}
  \caption{%
  A \reGP fit of $c_6$, where the relaxation
  thresholds have been selected by LOO-tCRPS.
  The observations~$\zn$ are shown on the $x$-axis, whereas 
  the ``relaxed'' observations~$\zn^{\star}$ are represented on the $y$-axis.
  The black squares represent interpolated observations and the blue points show relaxed observations.
  Moreover,
  the green lines represent the value zero, and
  the {black} lines represent $\pm \ t^{(g)}$, with $t^{(g)}$ chosen to be $t^{(0)}$ by the LOO-tCRPS.}
  \label{fig:g10_type_problem_slack}
\end{figure}

\section{Application to Bayesian Optimization}\label{sec:rgpi_bo}

\subsection{Efficient Global Optimization with Relaxation}\label{sec:ego_r}

The first motivation for introducing \reGP models is Bayesian
optimization, where obtaining good predictive distributions over
ranges corresponding to optimal values is a key issue. In this
article, we focus more specifically on the minimization problem
\begin{equation*}
  \min_{x\in\XX} f(x)\,,
\end{equation*}
where $f$ is a real-valued function defined on a compact set
$\XX\subset\RR^d$, but the methodology can be generalized
to constrained and/or multi-objective formulations.

Given $f$, our objective is to construct a sequence of evaluation
points ${x_1, x_2}\,\ldots \in\XX$ by choosing each point ${x_{n+1}}$ as
the maximizer of the expected improvement criterion~\eqref{eq:ei}
computed with respect to the \reGP predictive distribution, with a
relaxation {range} $\R_n = \left[ t_n, +\infty \right)$. %
More precisely, the sequence~${(x_n)}$ is constructed sequentially using
the rule
\begin{equation}
  \label{eq:ei-gp-r}
  {x_{n+1}} \;=\; \argmax_{x \in \XX}\; \EE_n\left( \left( m_n - \xi(x) \right)_{+} \right)\,,
\end{equation}
where $m_n = {f(x_1)\wedge \cdots \wedge f(x_n) }$, and $\EE_n$ is the
expectation under the \reGP predictive distribution with relaxation
{range}~$R_n$ and data~$\z_n = {(f(x_1),\,\ldots,\, f(x_n))\tr }$.

As in \cref{sec:choos-relax-set}, the relaxation threshold $t_n$
at iteration $n$ is chosen using the LOO-tCRPS
criterion~\eqref{eq:loo_tcrps} among candidate values
\begin{equation}
  \label{eq:thresholds-seq}
  t_n^{(0)} < t_n^{(1)} < \cdots < t_n^{{(G - 1)}}\,,
\end{equation}
where $t_n^{(0)}$ is the validation threshold, which delimits the
range of interest~$Q_n = \bigl( -\infty, t_n^{(0)} \bigr)$ used at
iteration~$n$. %
In the following, the optimization method just described will be
called efficient global optimization with relaxation (EGO-R), in
reference to the EGO name proposed by \cite{jones1998:_ego}.

Implementation specifics are detailed in \cref{sec:bench}. In
the next section, we show that using the EI criterion with a \reGP model
yields a convergent algorithm.

\subsection{Convergence of EGO-R with Fixed Parameters and Varying
  Threshold}\label{sec:theory_bo}

In this section, we extend the result of
\cite{vazquez_and_bect2010:_convergence_ei} and show the convergence of
the EGO-R algorithm, in the case where the predictive distributions
derive from a zero-mean Gaussian process with fixed covariance function.

We suppose that $\XX \subset \RR^d$ is a compact domain and that
$k: \XX \times \XX \to \RR$ is continuous, strictly positive-definite,
and has the NEB (no-empty ball) property
\citep{vazquez_and_bect2010:_convergence_ei}, which says that the
posterior variance cannot go to zero at a given point if there is no
evaluation points in a ball centered on this point.  In other words, the
NEB property requires that the posterior variance $\sigma_n^2(x)$ at
$x\in\XX$ remains bounded away from zero for any  $x$  not in the closure of
the sequence of points ${(x_n)}$ evaluated by the optimization algorithm. 
A stationary covariance function with smoothness $\alpha > 0$ satisfies
the NEB property \citep{vazquez_and_bect2010:_convergence_ei}, whereas
the squared-exponential covariance function does not
\citep{vazquez2010:_pointwise_consistency}.

\begin{proposition}\label[proposition]{prop:convergence_slack_ei_neb_varying_t}
  Let $k: \XX \times \XX \to \RR$ be a continuous strictly
  positive-definite covariance function that satisfies the NEB
  property, $\HH(\XX)$ the corresponding RKHS and $f \in \HH(\XX)$. %
  Let $n_0 > 0$. %
  Let ${\left( x_n \right)_{n \geq 1} }$ be a sequence in~$\XX$ such
  that, for each $n \geq n_0$, ${x_{n + 1}}$ is obtained
  by~\eqref{eq:ei-gp-r} with~$t_n > m_n$. %
  Then the sequence ${\left( x_n \right)_{n \geq 1} }$ is dense in~$\XX$.
\end{proposition}

\cref{prop:convergence_slack_ei_neb_varying_t} implies the
convergence of EGO-R with a fixed threshold $t > \min_{i\leq n_0} f({x_{i}})$.  In
this case, the theoretical insights from \cref{sec:theory_base}
suggest a faster convergence might be achieved due to the improved error
estimates~\eqref{eq:power_function_slack_cuvette}
and~\eqref{eq:rate_cuvette} in a neighborhood of the global minimum.

The convergence of EGO-R also holds in the case of a varying relaxation
{range} $\R_n = \left[t_n, + \infty\right)$, with $t_n > m_n$, and in
particular when $t_n$ is selected at each step using the LOO-tCRPS
criterion~\eqref{eq:loo_tcrps} with a validation
threshold~$t_n^{(0)} > m_n$.
In this case, 
the norm term in~\eqref{eq:power_function_slack_cuvette}
gets smaller if $\left( t_n \right)_{n \geq 1}$ is decreasing.

\subsection{Optimization Benchmark}\label{sec:bench}

In this section, we run numerical experiments to demonstrate the
interest of using EGO-R instead of EGO for minimization problems.
\cref{app:bench_ucb10} presents a similar study based on the
UCB  sampling criterion \citep{cox1992statistical, srinivas2010:_ucb}.
\cref{app:levelset} considers the task of
estimating an excursion set.

\subsubsection{Methodology}\label{sec:bench_optim_methodo}

In practice, we must choose the sequence of
thresholds~\eqref{eq:thresholds-seq}. The validation threshold
$t_n^{(0)}$ should be set above $m_n$ to ensure there is enough data to
carry out the validation. We propose {three} different heuristics
{to this end}: a) a
\emph{constant} heuristic, where $t_n^{(0)}$ remains constant throughout
the iterations and is set to an empirical quantile of an initial
data set constructed prior to running EGO-R, b) a \emph{concentration}
heuristic, where $t_n^{(0)}$ corresponds to an empirical quantile of $\zn$,
and c) a \emph{spatial}
heuristic, where~$t_n^{(0)}$ corresponds to a spatial quantile estimate of the function~$f$.
By spatial quantile, we mean a quantile of $f(U)$ with $U$ uniformly distributed
over the domain~$\XX$.

In the case of the constant heuristic, we set $t_n^{(0)}$ to the
$\alpha$-quantile of the function values on an initial design, which
is typically built to cover $\XX$ as uniformly as possible using, for example,
maximin Latin hypercube sampling \citep{mckay2000comparison}. In this
article, the
numerical experiments were conducted with $\alpha = 0.25$.

For the spatial heuristic,
the spatial quantile is estimated by building a
one-nearest-neighbor regression model trained on
the data~$(\xn, \, \zn)$, and then
taking an $\alpha$-quantile of its predictions
on a uniform sample on~$\XX$. The numerical value~$\alpha = 0.25$
is also used here.
This can be viewed as a refinement of the constant heuristic.

For the concentration heuristic, we consider the
$\alpha$-quantile of the values of $f$ at the points visited by the
algorithm (again with $\alpha = 0.25$). As the optimization algorithm
progresses, the evaluations are expected to cluster around the
global minimum. Thus, $t_n^{(0)}$ will approach the minimum
value, and the validation range $Q_n=(-\infty,\, t_n^{(0)})$ 
will shrink. {As a result}, since better predictive
distributions are expected in this range, a better convergence may be achieved.

{All three} heuristics are supported by the idealized
convergence result from the previous section.  Proposing alternative
{heuristics, or
conducting a theoretical comparison of their performances}, is out of the scope of this article.

For a given $t_n^{(0)}$, the candidate relaxation thresholds
$t_{n}^{(g)}$, $g={0},\,\ldots,\, {G - 1}$, are chosen such that $t_{n}^{(g)} - m_n$ ranges logarithmically
from~$t_n^{(0)} - m_n$ to
$\max f({x_i}) - m_n$ (with $G=10$ in the
experiments below). %
We adopt the convention that~$t_n^{(G - 1)} > \max f(x_i)$, so that
the $G$-th model is a standard GP.

To assess the performances of EGO-R with {these three} heuristics, we compare them to the standard EGO
algorithm. For all {four} algorithms, we use an initial design of
size {$n_0 = 10d$}, and we consider GPs with {a} constant mean {function} and a Matérn
covariance function with regularity $\nu = 5/2$. The maximization of
the sampling criteria~\eqref{eq:ei} and~\eqref{eq:ei-gp-r} is performed
using a sequential Monte Carlo approach
\citep{benassi12:_bayes_monte_carlo_, feliot2017:_jogo}.

The optimization algorithms are tested against a benchmark of test
functions from \cite{surjanovic13:_virtual_librar_simul_exper}
summarized in \cref{table:bench_problems}, with
{$n_{\mathrm{rep}} = 100$} (random) repetitions, and a budget of
$n_{\mathrm{tot}} = 300$ evaluations for each repetition.  This
benchmark is partly inspired by \cite{jones1998:_ego} and
\cite{merrill2021:_empirical_bo}. In particular, we also use a
log-version of the Goldstein-Price function as \cite{jones1998:_ego}.

\begin{table}
\caption{Optimization benchmark.}
\centering
\begin{tabular}{| c | c |}
\hline
Problem  &  $d$ \\
\hline
Branin & $2$  \\
\hline
Six-hump Camel & $2$ \\
\hline 
Three-hump Camel & $2$ \\
\hline
Hartman & $3$, $6$  \\
\hline
Ackley & $4$, $6$, $10$ \\
\hline
Rosenbrock & $4$, $6$, $10$  \\
\hline
Shekel & $5$, $7$, $10$ \\
\hline
Goldstein-Price & $2$ \\
\hline
Log-Goldstein-Price & $2$ \\
\hline
Cross-in-Tray & $2$ \\
\hline
Beale & $2$ \\
\hline
Dixon-Price & $4$, $6$, $10$ \\
\hline
Perm & $4$, $6$, $10$ \\
\hline
Michalewicz & $4$, $6$, $10$ \\
\hline
Zakharov & $4$, $6$, $10$ \\
\hline
\end{tabular}
\label{table:bench_problems}
\end{table}

To evaluate the algorithms, we construct for each test function a
list of target values, defined as spatial quantiles of the function
\citep[estimated with a subset simulation algorithm; see,
e.g.,][]{au2001:_estimation}. %
The probability levels associated with this list of spatial quantiles
define a non-linear scale that we use instead of raw function
values. %
We plot in this scale the 10\%/50\%/90\% quantiles of
the best function value found, as a function of~$n$. %
We also represent, for each probabilty level, 
the fraction of runs that reach the corresponding target.

\subsubsection{Findings}

\begin{figure}
\vspace{-0.8cm}
\begin{centering}
\resizebox{0.50\textwidth}{!}{\includegraphics{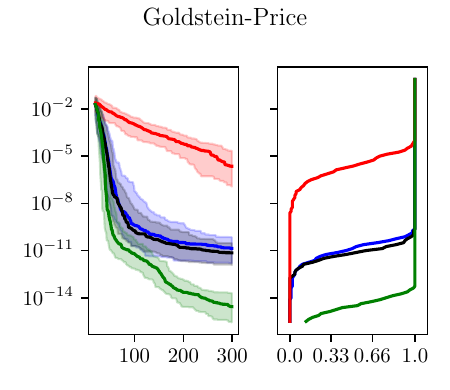}}%
\resizebox{0.50\textwidth}{!}{\includegraphics{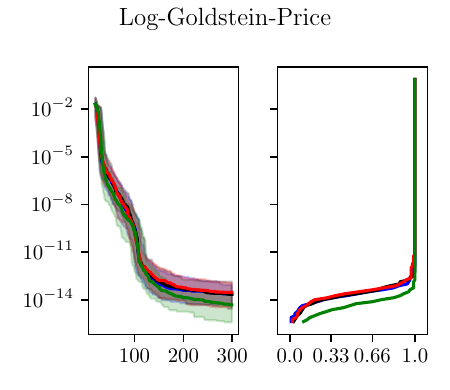}}\\[0em]
\resizebox{0.50\textwidth}{!}{\includegraphics{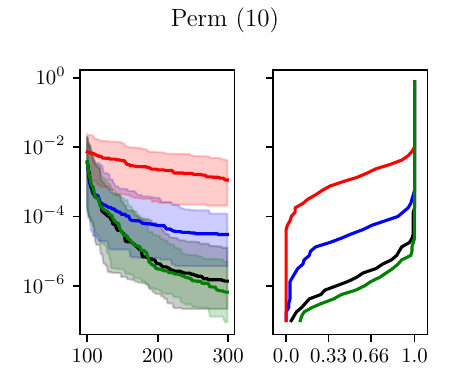}}%
\resizebox{0.50\textwidth}{!}{\includegraphics{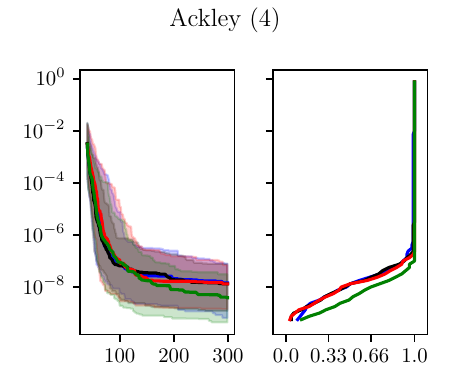}}\\[0em]
\resizebox{0.50\textwidth}{!}{\includegraphics{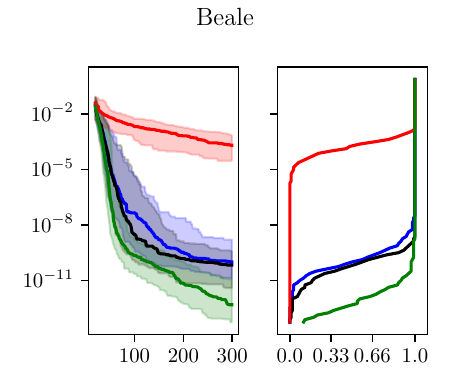}}%
\resizebox{0.50\textwidth}{!}{\includegraphics{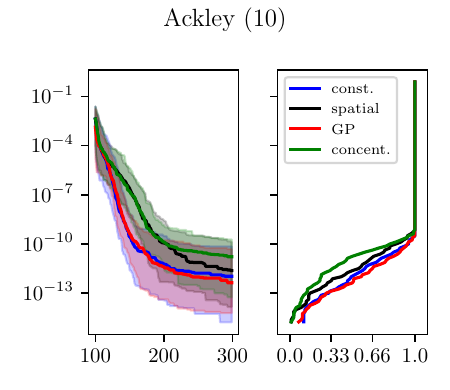}}%
\vspace{-1.0cm}
\end{centering}
\caption{%
    For each case, the left plot shows the evolution of quantiles of
    the best function value against~$n$, %
    and the right plot represents on the $x$-axis the fraction of runs
    that reach each target value. %
    Both plots use the probability level of the target as a scale for
    the $y$-axis. %
    Left plot: solid lines represent the median; shaded
    areas are deliminated by the~$10\%$ and~$90\%$ quantiles. %
    Both plots: EGO (red), %
    EGO-R + constant heuristic (blue), %
    EGO-R + concentration heuristic (green), %
    EGO-R + spatial heuristic (black). %
  }
\label{fig:results_exposition_figure}
\end{figure}

The full set of results is provided in \cref{app:bench}. In
\cref{fig:results_exposition_figure}, we present a representative subset of
these results.

First, observe in \cref{fig:results_exposition_figure} that the EGO-R
methods outperform EGO
significantly on functions that are difficult to model with stationary GPs,
such as Goldstein-Price, Perm ($10$), and Beale. %
In~these cases, relaxing the constraints for the
highest observations proves beneficial. %
Notice also that, among the three EGO-R variants, the one based on
the concentration heuristic converges faster in such situations. %
(Approximately half of the cases of our benchmark fall into this category;
see \cref{app:bench} for details.) %

In the other cases (Log-Goldstein-Price, Ackley ($4$) and Ackley
($10$) on \cref{fig:results_exposition_figure}) the EGO-R methods
perform similarly to EGO. %
This corresponds to functions where classical stationary GP modeling
already provides good predictions. %
In these cases, the LOO-tCRPS criterion for selecting the relaxation
range detects that the largest values also contribute to predicting near the
minimum, resulting in little to no relaxation being necessary. %
The concentration heuristic sometimes provides a small advantage on
such functions (see Log-Goldstein-Price and Ackley~($4$)), but
occasionally degrades performances compared to EGO and the other
EGO-R variants, %
as seen with the Ackley ($10$) function. %
A closer examination shows that the
concentration heuristic sometimes gets temporarily trapped in a
local minimum. %
We explain this by the fact that the \reGP model with the
concentration heuristic can become overly predictive in a small region
around the local minimum, but underestimate the function variations
elsewhere (the variance of the predictive distributions above
$t_n^{(0)}$ is too low, and the optimization algorithm does not
explore unknown regions sufficiently). %
In this sense, the constant and spatial heuristics can be considered
more conservative than the concentration heuristic.

{Finally}, it is useful to compare the performance of the EGO-R
algorithms on the Goldstein-Price function with the performance of the
EGO algorithm on the Log-Goldstein-Price function. Using \reGP modeling
allows for performance comparable to that achieved with a logarithmic
transformation, but in an automatic manner. This is illustrated in
\cref{fig:goldstein_transform}, where the (non-parametric) transform learned by \reGP
resembles a logarithmic transformation.

\begin{figure}
  \begin{center}
  \scalebox{1.0}{\input{figures/goldstein_transform.pgf}}
  \end{center}
  \caption{%
  A \reGP fit to the Goldstein-Price function with $n = 30$ points,
  with $\R = \left[10^3, \, + \infty\right)$.
  The observations $\zn$ are shown on the $x$-axis, whereas 
  the relaxed observations $\zn^{\star}$ are represented on the $y$-axis.}
  \label{fig:goldstein_transform}
\end{figure}

\subsubsection{{Running time of \reGP}}\label{sec:running_time}

Training a \reGP model takes longer than training a standard GP for
  several reasons. First, the optimization problem in~\eqref{eq:rgpi_problem} is an
  extension of~\eqref{eq:likelihood} and involves a higher dimension of the search space, as
  the number of relaxed observations increases, which likely increases
  the time required to solve it. Second, the automatic selection
  procedure described in \cref{sec:threshold_choice} involves testing $G$ relaxation
  ranges, making \reGP at least $G$ times more costly to compute.

  For~$G = 10$, \cref{fig:running_time} provides a quantitative assessment of the
  increase in computational time as a function of the number of
  evaluations~$n$, across four representative test cases. With automatic
  selection of the relaxation {range}, training a \reGP model is,
  roughly speaking, between $G = 10$ and~$1000$ times more expensive than
  training a standard GP.
  This trend holds for the other test functions
  in \cref{table:bench_problems} (figures omitted for brevity). Among the heuristics,
  the concentration heuristic is the most computationally expensive
  version of \reGP in this benchmark. This is because the number of
  observations outside the range of interest---and thus the maximum number
  of relaxed observations---grows linearly with~$n$. In contrast, for the
  spatial and constant heuristics, the computational overhead remains
  around a factor of $10$, which suggests that the primary cost increase
  comes from having to test~$G$ relaxation ranges.

Note that the procedure described in \cref{sec:choos-relax-set} involves an exhaustive search
over the set of relaxation range candidates.
Assessing the performance of \reGP with a smaller set of candidates, or using adaptive
strategies such as dichotomies, could help reduce computation time.
Note also that a factor~$G$ in computation time can be saved if the user already knows how to choose a relaxation range for his task. %
Another approach, left for future work, would be to solve~\eqref{eq:rgpi_problem}
with dedicated algorithms (see \cref{rem:numerics}).

Finally, it should be stressed that the computation time
is less of a concern for active learning tasks,
since evaluating the objective function---which can take several hours---is usually the time
bottleneck in this setting.

\begin{figure}
  \begin{centering}
\resizebox{0.48\textwidth}{!}{\input{figures/EI/running_time/goldsteinprice.pgf}}%
\resizebox{0.48\textwidth}{!}{\input{figures/EI/running_time/ackley6.pgf}}\\%
\resizebox{0.48\textwidth}{!}{\input{figures/EI/running_time/perm10.pgf}}%
\resizebox{0.48\textwidth}{!}{\input{figures/EI/running_time/zakharov10.pgf}}
\caption{%
Ratio between the time needed to train a \reGP model (including the
automatic selection of the relaxation range) and that required to train a GP during Bayesian optimization
with the EI criterion, across four representative test cases, as a function of $n$.
The curves show median ratios, while the shaded areas represent intervals bounded by the~$10\%$ and~$90\%$ quantiles.
Blue, green, and black represent \reGP
using the constant, concentration, and spatial heuristics, respectively.}
\label{fig:running_time}
\end{centering}
\end{figure}

\section{Conclusion}\label{sec:concl}

This article presents a new technique called reGP to build predictive
distributions for a function observed on a sequence of points. %
This technique can be applied when a user wants good predictive
distributions in a range of function values, for example below a given
threshold, and accepts degraded predictions outside this range. %
The technique relies on Gaussian process interpolation, and operates
by relaxing interpolation constraints outside the range of
interest. %
The relaxation range can be selected automatically, using a scoring
rule adapted to reGP models.
With respect to classical GP modeling, this goal-oriented technique
simply involves the choice of one additional key parameter, which is
the range of interest---i.e., the %
user only needs to specify a range of function values where good
predictions should be obtained. %

Such goal-oriented models can then be used in Bayesian sequential
search algorithms. %
Here we are {more specifically} interested in the problem of mono-objective optimization
and we propose to study the EI / EGO algorithm with such models. %
We provide theoretical guarantees for the convergence of the reGP-based
algorithm on the RKHS attached to the underlying GP covariance, %
and a numerical benchmark that shows very clear benefits of using
reGP models for the optimization of various functions.

A key element of the reGP approach is the definition of {the range of
interest, which is for instance} of the
form~$\left( -\infty, t^{(0)} \right)$ in a minimization problem. %
In some use cases the range will be provided by the user, but in
others it is desirable to set it automatically. %
{Three} simple heuristics have been proposed to
achieve this goal in our optimization benchmark, and it has been
observed that the choice of heuristic has an impact on the exploratory
behaviour of the resulting Bayesian optimization algorithm. %
Finding better heuristics, studying their properties, and assessing
their impact in Bayesian optimization applications, is an important
direction for future research.

More generally, the goal-oriented approach proposed in this article is
not limited to single-objective (Bayesian) optimization. %
The example from \cref{sec:tcrps_example} and the benchmark results of \cref{app:levelset} show that it is also
readily applicable, for instance, to level set estimation problems. %
\cref{app:noise} provides a proof of concept for extending the approach to noisy observations.
A thorough investigation of the best modeling choices and theoretical guarantees is left for future work.
Other extensions are possible but will require more work. %
Constrained and/or multi-objective optimization is another interesting
but challenging direction for future research: %
in this case the function of interest is multivariate (one objective
and several constraints, or several objectives), which requires
significant adaptations to the proposed methodology.

Another possible direction for future research would be to infer the
unknown function using the full predictive
distribution~\eqref{eq:joint_posterior}, of which \reGP is a mode
approximation. %
This distribution defines a Skew GP, for which various
inference methods have been proposed in the literature, using for instance
quasi Monte-Carlo sampling, linear elliptical slice sampling or Gibbs sampling %
\citep[see, e.g.,][]{da2012:_constrained, botev2017normal,
  benavoli2021unified, gessner2020integrals, takeno2023towards}. %
Adapting these ideas to the reGP setting may be relevant in a context
where uncertainty quantification is critical.

\acks{The authors thank the action editor and three anonymous
  reviewers for their insightful comments, which significantly
  improved both the content and the presentation of the article. This
  work was partially supported by the French Agence Nationale de la
  Recherche et de la Technologie (ANRT) through a CIFRE grant.}


\newpage

\appendix
\section{Properties of the Truncated CRPS}
\label{app:tcrps_expr}

We shall now write~\eqref{eq:tcrps} more explicitly for the case where
the range of interest is an interval $Q = (a, b)$,
$- \infty \leq a < b \leq +\infty$, %
and provide closed-form expressions for the case where, in addition,
the predictive distribution~$P$ is Gaussian.

\begin{remark}
  The value of the tCRPS for an interval~$Q = (a, b)$ remains
  unchanged if the interval is closed at one or both of its endpoints.
\end{remark}

\begin{remark}
  The value of the tCRPS for a finite (or countable) union of disjoint
  intervals follows readily from its values on intervals, since
  $Q \mapsto S_Q^{\tCRPS}(P,\,z)$ is $\sigma$-additive.
\end{remark}

We shall start by defining a quantity that shares similarities
with~\eqref{eq:ei}.
\begin{definition}
  $\EI_q^{\uparrow}(P, \,z) \;=\; \EE\left( \left( N_1 \vee \cdots
      \vee N_q - z \right)_+ \right)$ %
  with $N_j \stackrel{\mathrm{iid}}{\sim} P$.
\end{definition}

The following expressions hold for a general predictive
distribution~$P$.
\begin{proposition}\label[proposition]{prop:tcrps_general}
  Suppose that $P$ has a first order moment.
  \begin{itemize}
  \item Let $a, b \in \RR$ with $a \le b$. Then,
    \begin{align*}
      S_{a, \, b}^\tCRPS (P,\,z) \;=\;
      & \left( b \wedge z - a \right)_+
        \,+\, \EI^{\uparrow}_2(P,b) - \EI^{\uparrow}_2(P,a)
        \nonumber\\
      & \qquad \mskip 2mu - 2\, \one_{z \le b}\,
        \left( \EI^{\uparrow}_1(P,b) - \EI^{\uparrow}_1(P,a \vee z) \right).
    \end{align*}
  \item Let $b \in \RR$ and $N_1, N_2 \stackrel{\mathrm{iid}}{\sim} P$. Then,
    \begin{align*}
      S_{- \infty, \, b}^\tCRPS (P, \,z) \;=\;\,
      & b \wedge z \,+\, \EI^{\uparrow}_2(P,b) - \EE \left( N_1 \vee N_2 \right)
        \nonumber\\
      & - 2\, \one_{z \le b}\,
        \left( \EI^{\uparrow}_1(P,b) - \EI^{\uparrow}_1(P,z) \right).
    \end{align*}
  \item Finally, if $a \in \RR$, then
    \begin{align*}
      S_{a, \, +\infty}^\tCRPS (P, \,z) \;=\;\, S_{- \infty, \, -a}^\tCRPS (\underline{P} , \, -z), &  
    \end{align*}
    where $\underline{P}$ is the distribution of $- U$ if $U$ is
    $P$-distributed.
  \end{itemize}
\end{proposition}

Now, leveraging well-known analytic expressions \citep[see,
e.g.,][]{nadarajah_kotz2008:_exact_distribution_max_gaussians,
  chevalier_ginsbourger:_2013fast_computation_multi_points_ei}, we
have the following closed-form expressions in the Gaussian case.
\begin{proposition}{\citep{nadarajah_kotz2008:_exact_distribution_max_gaussians,
      chevalier_ginsbourger:_2013fast_computation_multi_points_ei}}\label[proposition]{prop:qei_gaussian} \mbox{}\\
  Suppose that $P = \mathcal{N}(\mu, \sigma^2)$ and let $\phi$ and
  $\Phi$ denote respectively the pdf and the cdf of the standard Gaussian
  distribution. Then
  \begin{itemize}
  \item
    $\EI_1^{\uparrow}(P, z) = \sigma\, h_1\left( \frac{\mu -
        z}{\sigma} \right)$, with
    \begin{equation*} h_1(t) = t\, \Phi (t) + \phi
      \left( t \right),
    \end{equation*}
  \item for $q \geq 2$, we have
    $\EI_q^{\uparrow}(P, z) = \sigma\, h_q\left( \frac{\mu -
        z}{\sigma} \right)$, where
    \begin{equation*}
      h_q(t) \;=\; q t\, \Phi_q(\delta_q^t; 0, D_q D_q\tr )
      \,+\, q\, \Phi \left( -t \right)^{q-1} \phi(t)
      \,+\, \frac{q(q -1)}{2 \sqrt{\pi}}\, \Phi_{q-1}(\delta_{q - 1}^t; 0, \tfrac{1}{2} B_q),
    \end{equation*}
    where $\Phi_q(\cdot; m, \Sigma)$ is the cdf of the multivariate
    $\mathcal{N}(m, \Sigma)$ distribution,
    \begin{equation*}
      B_q = 2\, \mathrm{diag}(0, \one_{q-2}\tr) + \one_{q-1} \one_{q-1}\tr,
    \end{equation*}
    $D_q$ is the matrix representing the linear map
    \begin{equation*}
      \RR^q \rightarrow \RR^q,\quad
      (y_1,\, \dots,\, y_q)\tr \mapsto (-y_1,\,  y_2 - y_1,\, y_3 - y_1,\, \dots,\, y_q - y_1)\tr,
    \end{equation*}
    and $\delta_q^{t} = (t, 0_{q-1}\tr)\tr$,
  \item finally for $q = 2$ we have
    \begin{equation*}
      \EE\left( N_1 \vee N_2 \right) = \mu + \frac{\sigma}{\sqrt{\pi}}.
    \end{equation*}
  \end{itemize}
\end{proposition}

The propriety of scoring rules is an important notion that formalizes
``well-calibration'' in the sense that a generating distribution must
be identified to be optimal on average.
\begin{definition}{\citep[see, e.g.,][]{gneiting:2007:scoring}}
  A scoring rule $S: \Pcal \times \RR \to \RR$ is said to be
  (strictly) proper with respect to $\Pcal$ if, for all
  $P_1, P_2 \in \Pcal$, the mapping $y \in \RR \mapsto S(P_1, \, y)$
  is $P_2$-quasi-integrable and the mapping
  \begin{equation*}
    P_1 \in \Pcal \mapsto S(P_1, \, P_2) := \EE_{U \sim P_2} \left( S \left( P_1 , \, U \right) \right)
  \end{equation*}
  admits $P_2$ as a (unique) minimizer.
\end{definition}

In the case of the truncated CRPS, simple calculations lead to
\citep{matheson1976:_scoring}:
\begin{equation*}
  S_Q^\tCRPS(P_1, \, P_2) \;=\; S_Q^\tCRPS(P_2, \, P_2) \,+\, \int_{Q}
  \left( F_{P_1}(u) - F_{P_2}(u) \right)^2\, \du.
\end{equation*}
It follows that $S_Q^\tCRPS$ is proper with respect to the class of all Borel
probability measures on~$\RR$ for any measurable $Q \subset \RR$, %
and is strictly proper with respect to the class of non-degenerate
Gaussian measures on~$\RR$ as soon as $Q$ has non-empty interior.

\clearpage

\section{{Theoretical Insights on \reGP}}\label{app:more_convergence}

This appendix provides further theoretical study of relaxed interpolation.
We start by recalling a well-known approximation result for the interpolation of
classes of functions outside the RKHS. 
Then, we give a corresponding adaptation for relaxed interpolation.
Finally, \cref{app:theory_params} presents results on approximation
and uncertainty quantification for \reGP with estimated parameters. The
notations are that of \cref{sec:theory_base}.

\subsection{{Escape Theorem for Gaussian Process Interpolation}}

In GP interpolation, error bounds are sometimes available for
interpolation of functions not belonging to the~RKHS.
Let~$q_n = \min_{i \neq j} \ns{x_i - x_j}/2$
be the separation distance
and~$\rho_n = h_n/q_n$ be the mesh ratio.
A sequence~$(x_n)_{ n\geq 1}$ is quasi-uniform if there exists a positive constant~$C_{\mathrm{qu}}$
such that~$q_n \leq h_n \leq C_{\mathrm{qu}} q_n$.
The following proposition is a reminder of the well-known ``escape''
theorem, which
provides error bounds
for standard Gaussian process interpolation
of classes of functions outside the RKHS attached to a kernel with regularity~$\alpha > 0$
(see Theorem~4.2 of \citealp{narcowich_et_al2006:_evasion_theorems}, and
Theorem~4.2 of \citealp{karvonen2020mleuq}, and Theorem~1 of \citealp{wynne_et_al__2023:_convergence_gp},
for recent extensions).
\begin{proposition}\label[proposition]{prop:standard_escape}
Let~$0 < \beta < \alpha < +\infty$ be such that~$f \in W_2^{\beta + d/2}(\XX)$
and assume that~$\XX$ satisfies \cref{assump:domain}.
If~$k$ has regularity~$\alpha$, then
\begin{equation*}
\ns{f - s_{\zn}}_{\Lset^{\infty}(\XX)}  \lesssim h_n^{\beta}  ( 1 + \rho_n^{\alpha - \beta} ) \ns{f}_{W_{2}^{\beta + d/2}(\XX) },
\end{equation*}
where~$\lesssim$ denotes inequality up to a constant, that does not depend
on $f$, $n$ or $(x_n)$. In particular, if~$(x_n)_{ n\geq 1}$
is quasi-uniform, we have
\begin{equation*}
\ns{f - s_{\zn}}_{\Lset^{\infty}(\XX)}  \lesssim n^{- \beta/d} \ns{f}_{W_{2}^{\beta + d/2}(\XX) }.
\end{equation*}
\end{proposition}

\subsection{{Convergence of Relaxed Interpolation of Functions Outside the RKHS}}

\cref{sec:convergence_regp} only considers functions belonging to the RKHS.
There are other situations in which \reGP is expected to offer advantages.
For the remainder of this appendix, assume~$f\colon \XX \to \RR$ is an arbitrary function.
For a relaxation {range}~$\R = \bigcup_{j = 1}^q R_j$, the subsets~$\XX_j \subset \XX$
are defined as in~\cref{sec:convergence_regp}.
\begin{definition}\label[definition]{def:relaxed_membership}
Let $\mathcal{F}$ be a set of functions~$\XX \to \RR$. We say that~$f$
is an~$R$-element of~$\mathcal{F}$ if
there exists a function~$g \in \mathcal{F}$ such that
$g$ and $f$ coincide  on~$\XX_0$
and $f(x) \in R_j \Leftrightarrow g(x) \in R_j$, for all~$j \geq 1$ and~$x \in \XX$.
\end{definition}
Naturally, if~$f \in \mathcal{F}$, then~$f$ is an~$R$-element of~$\mathcal{F}$.
Note also that~$f$ can be an~$R$-element of a space of continuous functions
and be discontinuous in the interior of an~$\XX_j$, for~$j \geq 1$.
When~$\mathcal{F} = \HH(\XX)$, this condition is equivalent to requiring
that the
set~$\HH_{\R,\,\XX}$, defined
by~\eqref{eq:continuous_constraints_convex}, is non-empty.

Suppose~$f \notin \HH(\XX)$ but~$f$ is an~$R$-element of~$\HH(\XX)$.
In this case, the error bounds~\eqref{eq:power_function} and~\eqref{eq:standard_linf_bound}
are invalid, whereas the conclusions of
\crefrange{prop:constrained_problem}{prop:h_bounds}
still hold.
(Indeed, observe that the \reGP predictor of~$f$ is that of a
function~$g$ given by \cref{def:relaxed_membership}.
The propositions apply to~$g$.)

Moreover, even if $f$ is not an~$R$-element of $\HH(\XX)$, error bounds can sometimes be stated on~$\XX_0$.
Recall that the space~$W_2^{\beta + d/2}(\XX)$ is an RKHS with a
continuous kernel for~$\beta > 0$.
If~$f$ is an~$R$-element of~$W_2^{\beta + d/2}(\XX)$,
then \cref{prop:constrained_problem}
ensures the existence of a minimum $W_2^{\beta + d/2}(\XX)$-norm
element~$s_{\R, \XX}^{(\beta)}$ of the set
of functions~$g \in W_2^{\beta + d/2}(\XX)$ coinciding
with~$f$ on~$\XX_0$ 
and such that~$f(x) \in R_j \Leftrightarrow g(x) \in R_j$, for all~$j \geq 1$ and~$x \in \XX$.
The following proposition is an adaptation
of \cref{prop:standard_escape}
for bounding the error of the \reGP predictor on~$\XX_0$.
It relies on membership in the sense of \cref{def:relaxed_membership}.
\begin{proposition}\label[proposition]{prop:regp_escape}
Suppose that~$\XX$ is bounded and measurable.
Let~$0 < \beta < \alpha < + \infty$ be such that~$f$ is an $R$-element of~$W_2^{\beta + d/2}(\XX)$
and~$B \subset \XX_0$ satisfy \cref{assump:domain}.
If~$k$ has regularity~$\alpha$, then
it holds that:
\begin{equation}\label{eq:escape_rate}
\ns{f - s_{R,\,n}}_{\Lset^{\infty}(B)}  \lesssim h_n^{\beta}  ( 1 + \rho_n^{\alpha - \beta} ) \ns{s_{\R, \XX}^{(\beta)}}_{W_{2}^{\beta + d/2}(\XX) },
\end{equation}
where~$\lesssim$ denotes inequality up to a constant, that does not depend
on $f$, $n$ or $(x_n)$.
In particular, if~$(x_n)_{ n\geq 1}$
is quasi-uniform, then it holds that:
\begin{equation}\label{eq:qu_escape_rate}
\ns{f - s_{R,\,n}}_{\Lset^{\infty}(B)}  \lesssim n^{- \beta/d} \ns{s_{\R, \XX}^{(\beta)}}_{W_{2}^{\beta + d/2}(\XX) }.
\end{equation}
\end{proposition}
Contrasting \cref{prop:regp_escape} with \cref{prop:standard_escape} suggests
an improved fit with \reGP in~$\XX_0$ for two reasons.
First, \cref{prop:regp_escape} applies
to a larger class of functions
since~$f$ can be an~$R$-element of~$W_2^{\beta + d/2}(\XX)$ without lying in~$W_{2}^{\beta + d/2}(\XX)$.
Furthermore, \cref{prop:finite_smooth_norm_reduction}
shows that~$\ns{s_{\R, \XX}^{(\beta)}}_{W_{2}^{\beta + d/2}(\XX) }$ can be arbitrarily 
smaller than~$\ns{f}_{W_{2}^{\beta + d/2}(\XX) }$ when~$f \in W_{2}^{\beta + d/2}(\XX)$.

\subsection{{\reGP with Estimated Parameters}}\label{app:theory_params}

The \reGP predictive distribution is obtained by conditioning the GP on the relaxed observations
and the model parameters given by~\eqref{eq:rgpi_problem}.
The results from \cref{sec:convergence_regp} do not take parameter selection into account.
This appendix presents adaptations of recent theoretical results about
parameter selection of Matérn covariance functions for Gaussian process interpolation of
deterministic functions.

We still assume a zero mean GP and use the notations from
\cref{sec:convergence_regp}, but in this appendix, we consider the RKHS
$\HH(\XX)$ with reproducing kernel $r$, not $k$, and the posterior variance will
be $\tau^2\sigma_n^2(\cdot)$, not $\sigma_n^2(\cdot)$ (recall
from~\cref{sec:gp} that the covariance function is
$k(\cdot,\, \cdot) = \tau^2 r(\cdot - \cdot)$).

%

First, we consider the maximum likelihood estimation of the variance
parameter~$\tau^2$.
Suppose that~$r$ is a Matérn correlation function with known regularity parameter~$\nu$
and range parameters~$(\rho_1, \dots, \rho_d)$.
Using the standard Gaussian process model, and
writing $\zn = (f(x_1), \dots, f(x_n))\tr$, it is well-known that the maximum likelihood estimate of~$\tau^2$
is given by:
$$
\hat{\tau}_n^2 = \frac{ \ns{s_{\zn}}_{\HH(\XX)}^2 }{n}.
$$

The estimated posterior variance at~$x \in \XX$ is~$\hat{\tau}_n^2 \sigma_n^2(x)$.
Suppose that~$(x_n)_{n \geq 1}$ is dense in~$\XX$.
It holds
that~$\ns{s_{\zn}}_{\HH(\XX)} \to \ns{f}_{\HH(\XX)}$ if $f \in \HH(\XX)$ \citep[see, e.g.,][Theorem 8.37]{iske2018approximation}.
Therefore, if~$f \in \HH(\XX)$ and~$x \in \XX$, then~\eqref{eq:power_function} implies that:
\begin{equation*}
\frac{\lvert f(x) - s_{\zn}(x) \lvert}{\hat{\tau}_n \sigma_n(x)} \leq C \sqrt{n}
\end{equation*}
for some~$C > 0$ and~$n$ large enough and using the convention~$0/0 = 1$.
In other words,
the GP can be overconfident by a factor of magnitude at most~$\sqrt{n}$~\citep{karvonen2020mleuq, karvonen2022error}.
If~$f \notin \HH(\XX)$, then~\eqref{eq:power_function} is not valid
and one can typically expect larger errors. %
\citet{karvonen2020mleuq} study how~$\ns{s_{\zn}}_{\HH(\XX)}$ diverges to compensate in this case.

Returning to relaxed Gaussian process interpolation with relaxation
{range}~$R$, notice that
the likelihood can be optimized with respect to the relaxed observation independently
of~$\tau^2$. Then, given the optimal relaxed observations~$\zn^{\star}$, the \reGP
variance estimate writes
$$
\hat{\tau}_{n, R}^2 = \frac{ \ns{s_{\R,\,n}}_{\HH(\XX)}^2 }{n},
$$
where~$s_{\R,\,n} = s_{\zn^{\star}}$
is the relaxed predictor.
Using~\eqref{eq:min_norm_extension} and~\eqref{eq:min_norm_slack},
it is easy to show that
$
\hat{\tau}_{n, \R}^2
\leq
\hat{\tau}_n^2
$.
The following proposition shows that the situations where~$\hat{\tau}_{n, \R}^2 = \mathcal{O}(n^{-1})$
are characterized by membership of~$f$ in~$\HH(\XX)$ in the sense of \cref{def:relaxed_membership}.
\begin{proposition}\label[proposition]{prop:diverging_variance}
Suppose that~$f$ is an~$R$-element of the space of continuous functions from~$\XX$ to~$\RR$
and that~$(x_n)_{n \geq 1}$ is dense in~$\XX$.
Using the notations of \cref{prop:constrained_problem}, it holds
that \hbox{$\ns{s_{\R,\,n}}_{\HH(\XX)} \to \ns{s_{\R, \XX}}_{\HH(\XX)}$}
if~$f$ is an $R$-element of~$\HH(\XX)$ and~$\ns{s_{\R,\,n}}_{\HH(\XX)} \to + \infty$
otherwise.
\end{proposition}
Coordinating \cref{prop:diverging_variance} with~\eqref{eq:power_function_slack_cuvette}
then shows that, for~$x \in \XX_0$, the \reGP predictor
can be overconfident by a factor of magnitude at most
$$
\sqrt{n} \frac{\sigma_{n, 0}(x)}{\sigma_n(x)},
$$
when~$f$ is an~$R$-element of~$\HH(\XX)$.
As mentioned in \cref{remark:regp_uq},
empirical and theoretical evidence about the screening effect
suggests that the ratio~$\sigma_{n, 0}(x)/\sigma_n(x)$ is asymptotically
close to one, for finite-smoothness covariance functions.
(Alternatively, matching bounds on power functions, such as Equation~\ref{eq:power_function_bounds}
and Theorem~4.4 from \citealp{karvonen2020mleuq}, can be used
to prove the existence of a constant~$C$ such that~$\sigma_{n, 0}(x) \leq C \sigma_n(x)$,
for~$n$ large enough,
in the case of a quasi-uniform sequence of points.)
In this case, \reGP can be overconfident by a factor of magnitude at most~$\sqrt{n}$
as for vanilla~GPs.

If~$f$ is not an~$R$-element of~$\HH(\XX)$, then~\eqref{eq:power_function_slack_cuvette}
does not hold anymore and \cref{prop:diverging_variance}
shows that~$\ns{s_{\R,\,n}}_{\HH(\XX)}$ diverges to compensate.
The following proposition is an adaptation of Proposition~4.5 from \citet{karvonen2020mleuq},
giving an upper bound of~$\hat{\tau}_{n, \R}^2$ for quasi-uniform sequences.
\begin{proposition}\label[proposition]{prop:diverging_variance_estimate}
Suppose that~$\XX$ is bounded and measurable, that~$(x_n)_{n \geq 1}$
is quasi-uniform,
and that there exists a~$\beta > 0$ such that~$f$ is an~$R$-element of~$W_2^{\beta + d/2}(\XX)$.
Then, with a Matérn covariance function with regularity~$\nu > \beta$, it holds that:
\begin{equation}\label{eq:diverging_variance_estimate}
\hat{\tau}_{n, \R}^2 \lesssim n^{2( \nu - \beta)/d - 1} \ns{s_{\R, \XX}^{(\beta)}}_{W_{2}^{\beta + d/2}(\XX)}^2,
\end{equation}
with~$s_{\R, \XX}^{(\beta)}$ defined in the paragraph preceding \cref{prop:regp_escape}.
\end{proposition}
Following \citet{2022_wang:_rigde}---but using a slightly different convention---we
define the smoothness~$\beta(f) = \sup \{ \beta > 0 \, : \, f \in W_{2}^{\beta + d/2}(\XX) \}$ of a function~$f$.
Given a relaxation {range}~$R$, we use the definition:
\begin{equation}\label{eq:def_relaxed_smoothness}
\beta_R(f) = \sup \left\lbrace \beta > 0 \, : \,  f \ \mathrm{ is \ an} \ R\mathrm{-element \ of } \ W_2^{\beta + d/2}(\XX) \right\rbrace,
\end{equation}
taking the relaxation into account.
It holds that~$\beta(f) = \beta_{\emptyset}(f) \leq \beta_R(f)$.

Let~$\nu > \beta_R(f) > 0$ and assume the supremum~\eqref{eq:def_relaxed_smoothness} defining~$\beta_R(f)$ is reached for simplicity.
The rate in the upper bound~\eqref{eq:diverging_variance_estimate} is optimized by taking~$\beta = \beta_R(f)$.
Under more restrictive assumptions on the spectrum and the support of~$f$,
\cite{karvonen2020mleuq} give matching lower bounds for traditional Gaussian process interpolation
and discuss uncertainty quantification for functions outside the~RKHS.
Obtaining a similar result for \reGP is more involved and thus left for future work.
Considering~\eqref{eq:qu_escape_rate} and Theorem~4.4 from \citet{karvonen2020mleuq},
it can be seen that a class of functions
such that the bound~\eqref{eq:diverging_variance_estimate} with~$\beta = \beta_R(f)$ is sharp
would also essentially be such that \reGP can be overconfident by a factor of magnitude
at most~$\sqrt{n}$ on~$\XX_0$ for quasi-uniform observations.

Finally, we consider the selection of the smoothness parameter~$\nu$
given fixed prior variance and range parameters.
For Gaussian process interpolation,
\citet{karvonen2023asymptotic} gives asymptotic bounds for maximum likelihood
estimates of~$\nu$ with quasi-uniform observations from a deterministic function.
The Matérn covariance function depends continuously on~$\nu$.
Therefore,
for~$0 < \nu_{\mathrm{min}} < \nu_{\mathrm{max}} < +\infty$
and~$\Theta = [ \nu_{\mathrm{min}}, \, \nu_{\mathrm{max}} ] $,
the minimum of~\eqref{eq:rgpi_problem}
is reached on~$\Theta \times C_{\R,\,n}$
according to the paragraph following \cref{def:rgpi}.
For each~$n$, let~$(\hat{\nu}_n^R, \, \zn^{\star})$ be such a minimizer,
with ties broken arbitrarily.
The following asymptotic lower bound
is an adaptation of the first statement of Theorem~3.12 from \citet{karvonen2023asymptotic}.%
\begin{proposition}\label[proposition]{prop:lower_bound_smoothness}%
Let~$0 < \nu_{\mathrm{min}} < \nu_{\mathrm{max}} < +\infty$ and $\Theta = [ \nu_{\mathrm{min}}, \, \nu_{\mathrm{max}} ]$
and assume~$\beta_R(f) + d/2 \in \Theta$,
with~$\beta_R(f)$ given by~\eqref{eq:def_relaxed_smoothness}.
If~$\XX$ satisfies \cref{assump:domain} and the sequence~$(x_n)_{n \geq 1}$ is quasi-uniform,
it holds that~$\liminf \hat{\nu}_n^R \geq \beta_R(f) + d/2$.%
\end{proposition}%
A smoothness parameter estimate based on~\reGP%
is asymptotically strictly larger than~$\beta_R(f)$.
Consequently, the function~$f$ is not an~$R$-element of the~RKHS.
This echoes the results obtained for Gaussian process interpolation in the previous reference.
These results must be seen considering that parameters are selected using
the likelihood of a model whose sample paths are almost surely not contained in
the RKHS~\citep[see, e.g.,][]{lukic2001stochastic}.
\citet[Section~6]{petit2025asymptotic}
studies asymptotically the effect of a joint estimation of~$\nu$ and~$\tau^2$
in a simplified Gaussian process interpolation framework.

Assume the supremum~\eqref{eq:def_relaxed_smoothness} is reached for simplicity and~$0 < \beta_R(f) < +\infty$.
With a Matérn covariance function with regularity~$\nu > 0$, quasi-uniform observations,
and~$B \subset \XX_0$ satisfying \cref{assump:domain},
the upper bounds~\eqref{eq:rate_cuvette}  and~\eqref{eq:qu_escape_rate}
yield:
\begin{equation}\label{eq:error_regp_nu}
  \left\{\; \begin{aligned}
    \ns{f - s_{R,\,n}}_{\Lset^{\infty}(B)} & \lesssim n^{-\nu / d} & \ & \mathrm{if} \ \nu \leq \beta_R(f), \\
    \ns{f - s_{R,\,n}}_{\Lset^{\infty}(B)} & \lesssim n^{-\beta_R(f) / d} & \ & \mathrm{otherwise.} \\
    \end{aligned} \right.
\end{equation}
As with standard Gaussian processes,
this
suggests that,
for quasi-uniform observations, the most
accurate \reGP models on~$\XX_0$ are obtained by taking~$\nu \geq \beta_R(f)$. 
\cref{prop:lower_bound_smoothness} shows that \reGP
 selects such models automatically.

The previous remark also applies with~$R = \emptyset$, recovering
a conclusion of \citet[Section~4.1]{karvonen2023asymptotic} for Gaussian process interpolation.
Consequently, the inequality~$\beta(f) = \beta_{\emptyset}(f) \leq \beta_R(f)$ and~\eqref{eq:error_regp_nu}
suggest an improved fit with \reGP compared to a traditional Gaussian process
when both models are used with estimated smoothness parameters.

Finally, note that the quasi-uniformity assumption is rather strong and prevents applications to active learning tasks
\citep[see][Section~5.2, for a possible application to a Bayesian optimization policy
designed to ensure quasi-uniformity]{wynne_et_al__2023:_convergence_gp}.

\clearpage

\section{Proofs}
\label{app:proofs}

\begin{lemma}{\citep[][Section 1.5]{aronszajn1950:_rkhs}}\label[lemma]{lem:extension}
  Let $k\colon \XX \times \XX \to \RR$ be a positive-definite covariance
  function, $\UU \subset \XX$, and $\HH(\UU)$ be the RKHS attached
  to the restriction of $k$ to $\UU \times \UU$. The RKHS
  $\HH(\UU)$ is the space of restrictions of functions from $\HH(\XX)$
  and the norm of $g \in \HH(\UU)$ is given by
\begin{equation*}
\inf_{h \in \HH(\XX), \ h_{\lvert \UU} = g} \ns{h}_{\HH(\XX)}.
\end{equation*}
\end{lemma}

\noindent
{\bf Proof of \cref{prop:finite_constrained_problem}}.
First the existence and the uniqueness of the solution are given by the first
statement of \cref{prop:constrained_problem}
(with $\HH_{\R,\,n} = \HH_{\R,\, \{x_1, \dots, x_n\}}$).

Furthermore let $\underline{z} \in \RR^n$ and write
$\underline{\alpha} = K_n^{-1} \underline{z}$, the reproducing
property~\eqref{eq:reproducing_prop} gives
\begin{equation}\label{eq:norm_blup}
  \ns{s_{\underline{z}}}_{\HH(\XX)}^2  = \underline{\alpha}\tr K_n \underline{\alpha} = \underline{z}\tr K_n^{-1} \underline{z},
\end{equation}
and therefore
$$
\min_{h \in \HH_{\R,\,n}} \ns{h}_{\HH(\XX)}^2 =
\inf_{\underline{z} \in C_{\R,\,n}} \min_{\ h \in \HH(\XX), \, h(\xn) = \underline{z}}
\ns{h}_{\HH(\XX)}^2 = \inf_{\underline{z} \in C_{\R,\,n}}
\underline{z}\tr K_n^{-1} \underline{z},
$$
where the last infimum is uniquely reached by the evaluation of the solution on
$\xn$.\hfill\BlackBox
\\

\noindent {\bf Proof of \cref{prop:joint_posterior}}.  Write
$K_{m, \, n}$ for the covariance matrix of the random vector
$\left( {\underline{Z}_m^{\prime}}\tr, \, \Zn\tr \right)\tr$.  Using the
equalities~\eqref{eq:likelihood} and~\eqref{eq:norm_blup}, and a slight
abuse of notation by dropping irrelevant constants with respect to
$\underline{z}^{\prime}$ and $\underline{z} \in C_{\R,\,n}$, we have
$$
- 2 \ln \left( p \left( \underline{z}^{\prime}, \, \underline{z} \, \lvert \, \Zn \in C_{\R,\,n} \right) \right)
=
 \left( {\underline{z}^{\prime}}\tr, \, \underline{z}\tr \right) K_{m, \, n}^{-1} \left( {\underline{z}^{\prime}}\tr, \, \underline{z}\tr \right)\tr
=
\min_{h \in \HH(\XX), \, h(\xn) = \underline{z}, \, 
h(\underline{x}_{m}^{\prime} ) = \underline{z}^{\prime} }
\ns{h}_{\HH(\XX)}^2.
$$
This gives
$$
\inf_{\underline{z}^{\prime} \in \RR^{m}, \, \underline{z} \in C_{\R,\,n}} 
- 2 \ln \left( p \left( \underline{z}^{\prime}, \,
    \underline{z} \, \lvert \, \Zn \in C_{\R,\,n} \right) \right) =
\min_{h \in \HH(\XX), \, h(\xn) \in C_{\R,\,n} }
\ns{h}_{\HH(\XX)}^2,
$$
which is reached by taking $\underline{z} = \zn^{\star}$
and $\underline{z}^{\prime} = \left(s_{\zn^{\star}}(x_{1}^{\prime} ), \dots,
s_{\zn^{\star}}(x_{m}^{\prime} ) \right)\tr$.\hfill\BlackBox
\\

\noindent
{\bf Proof of \cref{prop:power_function_bounds}}.
First, one has
$$
\sup_{x \in \XX} \sigma_n(x) = \sup_{x \in \XX} \sup_{\ns{f}_{\HH(\XX)}
  = 1} \abs{f(x) - s_{\zn}(x)} = \sup_{\ns{f}_{\HH(\XX)} = 1}
\ns{f - s_{\zn}}_{\Lset^\infty(\XX)}.
$$

Now, let $f \in \HH(\XX)$ such that $\ns{f}_{\HH(\XX)} = 1$,
and $\XX^o$ be the interior of $\XX$.
The boundary of~$\XX^o$ is the one of~$\XX$ under \cref{assump:domain},
and the Sobolev space~$W_2^{\alpha + d/2}(\XX^o)$
defined by~\eqref{eq:restriction_sobolev_norm}
is norm-equivalent to the Sobolev–Slobodeckij space
(see, e.g., Proposition~3.4 of \citealp{di2012:_frac_sobolev}, for a statement
on $\RR^d$ and Theorem~1.4.3.1 of \citealp{grisvard1985:_elliptic}, for the existence of
an extension operator).

Then, one can
apply Theorem~4.1 from~\citet{arcangeli2007:_error_bounds}
to~$f - s_{\zn}$ restricted to $\XX^o$ to show that, for $h_n$ lower than some positive~$h_0$
(not depending on $f$ or $\left(x_n\right)_{n \geq 1}$), we have:
\begin{equation}\label{eq:arcangeli_inequalities}
\ns{f - s_{\zn}}_{\Lset^\infty(\XX)}
= \ns{f - s_{\zn}}_{\Lset^\infty(\XX^o)}
\lesssim h_n^{\alpha} \ns{f - s_{\zn}}_{W_2^{\alpha {+ d/2}}(\XX^o)}
\lesssim h_n^{\alpha} \ns{f - s_{\zn}}_{\HH(\XX)}
\lesssim h_n^{\alpha}
\end{equation}
by continuity of $f - s_{\zn}$,
since $\ns{\cdot}_{W_2^{\alpha + d/2}(\XX^o)} \leq \ns{\cdot}_{W_2^{\alpha + d/2}(\XX)}$
due to the definition~\eqref{eq:restriction_sobolev_norm},
$W_2^{\alpha + d/2}(\XX)$ being norm equivalent to $\HH(\XX)$,
and because of the projection interpretation of~$s_{\zn}$
\citep[see, e.g.,][Theorem 13.1]{wendland04:_scatt}.
Finally, one can get rid of the condition $h_{n} \leq h_{0}$ for simplicity by {possibly} increasing the constant, since
$
\sup_{x \in \XX} \sigma_n(x) \leq \sup_{x \in \XX} \sqrt{k(x, \, x)} < + \infty
$ by compacity.\hfill\BlackBox
\\

\noindent {\bf Proof of \cref{prop:constrained_problem}}.
First observe that $\HH_{\R,\,\UU}$ is not empty since it contains
$f$.  Furthermore, it is easy to verify that $\HH_{\R,\,\UU}$ is
convex and that it is closed since pointwise evaluation functionals are continuous on an RKHS.
The problem is then the one of projecting the null function on a convex closed subset; hence the
existence and the uniqueness.

Then, the function $s_{\R,\,n}$ is the projection of the null function on the closed
convex set~$\HH_{\R,\,n}$.
Moreover, the sequence $\left( \HH_{\R,\,n} \right)_{n \geq 1}$ is non-increasing,
so the sequence $\left( s_{\R,\,n} \right)_{n \geq 1}$ converges in~$\HH(\XX)$
to the projection of the null function on~$\bigcap_{n \geq 1} \HH_{\R,\,n} $
\citep[see, e.g.,][Exercice 5.5]{brezis2011:_functional},
i.e., the solution of~\eqref{eq:continuous_extension}, with $\UU = \{ x_n \}$.
But this last solution is also the solution on the closure since it satisfies the constraints by continuity.\hfill\BlackBox
\\

\noindent {\bf Proof of \cref{prop:convergence_bounds_model_cuvette}}.
Define $\xnzero$ and $\znzero$ to be data points within $\XX_0$, and $s_{\xnzero, \, \znzero}$
the associated (interpolation) predictor, i.e., the solution of~\eqref{eq:min_norm_extension}.
{Take the null function if there are no observations within~$B$.}
Observing that $s_{\xnzero, \, \znzero}$ interpolates $s_{\R,\,n} $, we have:
\begin{align*}
  \bigl| f(x) - s_{\R,\,n} (x) \bigr|
  & \;\leq\; \bigl| f(x) - s_{\xnzero, \, \znzero}(x) \bigr|
    + \bigl| s_{\xnzero, \, \znzero}(x) - s_{\R,\,n} (x) \bigr|\\
  & \;\leq\; \sigma_{n, \, 0}(x)\, \ns{f}_{\HH(\XX_0)}
    + \sigma_{n, \, 0}(x)\, \ns{s_{\R,\,n}}_{\HH(\XX_0)}\\
  & \;\leq\; 2 \sigma_{n, 0}(x)\, \ns{s_{\R,\,\XX}}_{\HH(\XX)},
\end{align*}
since $f$ coincides with $s_{\R,\,\XX}$ on $\XX_0$, $s_{\R,\,\XX} \in \HH_{\R,\,n}$,
and by \cref{lem:extension}.\hfill\BlackBox
\\

\begin{lemma}\label[lemma]{lem:finite_diff_rate_dense_design}
  Use the notation~\eqref{eq:pointwise_func}.
  If $k$ has smoothness $\alpha > 0$, then there exists $h_0 > 0$ depending
  only on $\alpha$ such that, for all $x, y \in \XX$ {satisfying} $\ns{x - y} \leq h_0$, we have
\begin{equation*}
  \ns{\delta_{y}  - \delta_{x}}_{\HH^{\star}(\XX)}
  \lesssim \ns{x - y}^{\alpha}, \ \mathrm{for} \ \alpha < 1,
\end{equation*}
\begin{equation*}
  \ns{\delta_{y}  - \delta_{x}}_{\HH^{\star}(\XX)} \lesssim \sqrt{\abs{\ln(\ns{x - y})}}\, \ns{x - y} , \ \mathrm{for} \ \alpha = 1,
\end{equation*}
and 
\begin{equation*}
\ns{\delta_{y}  - \delta_{x}}_{\HH^{\star}(\XX)}  \lesssim \ns{x - y}, \ \mathrm{for} \ \alpha > 1.
\end{equation*}
\end{lemma}
\noindent {\bf Proof}.  Since equivalent norms give equivalent operator
norms on the topological dual of a normed space, it suffices to show the
result for a unit-variance isotropic Matérn covariance
function~\eqref{eq:matern-cov} of regularity~$\alpha$.

In this case, we have
$$
\ns{\delta_{y}  - \delta_{x}}_{\HH^{\star}(\XX)}^2
=
k(x, \, x) + k(y, \, y) - 2 k(x, \, y)
= 2 \left(1 - r_{\alpha}(\ns{x - y}) \right),
$$
with $r_{\alpha}$ the corresponding isotropic correlation function. Standard
results on principal irregular terms \citep[see, e.g.,][Chapter
2.7]{stein1999:_interpolation_of_spatial_data} give the
results.\hfill\BlackBox
\\

\begin{lemma}\label[lemma]{lem:bound_h}
Let $B \subset \XX$ {satisfying} \cref{assump:domain} and $h_{n, \, B}$ be the fill
distance of $\XX_{n, \, B} = \{x_1, \dots, x_n\} \cap B$ within~$B$, with the convention $h_{n, \, B} = \mathrm{diam}(B)$
if $\XX_{n, \, B}$ is empty. Then, $h_{n, \, B} \lesssim h_n$.
\end{lemma}
\noindent
{\bf Proof}.
The idea of the proof is given by \citet[Lemma 11.31]{wendland04:_scatt}, but it is interlinked with
a much more sophisticated construction, so we provide a specific version here for completeness.
\citet[][Section 4.11]{adams2003:_sobolev} show that $B$ {satisfies} a cone condition
with radius $\rho > 0$ and angle $\phi \in \left(0, \, \pi/2\right)$.
If $\XX_{n, \, B}$ is not empty, then the compacity of $B$
ensures the existence of an~$x \in B$ such that $d(x, \XX_{n, \, B}) = h_{n, \, B}$.
(If $\XX_{n, \, B}$ is empty, then the rest of the proof is also valid taking an arbitrary~$x \in B$.)

A cone $C$ originating from $x$ with angle $\phi$ and radius $\delta = \min(h_{n, \, B}, \, \rho)$
is contained in~$B$ and its interior do not contains observations.
Furthermore, \citet[][Lemma 3.7]{wendland04:_scatt} shows that there exists a $y \in C$ such that
the open ball $B \left( y, \, \delta \sin(\phi) \left( 1 + \sin(\phi) \right)^{-1} \right)$ is subset
of $C$, and therefore contains no observations as well. Thus,
$h_n \geq \delta \sin(\phi) (1 + \sin(\phi))^{-1}$.
Now, if~$h_{n, \, B} \leq \rho$, then the desired result follows.
If not, the result holds as well since $h_{n, \, B} \leq \mathrm{diam}(B)$.\hfill\BlackBox
\\

\noindent {\bf Proof of \cref{prop:h_bounds}}.
For the first assertion,
suppose first that~$B$ contains observations and
let~$\sigma_{n, \, B}$ be the power function using only the observations
within~$B$.
It can be shown that the restriction of the kernel to~$B$ also has regularity~$\alpha$
using~\eqref{eq:restriction_sobolev_norm}
and \cref{lem:extension}.
Using \cref{prop:convergence_bounds_model_cuvette},
the inequality $\sigma_{n, \, 0} \leq \sigma_{n, \, B}$ given by the projection residuals interpretation,
and applying \cref{prop:power_function_bounds} to~$B$ yields
a bound depending on the fill distance $h_{n, \, B}$ of $\{x_1, \dots, x_n\} \cap B$ within~$B$.
Finally, \cref{lem:bound_h} allows us to conclude.
Dealing with the case where~$B$ contains no observations
using~\eqref{eq:reproducing_prop} makes no difficulty.

Regarding the second assertion, $f$ is continuous so the sets $\XX_j$
are compact for $j \geq 1$. In addition, they are disjoint so
$$\delta = \min_{1 \leq j < p} \inf_{\, x \in \XX_j, \, y \in \XX_p} \ns{x - y} > 0.$$
Suppose now that $h_n < \delta$ and let $j \geq 1$, $x \in \XX_j$
and $1 \leq i \leq n$ the index of the closest $x_i$ to~$x$.
By definition, $\ns{x - x_i} \leq h_n$ and therefore $x_i \in \XX_0 \cup \XX_j$.
Let $\HH^{\star}(\XX)$ be the (topological) dual of~$\HH(\XX)$
and
\begin{equation}\label{eq:pointwise_func}
\delta_y\colon h \in \HH(\XX) \mapsto h(y),
\end{equation}
which lies in~$\HH^{\star}(\XX)$ for all $y \in \XX$.
Then using the reproducing property~\eqref{eq:reproducing_prop}, we have
$$
\left| s_{\R,\,n} (x_i) - s_{\R,\,n} (x) \right|
\leq \ns{\delta_{x_i} - \delta_x}_{\HH^{\star}(\XX)} \ns{s_{\R,\,n} }_{\HH(\XX)}
\leq \ns{\delta_{x_i} - \delta_x}_{\HH^{\star}(\XX)} \ns{s_{\R,\,\XX}}_{\HH(\XX)},
$$
and therefore
\begin{equation*}\label{eq:power_function_slack_relaxation}
d(s_{\R,\,n} (x), \, R_j)
\leq \ns{\delta_{x_i}  - \delta_{x}}_{\HH^{\star}(\XX)}  \ns{s_{\R,\,\XX}}_{\HH(\XX)} + d(s_{\R,\,n} (x_i), \, R_j).
\end{equation*}
Now, if $x_i \in \XX_j$,
then~$d(s_{\R,\,n} (x_i), \, R_j) = 0$.
Otherwise $x_i \in \XX_0$ necessarily and, then, using the fact that $s_{\R,\,\XX}(x) \in R_j$, we have:
$$
d(s_{\R,\,n} (x_i), \, R_j)
\leq \abs{ s_{\R,\,n} (x_i) - s_{\R,\,\XX}(x) }
= \abs{ s_{\R,\,\XX}(x_i) - s_{\R,\,\XX}(x) }
\leq \ns{\delta_{x_i} - \delta_x}_{\HH^{\star}(\XX)} \ns{s_{\R,\,\XX}}_{\HH(\XX)}.
$$
So one can use
\cref{lem:finite_diff_rate_dense_design} along with the previous
statements to conclude if $h_n < \min \left( \delta, \, h_0 \right)$.
(It may be necessary to reduce~$\min \left( \delta, \, h_0 \right)$
to use the monotonicity of~$x \mapsto x \sqrt{-\ln(x)}$,
for~$x > 0$ close to zero.)

Finally, treating the case $h_n \geq \min \left( \delta, \, h_0 \right)$ is straightforward using
the fact that $\sup_{x \in \XX} \sqrt{k\left(x, \, x\right)}$ is finite thanks to the compacity
of $\XX$ and
$d(s_{\R,\,n} (x), \, R_j)
\leq \abs{ s_{\R,\,n} (x) - s_{\R,\,\XX}(x) }$ for~$j \geq 1$ and $x \in \XX_j$.\hfill\BlackBox
\\

\begin{lemma}\label[lemma]{lem:prod_sobolev_test_function}
If $g, \, h \in W_2^{\gamma}(\XX)$ for $\gamma > d/2$, then $g h \in W_2^{\gamma}(\XX)$.
\end{lemma}
\noindent
{\bf Proof}.
By the definition~\eqref{eq:restriction_sobolev_norm} of $W_2^{\gamma}(\XX)$, the functions
$g$ and $h$ can be extended into functions on~$\RR^d$,
having their product in~$W_2^{\gamma}(\RR^d)$ \citep[Theorem 2.1]{strichartz1967:_multipliers}.
Taking the restrictions shows the desired result.\hfill\BlackBox
\\

\noindent {\bf Proof of \cref{prop:finite_smooth_norm_reduction}}.  We use a bump function
argument.  Let $B(x_0, r) \subset \XX_j$ (with $r > 0$) be an open ball.
There exists a $C^{\infty}$ function $\phi\colon \RR^d \to \RR$ such
that
$$
    \left\{
      \begin{array}{lll}
          0 \leq \phi \leq 1, & & \\
          \phi(x)= 1  & \mathrm{only \ if} & x = 0, \\
          \phi(x)= 0  & \mathrm{if} & x \in \XX \setminus B(0, r). \\
    \end{array}
    \right.
$$
Let $c \in R_j \setminus \{ f(x_0) \}$,
$\phi_n = \phi \left( n \left( \cdot - x_0 \right) \right)$ as a function
on $\XX$, and
$f_n = \left(1 - \phi_n \right) f + c \phi_n$, for $n \geq 1$.
We have~$\phi_n \in W_2^{\alpha + d/2}(\RR^d)$ as a function on~$\RR^d$,
so it belongs to $W_2^{\alpha + d/2}(\XX)$ as a function on $\XX$, and
\cref{lem:prod_sobolev_test_function} ensures that
$f_n \in \HH(\XX)$.  Moreover, it is easy to check that
$f_n \in \HH_{\R,\, \XX}$.  Observe that the sequence
$(f_n)_{n \geq 1}$ converges pointwise to a discontinuous function that
lies thus outside $\HH(\XX)$.

Suppose now that $\ns{f_n}_{\HH(\XX)} \nrightarrow +\infty$. Then, one can extract a
bounded subsequence of norms
and a classical weak compacity argument would yield a weakly convergent subsequence, which is impossible
since the pointwise limit is not in $\HH(\XX)$.\hfill\BlackBox
\\

\begin{lemma}\label[lemma]{lem:model_constit_varying_t}
Use the notations from the proof of \cref{prop:convergence_slack_ei_neb_varying_t}
and let $(y_n)_{n \geq 1}$ be a sequence in~$\XX$. Assume that the sequence
$(y_n)_{n \geq 1}$ is convergent, denote by $y^{\star}$ its limit and
assume that $y^{\star}$ is an adherent point of the set $\{ x_n \}$.
Let $t_{\infty} = \liminf t_n$, then
\begin{itemize}
\item $s_{n}(y_n) \to f(y^{\star})$ if $f(y^{\star}) < t_{\infty}$,
\item $\liminf s_{n}(y_n) \geq t_{\infty}$, otherwise.
\end{itemize}
In particular, we have
\begin{equation*}
\liminf s_{n}(y_n) \geq \min( f(y^{\star}), \ t_{\infty}).
\end{equation*}
\end{lemma}
\noindent {\bf Proof}.
Suppose that $y^{\star} \notin \{ x_n \}$. Then, let $\left( x_{\phi(n)} \right)_{n \geq 1}$ be a subsequence
converging to $y^{\star}$ and let
$\psi(n) = \max \{ \phi(k), \phi(k) \leq n \}$.
We proceed as in \cref{prop:h_bounds} to have:
\begin{equation}\label{eq:decoupage_seuil_variable}
\left| s_{n}(x_{\psi(n)}) - s_{n}(y_n) \right|
\leq \ns{\delta_{x_{\psi(n)}} - \delta_{y_n}}_{\HH^{\star}(\XX)} \ns{s_{n}}_{\HH(\XX)} \to 0,
\end{equation}
thanks to the continuity of~$k$ and the inequality $\ns{s_{n}}_{\HH(\XX)} \leq \ns{f}_{\HH(\XX)}$.

Finally, we have
$s_{n}(x_{\psi(n)}) \geq \min \left( f \left( x_{\psi(n)}
  \right), \, t_n \right)$ by construction and therefore
$\liminf s_{n}(x_{\psi(n)}) \geq \min \left( f \left( y^{\star}
  \right), \, t_{\infty} \right)$, which gives the second
assertion thanks to~\eqref{eq:decoupage_seuil_variable}. Observe that $f(x_{\psi(n)}) < t_n$ ultimately if
$f(y^{\star}) < t_{\infty}$ for the first assertion.

If $y^{\star} \in \{ x_n \}$, then the result follows similarly.\hfill\BlackBox
\\

\begin{lemma}\label[lemma]{lem:liminf_ei_varying_t}
Using the notations from the proof of \cref{prop:convergence_slack_ei_neb_varying_t}
and writing $v_n = \sup_{x \in \XX} \rho_{n, \, t_n}(x)$,
we have $\liminf v_n = 0$.
\end{lemma}
\noindent {\bf Proof}.  This is an adaptation of Lemma~12 from
\citet{vazquez_and_bect2010:_convergence_ei}.

Let $y^{\star}$ be a cluster point of $\left(x_n\right)_{n \geq 1}$ and let $\left( x_{\phi(n)} \right)_{n \geq 1}$ be
a subsequence converging to $y^{\star}$.  We are going to prove that
$v_{\phi(n)-1} \to 0$.  Define
$$
z_{\phi(n) - 1} = m_{\phi(n) - 1} - s_{\phi(n) - 1} (x_{\phi(n)}).
$$
Then, \cref{lem:model_constit_varying_t} gives\footnote{
\cref{lem:model_constit_varying_t} yields
$\liminf s_{n - 1} (x_{\psi(n)}) \geq \min( f(y^{\star}),
\ t_{\infty})$ with $\psi(n) = \max \{ \phi(k), \phi(k) \leq n \}$, and the claim follows by extracting a $\phi$-subsequence.
}
$\liminf s_{\phi(n) - 1} (x_{\phi(n)}) \geq \min( f(y^{\star}),
\ t_{\infty})$. Moreover we have
$$m_{\phi(n) - 1} \leq \min \left( f \left( x_{\phi(n - 1)} \right), \ t_{\phi(n) -1} \right),$$
since $\phi(n - 1) \leq \phi(n) - 1$,
so
$\lim m_{\phi(n) - 1} \leq \min( f(y^{\star}), \
t_{\infty})$ holds because
$\left( m_{\phi(n) - 1} \right)_{n \geq 1}$ is
non-increasing. The previous arguments show that
$ \limsup z_{\phi(n) - 1} \leq 0 $.

Moreover, one can use Proposition~10 from
\citet[$(i) \Rightarrow (ii)$]{vazquez_and_bect2010:_convergence_ei} similarly to show that $\sigma_{\phi(n) - 1}^2(x_{\phi(n)}) \to 0$
and therefore
$$v_{\phi(n)-1} = \gamma \left( z_{\phi(n) - 1}, \ \sigma_{\phi(n) - 1}^2(x_{\phi(n)}) \right)
\leq \gamma \left( \sup_{k \geq n} z_{\phi(k) - 1}, \ \sigma_{\phi(n) - 1}^2(x_{\phi(n)}) \right) \to 0,$$
since $\gamma$ is non-decreasing with respect to its first argument
and continuous.\hfill\BlackBox
\\

\noindent {\bf Proof of \cref{prop:convergence_slack_ei_neb_varying_t}}. %
This is an adaptation of Theorem~6 from
\citet{vazquez_and_bect2010:_convergence_ei}. %
Write $s_n = s_{\R_n,\,n}$ for the \reGP predictor at
step $n$ to avoid cumbersome notations. %
Then, for \hbox{$x \in \XX$},
write~$\rho_{n,\, t_n}(x) = \gamma (m_n - s_n(x), \, \sigma_n^2(x))$ for
the expected improvement under the \reGP predictive distribution,
with~$\gamma$ the function defined in \cref{prop:ei}.

Suppose that there exists some $x_0 \in \XX$ such that $\sigma_n^2(x_0) \geq C_1 > 0$.
The sequence $\left(m_n\right)_{n \geq 1}$ converges
and the reproducing property~\eqref{eq:reproducing_prop} yields
$$
\abs{s_{n}(x_0)} \leq \sqrt{k(x_0, \ x_0)} \ns{s_{n}}_{\HH(\XX)} \leq \sqrt{k(x_0, \ x_0)} \ns{f}_{\HH(\XX)},
$$
so the sequence $ \left( \left| m_n - s_{n}(x_0) \right| \right)_{n \geq 1}$
is bounded by, say $C_2$. We {then have}
$$
v_n = \sup_{x \in \XX} \rho_{n,\, t_n}(x) \geq \gamma \left( m_n - s_{n}(x_0),
    \sigma_n^2(x_0) \right) \geq \gamma \left(- C_2, C_1
\right) > 0
$$
by \cref{prop:ei}. But this yields a contradiction with
\cref{lem:liminf_ei_varying_t}, so the decreasing
sequence~$\left( \sigma_n^2 \right)_{n \geq 1}$ converges pointwise on
$\XX$ to zero. Proposition~10 from
\cite{vazquez_and_bect2010:_convergence_ei} then implies that every
$x \in \XX$ is adherent to $\{ x_n \}$.\hfill\BlackBox
\\

\begin{lemma}\label[lemma]{lem:tcrps_general}
Assume that $b$ is finite, and that either $a$ is finite too or
$\int_{-\infty}^0 F_P(u)\, \du = \int_{-\infty}^0 |u|\, P(\du)$ is
finite.
Then
\begin{align*}
  S_{a, \, b}^\tCRPS(P, \,z) & \;=\; (b - a \vee z)_+ + R_2(a,b) - 2 R_1(a \vee z, b),
\end{align*}
where
\begin{equation*}
  R_q(a,b) \;=\;  \int_{-\infty}^{+\infty} \one_{a \le u \le b}\, F_P(u)^q\, \du.
\end{equation*}
\end{lemma}
\noindent
{\bf Proof}.
\begin{align*}
  S_{a, \, b}^\tCRPS(P, \,z)
  & \;=\; \int_{-\infty}^{+\infty} \one_{a \le u \le b}\, \left( \one_{z \le u} - F_P(u) \right)^2\, \du\\
  & \;=\; \int_{-\infty}^{+\infty} \one_{a \le u \le b}\, \left( \one_{z \le u} + F_P(u)^2 - 2\, \one_{z \le u}\, F_P(u) \right)\, \du\\
  & \;=\; \int_{-\infty}^{+\infty} \left( \one_{a \vee z \le u \le b}\,  + \one_{a \le u \le b}\, F_P(u)^2 - 2\, \one_{a \vee z \le u \le b}\, F_P(u) \right)\, \du\\
  & \;=\; (b - a \vee z)_+ + R_2(a,b) - 2 R_1(a \vee z, b).
\end{align*}
\hfill\BlackBox
\\

\begin{lemma} \label[lemma]{lem:Iq-EIup}
  Let $a, b \in \RR$ with $a \le b$. Let $q \ge 1$.  Then
  \begin{equation*}
    R_q(a,b) \;=\; b - a + \EI^{\uparrow}_q(P,b) - \EI^{\uparrow}_q(P,a).
  \end{equation*}
\end{lemma}
\noindent
{\bf Proof}.
  \begin{align*}
    R_q(a,b)
    & \;=\;  \int \one_{a \le u \le b}\, F_P(u)^q\, \du\\
    & \;=\;  \int \one_{a \le u \le b}\, \prod_{j=1}^q \EE\left( \one_{N_j \le u}\right)\, \du
      \qquad \text{with } N_j \stackrel{\mathrm{iid}}{\sim} P\\
    & \;=\; \EE\left( \int \one_{a \vee N_1 \vee \cdots \vee N_q \le u \le b} \,\du\right)\\
    & \;=\; \EE\left( \left( b - a \vee N_1 \vee \cdots \vee N_q \right)_+ \right)\\
    & \;=\; b - a + \EI^{\uparrow}_q(P,b) - \EI^{\uparrow}_q(P,a).
  \end{align*}\hfill\BlackBox
\\

\noindent
{\bf Proof of \cref{prop:tcrps_general}}.
The first result is given by
\cref{lem:tcrps_general} and \cref{lem:Iq-EIup}.

Then using the dominated convergence theorem, it is easy to see that,
when $a \to -\infty$
\begin{equation*}
  \EI^{\uparrow}_q(P,a) \;=\; \EE\left( N_1 \vee \cdots \vee N_q \right) - a + o(1),
\end{equation*}
and therefore
\begin{equation*}
  R_q(-\infty,b)
  \;=\; \lim_{a \to -\infty} R_q(a,b)
  \;=\; b + \EI^{\uparrow}_q(P,b) - \EE\left( N_1 \vee \cdots \vee N_q \right),
\end{equation*}
which gives the second statement.

Finally, a change of variable gives
$$
\int_{a}^{+\infty} \left(F_P(u)- \one_{z \leq u} \right)^2 \du
= \int_{-\infty}^{-a} \left(F_{\underline{P}}(u) - P(U = -u) - \one_{- z < u} \right)^2 \du,
$$
and the last statement follows by observing that a probability measure
admits at most a countable number of atoms.\hfill\BlackBox
\\

\noindent {\bf Proof of \cref{prop:standard_escape}}.
The proof follows closely those of Theorem~4.2 from \citet{narcowich_et_al2006:_evasion_theorems},
Theorem~4.2 from \citet{karvonen2020mleuq}, and Theorem~1 from \citet{wynne_et_al__2023:_convergence_gp}.
It is therefore not entirely repeated for brevity.
Note that Theorem~4.1 from \citet{arcangeli2007:_error_bounds} must be used with~$q = + \infty$ and~$q = 2$
but requires~$\XX$ to be open.
Under \cref{assump:domain},
the adaptation for~$q = + \infty$ is given by~\eqref{eq:arcangeli_inequalities}
in the proof of \cref{prop:power_function_bounds}.
For~$q = 2$, use the fact that a locally Lipschitz boundary has zero measure.

The proofs in the references give the result for~$h_n$ lower than some~$h_0$ (not dependent of~$(x_n)_{n \geq 1}$ and~$f$).
Removing this condition by possibly increasing the constant makes no difficulty.

The fact that~$h_n \lesssim n^{-1/d}$ for a quasi-uniform sequence \citep[see, e.g.,][Proposition~14.1]{wendland04:_scatt}
gives the second statement.\hfill\BlackBox
\\

\noindent {\bf Proof of \cref{prop:regp_escape}}.
Assume there are observations~$(\xnB, \, \znB)$ within~$B$ and
let $s_{\xnB, \, \znB}$
be the associated (interpolation) predictor.
We have:
$$
\ns{f - s_{R,\,n}}_{\Lset^{\infty}(B)} \leq \ns{f - s_{\xnB, \, \znB}}_{\Lset^{\infty}(B)} + \ns{s_{\xnB, \, \znB} - s_{R,\,n}}_{\Lset^{\infty}(B)}.
$$
Use the notations of \cref{lem:bound_h}, let~$q_{n, \, B}$ be the separation distance
of points within~$B$ (with the convention~$q_{n, \, B} = \mathrm{diam}(\XX)$ if~$\XX_{n, \, B}$ has less than two distinct elements),
and let~$\rho_{n, \, B} = h_{n, \, B}/q_{n, \, B}$.
It holds that~$f_{\lvert B} \in W_{2}^{\beta + d/2}(B)$ since~$B \subset \XX_0$
and~$f$ is an~$R$-element of~$W_{2}^{\beta + d/2}(\XX)$.
It can be shown that the restriction of the kernel to~$B$ also has regularity~$\alpha$
using~\eqref{eq:restriction_sobolev_norm}
and \cref{lem:extension}.
Therefore, we can apply \cref{prop:standard_escape} to~$f_{\lvert B}$
to get:
$$
\ns{f - s_{\xnB, \, \znB}}_{\Lset^{\infty}(B)}
\lesssim h_{n, \, B}^{\beta} ( 1 + \rho_{n, \, B}^{\alpha - \beta}) \ns{f}_{W_2^{\beta + d/2}(B)}
\lesssim h_{n}^{\beta} ( 1 + \rho_{n}^{\alpha - \beta}) \ns{s_{\R, \XX}^{(\beta)}}_{W_{2}^{\beta + d/2}(\XX) },
$$
where we used \cref{lem:bound_h}, the inequality~$q_n \leq q_{n, \, B}$, the definition~\eqref{eq:restriction_sobolev_norm},
and the fact that~$s_{\R, \XX}^{(\beta)}$ coincides with~$f$ on~$B$.

Consider the restrictions on~$B$ and observe that~$s_{R,\,n} \in \HH(B)$ and that $s_{\xnB, \, \znB}$ is the associated interpolation predictor.
One can combine~\eqref{eq:power_function} and \cref{prop:power_function_bounds} to get
$$
\ns{s_{\xnB, \, \znB} - s_{R,\,n}}_{\Lset^{\infty}(B)}
\lesssim h_{n, \, B}^{\alpha} \ns{s_{R,\,n}}_{\HH(B)}
\lesssim h_{n}^{\alpha} \ns{s_{R,\,n}}_{\HH(\XX)}
$$
using \cref{lem:extension,lem:bound_h}.

Observe that the proof of Lemma~A.1. from~\citet{karvonen2020mleuq}
is valid without restrictions on the domain~$\XX$.
Therefore, there exists a constant~$C_\beta > 0$ such that, for each~$n$,
there exists a function~$u_{n}^{(\beta)} \in W_2^{\alpha + d/2}(\XX)$
coinciding with~$s_{\R, \XX}^{(\beta)}$ on the design~$\xn$ and
satisfying
\begin{equation*}
\ns{u_{n}^{(\beta)}}_{W_2^{\alpha + d/2}(\XX)}
\leq C_\beta q_n^{\beta - \alpha} \ns{s_{\R, \XX}^{(\beta)}}_{W_2^{\beta + d/2}(\XX)}.
\end{equation*}
To conclude, observe that~$u_{n}^{(\beta)} \in \HH_{\R,\,n}$ and thus
\begin{equation}\label{eq:bound_norm}
\ns{s_{R,\,n}}_{\HH(\XX)} \leq
\ns{u_{n}^{(\beta)}}_{\HH(\XX)}
\lesssim \ns{u_{n}^{(\beta)}}_{W_2^{\alpha + d/2}(\XX)}
\lesssim q_n^{\beta - \alpha} \ns{s_{\R, \XX}^{(\beta)}}_{W_2^{\beta + d/2}(\XX)}.
\end{equation}

The previous display is also true when~$B$ does not contain observations.
Coordination with~\eqref{eq:reproducing_prop} and Lemma~\ref{lem:bound_h}
shows that~\eqref{eq:escape_rate} also holds in this case.

The fact that~$h_n \lesssim n^{-1/d}$ for a quasi-uniform sequence \citep[see, e.g.,][Proposition~14.1]{wendland04:_scatt}
gives the second statement.\hfill\BlackBox
\\

\noindent {\bf Proof of \cref{prop:diverging_variance}}.
Let~$g \in \HH(\XX)$ make~$f$ an~$R$-element of~$\HH(\XX)$.
Applying \cref{prop:constrained_problem} to~$g$
proves one implication.

Conversely, the sequence~$(\ns{s_{\R,\,n}}_{\HH(\XX)})_{n \geq 1}$ is non-decreasing.
Suppose that it is bounded.
One can then extract a weakly convergent subsequence
from~$(s_{\R,\,n})_{n \geq 1}$.
The sets~$\HH_{\R,\,n}$ are closed and convex and hence weakly closed. 
The weakly convergent subsequence is ultimately in each of these sets
so its limit lies in their intersection.
It can be checked that the existence of a continuous function coinciding with~$f$
on~$\XX_0$ and defining the same~$\XX_j$s ensures that this
intersection is equal to~$\HH_{\R,\,\XX}$.\hfill\BlackBox
\\

\noindent {\bf Proof of \cref{prop:diverging_variance_estimate}}.
The result is given by~\eqref{eq:bound_norm} with~$\alpha = \nu$
and~$q_n \lesssim n^{-1/d}$ by quasi-uniformity \citep[see, e.g.,][Proposition~14.1]{wendland04:_scatt}.\hfill\BlackBox
\\

\noindent {\bf Proof of \cref{prop:lower_bound_smoothness}}.
The proof follows the ideas of~\citet[Theorem~3.11 and Theorem~3.12]{karvonen2023asymptotic}.
Write~$K_{n, \nu}$ to underline the dependence of the covariance matrix on the parameter~$\nu$.
Let~$\epsilon > 0$ be small enough and write
$$\mathcal{M}(\nu) = \min_{\underline{z} \in C_{\R,\,n}} \mathcal{L} ( \nu; \, \underline{z} )
\propto \text{constant} + \log \left( \det \left( K_{n, \nu} \right) \right) + \min_{\underline{z} \in C_{\R,\,n}} \underline{z}\tr K_{n, \nu}^{-1} \underline{z}.$$
For~$\tau = \beta_R(f) + d/2 - \epsilon/2$,
we are going to prove
that~$\inf_{\nu_{\mathrm{min}} \leq \nu \leq \beta_R(f) + d/2 - \epsilon} \mathcal{M}(\nu) - \mathcal{M}(\tau)$
is ultimately positive.

First, observe that Proposition~3.7 of \citet{karvonen2023asymptotic}
applies in our setting (by potentially transforming~$\XX$
to deal with an isotropic covariance function).
Our \cref{prop:power_function_bounds} adapts
Proposition~3.6 from the previous reference in the setting of \cref{assump:domain}.
Proceed as in the reference by using these two propositions and Stirling’s formula to get
$$
\inf_{\nu_{\mathrm{min}} \leq \nu \leq \beta_R(f) + d/2 - \epsilon} 
\log ( \det ( K_{n, \nu} ) ) - \log ( \det ( K_{n, \tau} ) ) 
\geq \mathcal{O}(n) + \frac{\epsilon}{d} \log(n!)
\sim \frac{\epsilon}{d} n \log(n).
$$
To conclude, observe that~$f$ is an~$R$-element of~$W_2^{\beta_R(f) -\epsilon/2  + d/2}(\XX)$
by definition, so~\eqref{eq:bound_norm} and quasi-uniformity yields:
$$
\min_{\underline{z} \in C_{\R,\,n}} \underline{z}\tr K_{n, \tau}^{-1} \underline{z}
\lesssim n.
$$\hfill\BlackBox
\\

\clearpage

\section{{EI Benchmark Results}}
\label{app:bench}

\begin{figure}[h!]
\begin{centering}
\resizebox{0.50\textwidth}{!}{\includegraphics{figures/EI/ackley4.pdf}}%
\resizebox{0.50\textwidth}{!}{\includegraphics{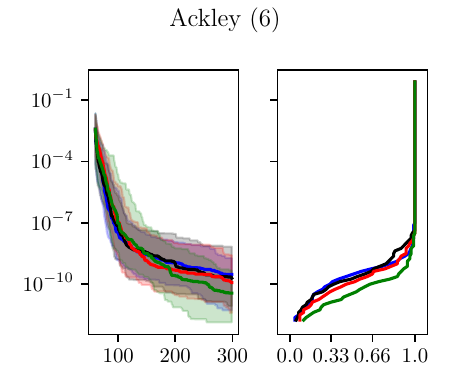}}\\%
\resizebox{0.50\textwidth}{!}{\includegraphics{figures/EI/ackley10.pdf}}%
\resizebox{0.50\textwidth}{!}{\includegraphics{figures/EI/beale.pdf}}\\%
\resizebox{0.50\textwidth}{!}{\includegraphics{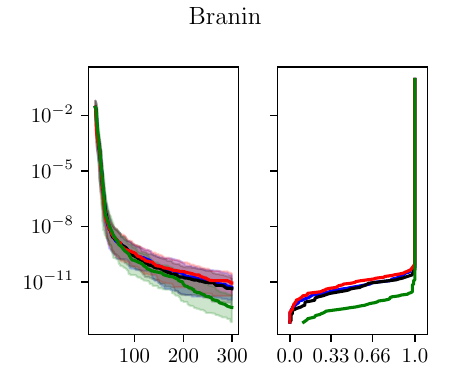}}%
\resizebox{0.50\textwidth}{!}{\includegraphics{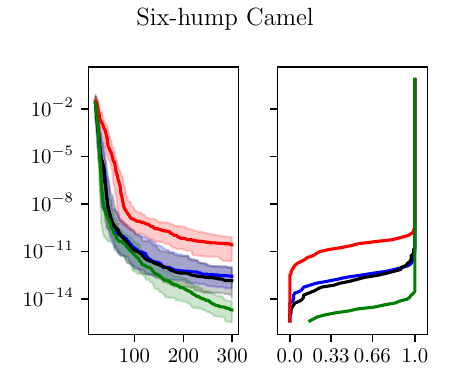}}
\end{centering}
\vspace{-0.75cm}
\caption{EGO and EGO-R results for a subset of
the test functions from \cref{table:bench_problems}.
Same legend as in \cref{fig:results_exposition_figure}.}
\label{fig:ei_1}
\end{figure}

\begin{figure}
\begin{centering}
\resizebox{0.50\textwidth}{!}{\includegraphics{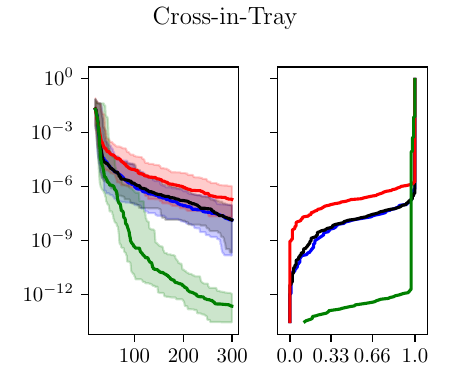}}%
\resizebox{0.50\textwidth}{!}{\includegraphics{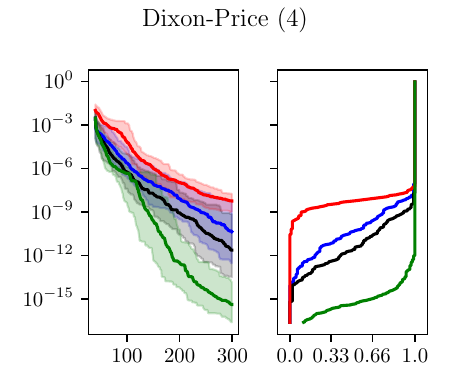}}\\%
\resizebox{0.50\textwidth}{!}{\includegraphics{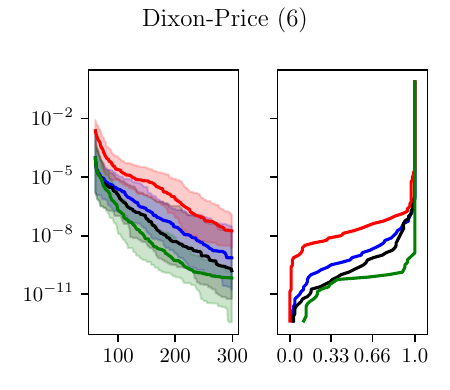}}%
\resizebox{0.50\textwidth}{!}{\includegraphics{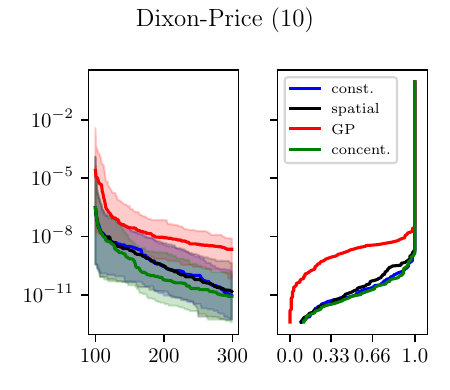}}\\%
\resizebox{0.50\textwidth}{!}{\includegraphics{figures/EI/goldstein_price_log.pdf}}%
\resizebox{0.50\textwidth}{!}{\includegraphics{figures/EI/goldsteinprice.pdf}}
\end{centering}
\vspace{-0.75cm}
\caption{EGO and EGO-R results for a subset of
the test functions from \cref{table:bench_problems}.
Same legend as in \cref{fig:results_exposition_figure}.}
\label{fig:ei_2}
\end{figure}

\begin{figure}
\begin{centering}
\resizebox{0.50\textwidth}{!}{\includegraphics{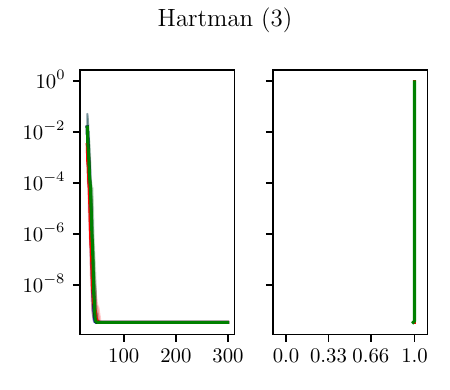}}%
\resizebox{0.50\textwidth}{!}{\includegraphics{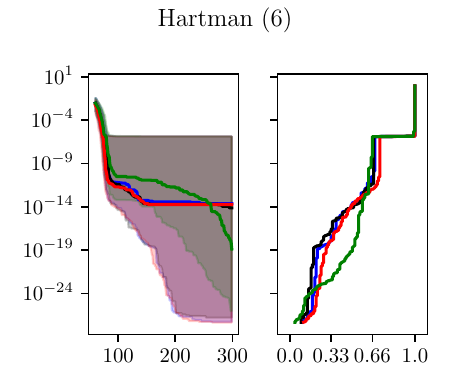}}\\%
\resizebox{0.50\textwidth}{!}{\includegraphics{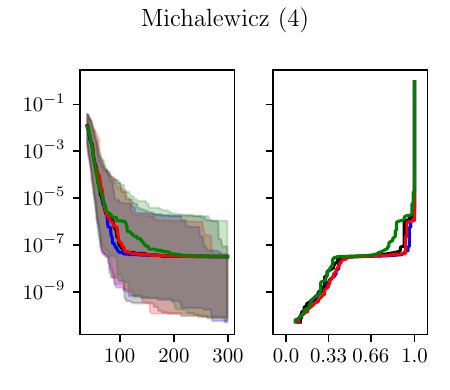}}%
\resizebox{0.50\textwidth}{!}{\includegraphics{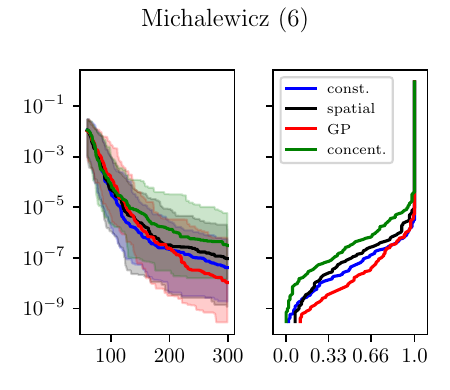}}\\%
\resizebox{0.50\textwidth}{!}{\includegraphics{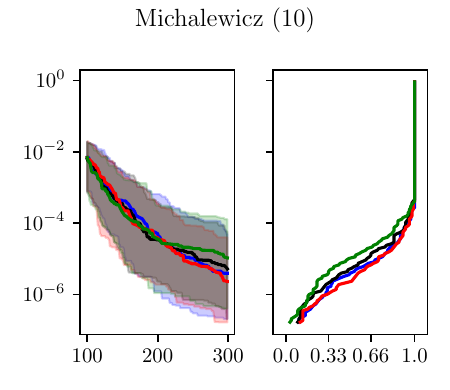}}%
\resizebox{0.50\textwidth}{!}{\includegraphics{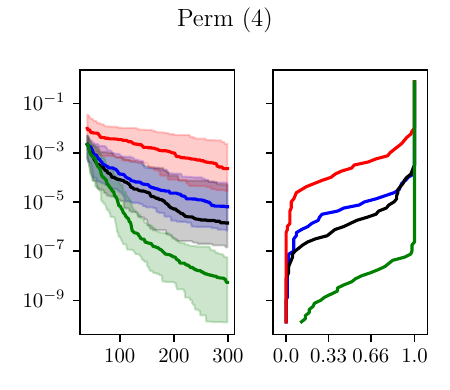}}
\end{centering}
\vspace{-0.75cm}
\caption{EGO and EGO-R results for a subset of
the test functions from \cref{table:bench_problems}.
Same legend as in \cref{fig:results_exposition_figure}.}
\label{fig:ei_3}
\end{figure}

\begin{figure}
\begin{centering}
\resizebox{0.50\textwidth}{!}{\includegraphics{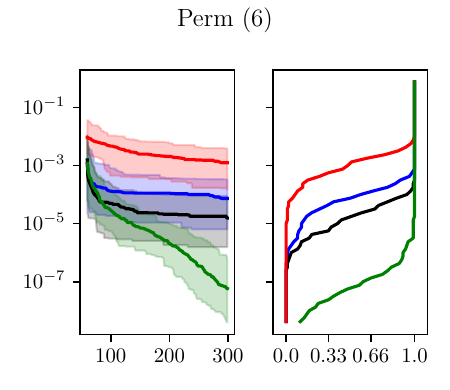}}%
\resizebox{0.50\textwidth}{!}{\includegraphics{figures/EI/perm10.pdf}}\\%
\resizebox{0.50\textwidth}{!}{\includegraphics{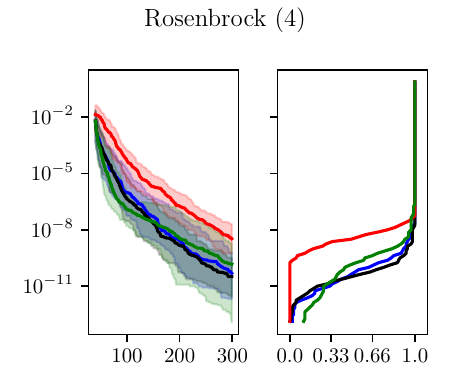}}%
\resizebox{0.50\textwidth}{!}{\includegraphics{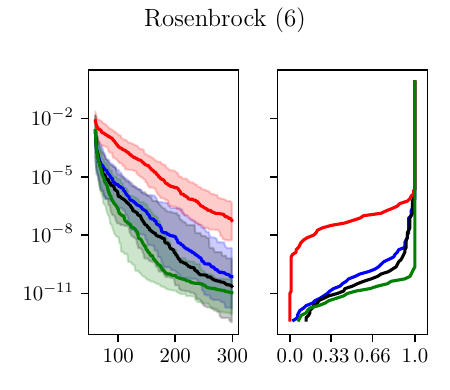}}\\%
\resizebox{0.50\textwidth}{!}{\includegraphics{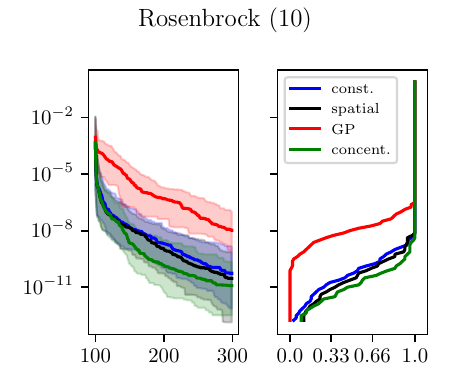}}%
\resizebox{0.50\textwidth}{!}{\includegraphics{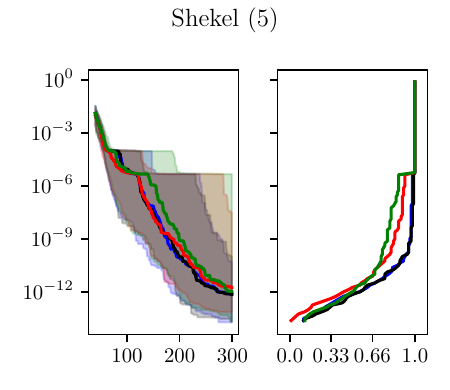}}%
\end{centering}
\vspace{-0.75cm}
\caption{EGO and EGO-R results for a subset of
the test functions from \cref{table:bench_problems}.
Same legend as in \cref{fig:results_exposition_figure}.}
\label{fig:ei_4}
\end{figure}

\begin{figure}
\begin{centering}
\includegraphics{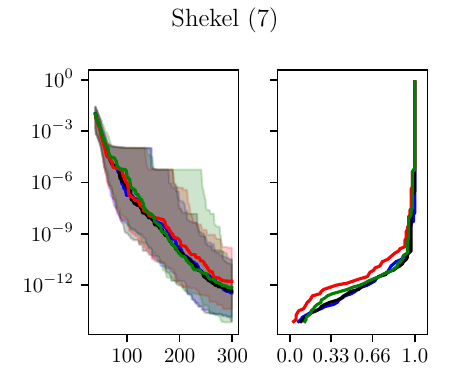}
\includegraphics{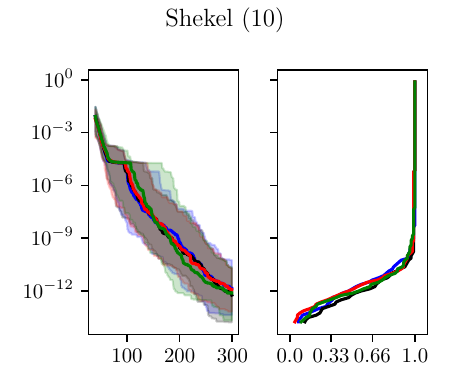}
\includegraphics{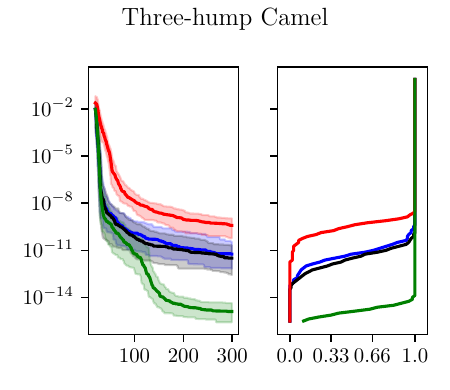}
\includegraphics{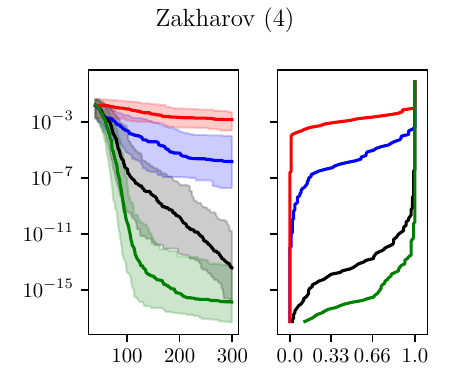}
\includegraphics{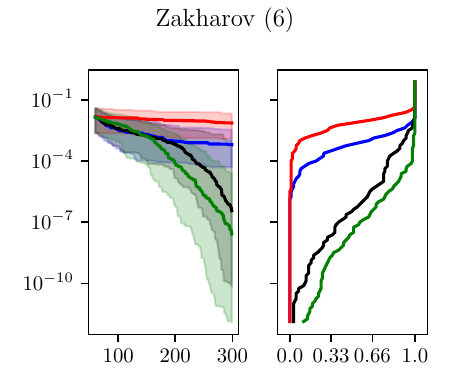}
\includegraphics{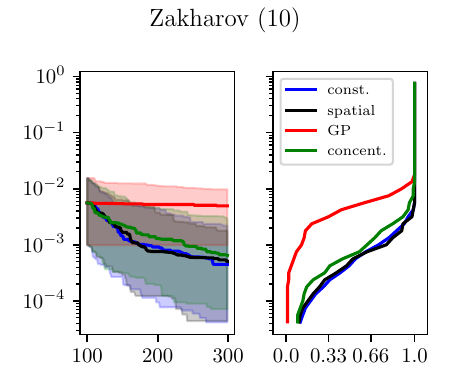}
\end{centering}
\vspace{-0.75cm}
\caption{EGO and EGO-R results for a subset of
the test functions from \cref{table:bench_problems}.
Same legend as in \cref{fig:results_exposition_figure}.}
\label{fig:ei_5}
\end{figure}

The results are provided in \crefrange{fig:ei_1}{fig:ei_5},
for all the test functions from \cref{table:bench_problems},
using the same format as in \cref{fig:results_exposition_figure}.

Observe that all the EGO-R methods yield (sometimes very) substantial
improvements on
Beale, Goldstein-Price,
Six-hump Camel, Dixon-Price~$(4)$, 
all instances of Perm and Rosenbrock, 
Three-hump Camel,
and Zakharov~$(4)$ and~$(6)$.
Focusing on these test functions, the concentration heuristic appears to give the best results, while
the constant heuristic brings the least improvement.
The spatial heuristic gives intermediate results between the two,
sometimes giving results similar to the concentration heuristic and,
at other times, results similar to the constant heuristic.

Taking variability (colored areas) into account, EGO-R with the constant and spatial heuristics performs only slightly better than EGO on
Cross-in-Tray and Dixon-Price~$(6)$ and~$(10)$.
The concentration heuristic, however, brings substantial improvements on these test functions, though. 
All EGO-R methods perform
as EGO on Branin, Ackley~$(4)$ and~$(6)$, Log-Goldstein-Price,
Michalewicz~$(4)$ and~$(10)$,
all the Shekel instances,
Hartman~$(3)$,
and Zakharov~$(10)$. 

The performance of the EGO-R methods is rarely worse than EGO, and never dramatically so.
Nevertheless, the concentration heuristic
yield substantially higher median empirical best minima on Ackley~$(10)$, and Michalewicz~$(6)$ and~$(10)$.
This is also the case of the spatial heuristic on Ackley~$(10)$.

\begin{remark}
The results are spread out for the Hartman~$(6)$ function.
Investigations have shown this is due to a fraction
of runs getting trapped in local minima.
The proportions of runs reaching, for~$n = 300$, an empirical best evaluation close
to the global minimum are~$72\%$,~$68\%$,~$67\%$, and~$66\%$
for EGO and EGO-R with the constant, spatial, and concentration heuristics, respectively.
(To~measure the closeness to the global minimum,
a list~$x_1^\star, \dots, x_N^\star$ of local minima was estimated
by manually analyzing the output of repetitions of a random-start local optimizer.
Then, a function value~$f(x)$ is said to be close to the global minimum
if~$\lvert f(x) -  f(x_i^\star)\lvert$ is minimized by~$\argmin_i f(x_i^\star)$.)  
\end{remark}

\clearpage

\section{{UCB Benchmark Results}}
\label{app:bench_ucb10}

The Upper Confidence Bound \citep[UCB,][]{cox1992statistical, srinivas2010:_ucb}
is an other sampling criterion for function minimization.
Given a parameter~$\beta > 0$, the function~$f$ is minimized
by choosing sequentially~$x_{n +1}$ minimizing
\begin{equation}\label{eq:ucb}
\mu_n(x) - \beta \sigma_n(x),
\end{equation}
with respect to $x \in \XX$.
In other words, the UCB algorithm minimizes
an optimistic quantile of the Gaussian predictive distribution
to select the next observation. Large values of~$\beta$ favor exploration.

Varying~$\beta$ slowly with~$n$ helps to prove
convergence \citep{srinivas2010:_ucb}.
Nevertheless, a few experiments (not shown here) suggest that, overall,
on the test functions from \cref{table:bench_problems},
the best vanilla GP baselines were
obtained with a fixed value.
Consequently, this finite-sample study was restricted to a fixed value of~$\beta$
as in \citet{cox1992statistical} and~\citet{picheny2013benchmark}.
The value was chosen so that~\eqref{eq:ucb} corresponds to the~$10\%$-quantile of the predictive distribution.

\begin{figure}
\vspace{-0.8cm}
\begin{centering}
\includegraphics{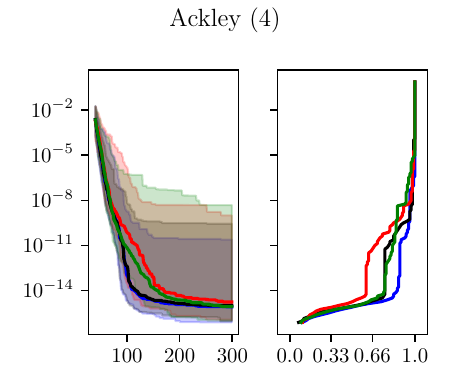}
\includegraphics{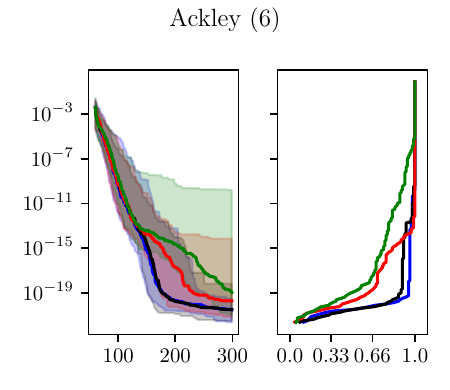}
\includegraphics{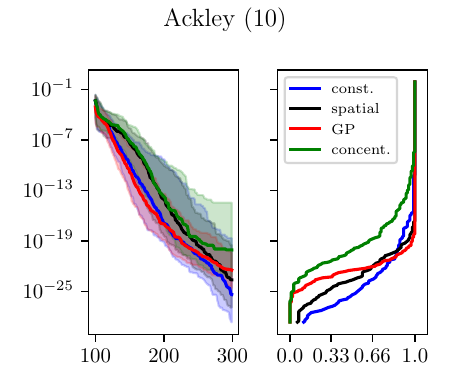}
\includegraphics{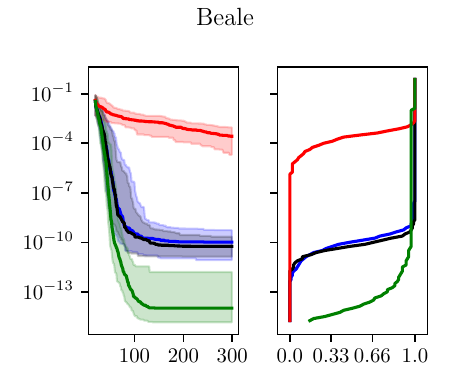}
\includegraphics{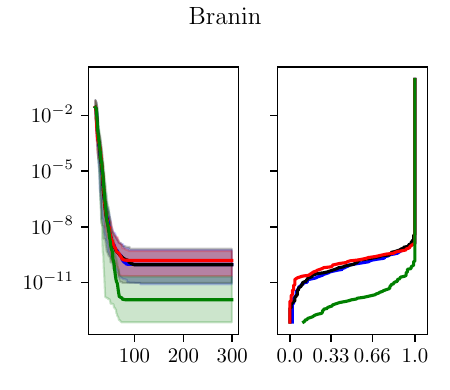}
\includegraphics{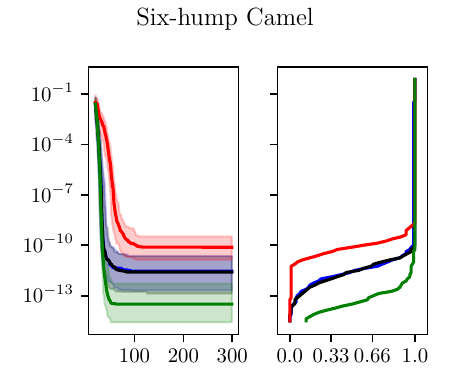}
\end{centering}
\vspace{-1.1cm}
\caption{%
UCB results for a subset of
the test functions from \cref{table:bench_problems}.
Same legend as in \cref{fig:results_exposition_figure}.
The red color stands for the UCB algorithm with vanilla GPs and
the blue, green, and black colors for UCB with reGP, using the constant,
concentration, and spatial heuristics respectively.}
\label{fig:ucb10_1}
\end{figure}

\begin{figure}[h!]
\begin{centering}
\includegraphics{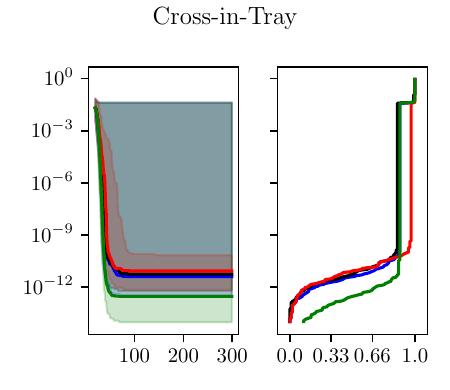}
\includegraphics{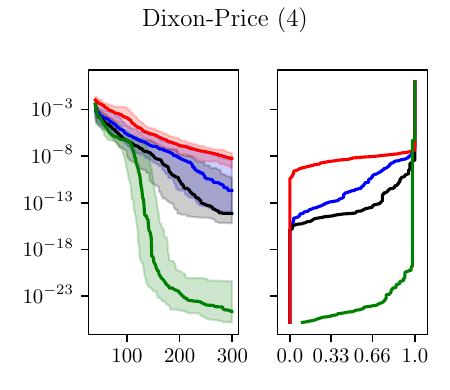}
\includegraphics{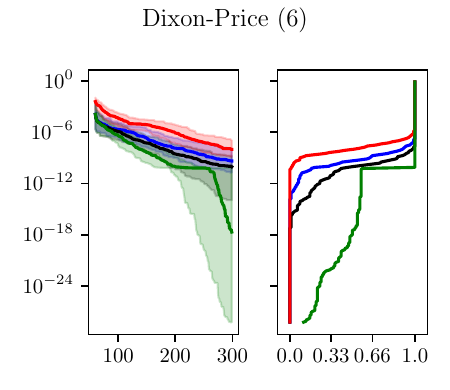}
\includegraphics{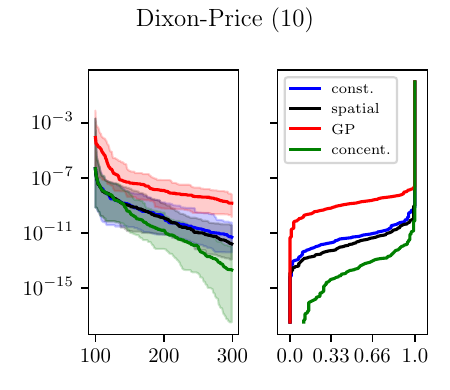}
\includegraphics{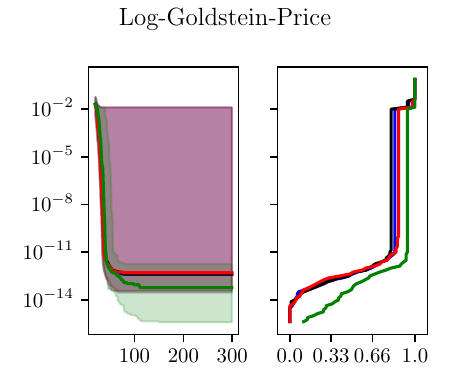}
\includegraphics{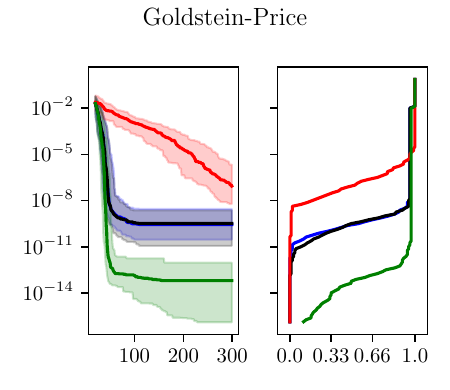}
\end{centering}
\vspace{-0.75cm}
\caption{UCB results for a subset of
the test functions from \cref{table:bench_problems}.
Same legend as in \cref{fig:ucb10_1}.}
\label{fig:ucb10_2}
\end{figure}

\begin{figure}
\begin{centering}
\includegraphics{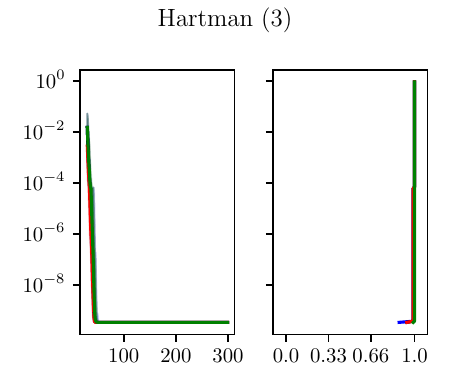}
\includegraphics{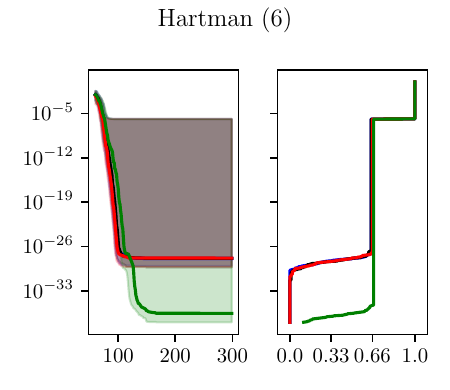}
\includegraphics{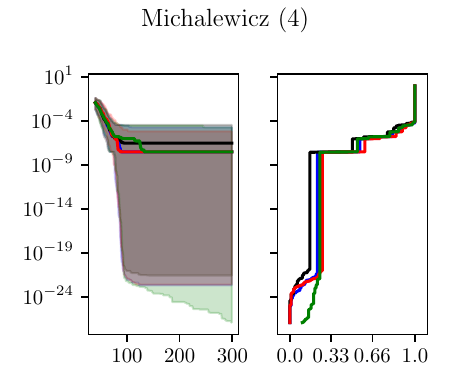}
\includegraphics{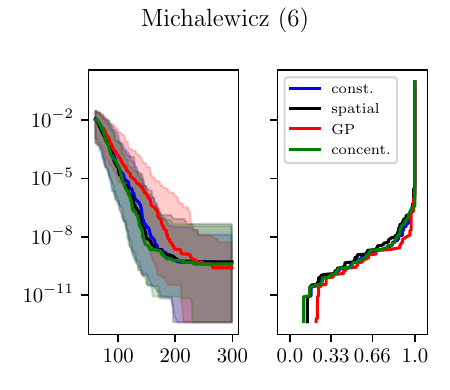}
\includegraphics{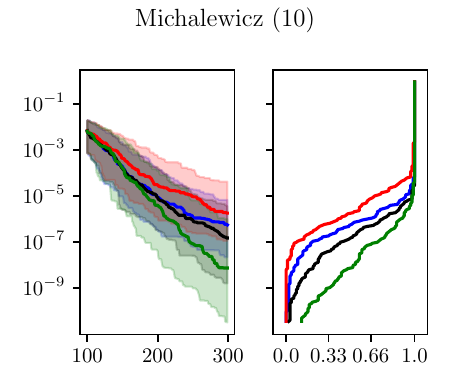}
\includegraphics{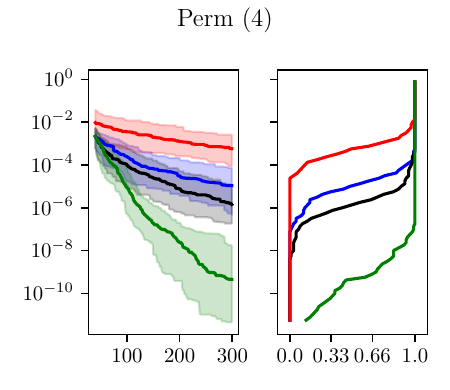}
\end{centering}
\vspace{-0.75cm}
\caption{UCB results for a subset of
the test functions from \cref{table:bench_problems}.
Same legend as in \cref{fig:ucb10_1}.}
\label{fig:ucb10_3}
\end{figure}

\begin{figure}
\begin{centering}
\includegraphics{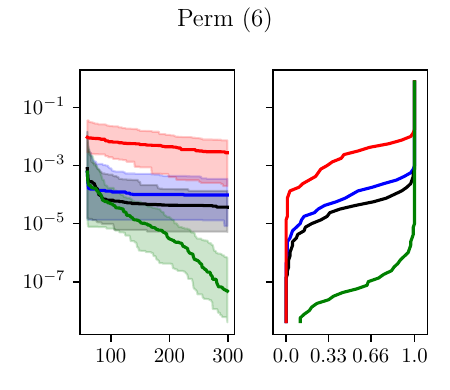}
\includegraphics{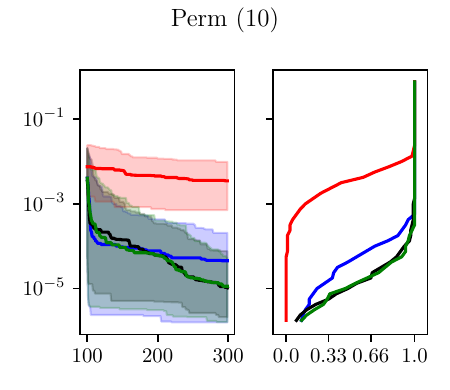}
\includegraphics{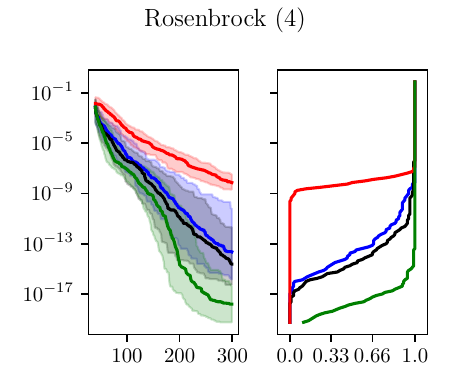}
\includegraphics{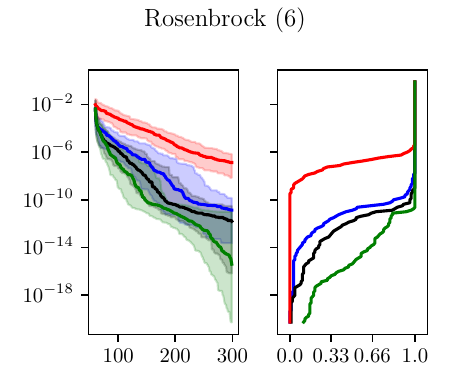}
\includegraphics{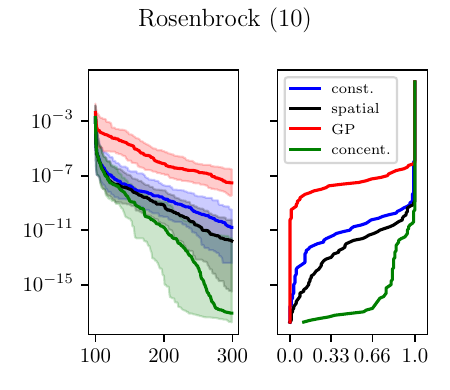}
\includegraphics{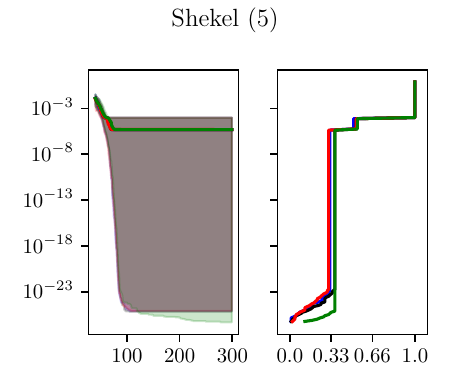}
\end{centering}
\vspace{-0.75cm}
\caption{UCB results for a subset of
the test functions from \cref{table:bench_problems}.
Same legend as in \cref{fig:ucb10_1}.}
\label{fig:ucb10_4}
\end{figure}

\begin{figure}
\begin{centering}
\includegraphics{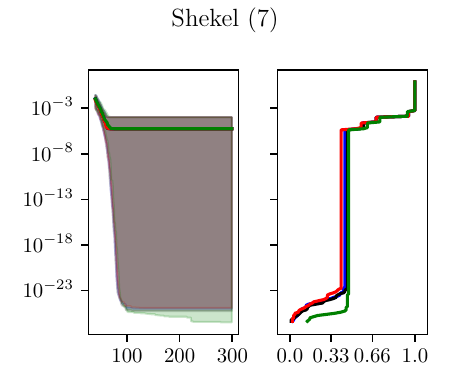}
\includegraphics{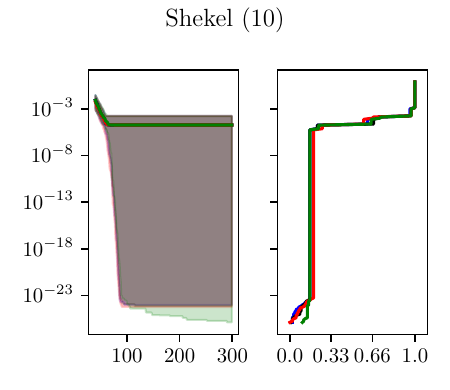}
\includegraphics{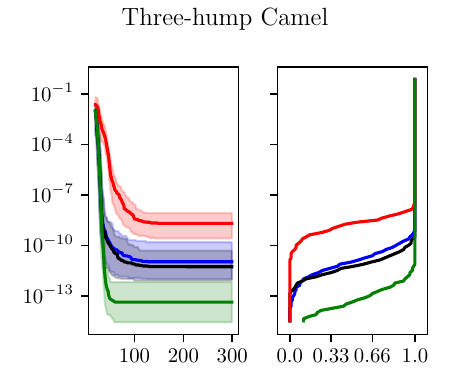}
\includegraphics{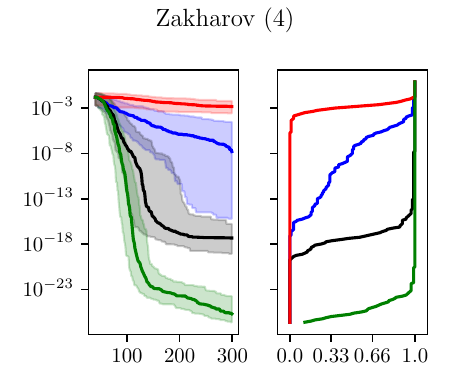}
\includegraphics{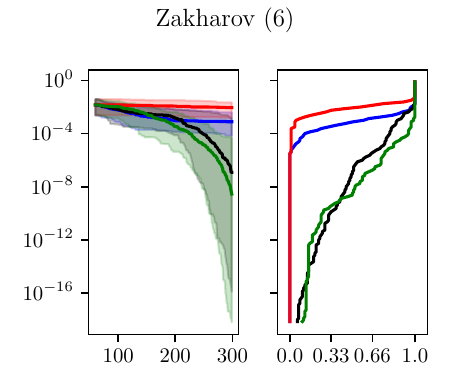}
\includegraphics{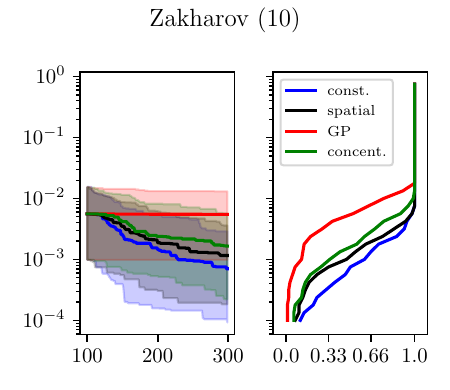}
\end{centering}
\vspace{-0.75cm}
\caption{UCB results for a subset of
the test functions from \cref{table:bench_problems}.
Same legend as in \cref{fig:ucb10_1}.}
\label{fig:ucb10_5}
\end{figure}

As with the EI criterion,
the application of \reGP to the UCB
algorithm simply involves plugging the \reGP predictive distribution
into the criterion~\eqref{eq:ucb}.
For choosing the validation threshold~$t_n^{(0)}$,
we will again compare the three heuristics introduced in \cref{sec:bench_optim_methodo},
namely constant, concentration, and spatial.
(Given~$t_n^{(0)}$, the relaxation range candidates will also be defined as in
\cref{sec:bench_optim_methodo}.) %
More generally, the assessment was carried out using the methodology described in \cref{sec:bench_optim_methodo}.
The criterion~\eqref{eq:ucb} was optimized using a subset simulation
algorithm.
The results are shown in \crefrange{fig:ucb10_1}{fig:ucb10_5}.

When applied to UCB, all \reGP variants outperform the standard UCB algorithm on
Beale, Six-hump Camel, Dixon-Price~$(4)$ and~$(10)$, Goldstein-Price,
all instances of Perm and Rosenbrock,
Three-hump Camel, 
Zakharov~$(4)$ and~$(6)$.
Furthermore,
the concentration heuristic yields the best results on these test functions.

The spatial and constant heuristics perform as standard GPs on
Branin, and Dixon-Price~$(6)$. Only the concentration heuristic
brings a substantial benefit.
However, GPs and all \reGP variants give similar results on
Ackley~$(4)$, Michalewicz~$(6)$ and~$(10)$, Hartman~$(3)$, and Zakharov~$(10)$.
The concentration heuristic underperforms on Ackley~$(6)$ and~$(10)$.

The results are spread out for all algorithms on
Cross-in-Tray, Log-Goldstein-Price, Hartman~$(6)$,
Michalewicz~$(4)$, and Shekel~$(5)$,
$(7)$, and~$(10)$.
As with the Hartman~$(6)$ function with EI, this is caused
by local minima.
This is a known feature of the standard UCB algorithm with fixed~$\beta$
\citep[see][Section~5]{jones2001taxonomy}.
(However, as mentioned earlier, this produced the best baselines, overall, in this study.)
\cref{table:ucb_local_minima_props}
shows the fractions of runs getting close to the global minima on these
test functions.
Observe that the \reGP variants are not routinely more often trapped
in local minima.
Furthermore, the figures show that the concentration heuristic
still benefits runs reaching the global minimum.

\begin{table}
\caption{Proportion of UCB runs reaching, for $n = 300$, an empirical best evaluation close
to the global minimum. These proportions are estimated as described
in \cref{app:bench}.}
\centering
\begin{tabular}{| c | c | c | c | c |}
\hline
Problem  & GP & constant & spatial & concentration \\
\hline
Hartman~$(6)$ & $66\%$ & $65\%$ & $65\%$ & $67\%$  \\
\hline
Shekel~$5$ & $31\%$ & $32\%$ & $36\%$ & $36\%$ \\
\hline
Shekel~$7$ & $41\%$ & $44\%$ & $46\%$ & $47\%$ \\
\hline
Shekel~$10$ & $19\%$ & $16\%$ & $16\%$ & $16\%$ \\
\hline
Log-Goldstein-Price & $87\%$ & $84\%$ & $81\%$ & $94\%$ \\
\hline
Cross-in-Tray & $97\%$ & $88\%$ & $86\%$ & $88\%$ \\
\hline
Michalewicz~$(4)$ & $26\%$ & $22\%$ & $16\%$ & $24\%$ \\
\hline
\end{tabular}
\label{table:ucb_local_minima_props}
\end{table}

\clearpage

\clearpage

\section{Application to the Estimation of an Excursion Set}\label{app:levelset}

Consider the problem of estimating the set~$\{ x \in \XX, \, f(x) \leq u\}$,
for some $u \in \RR$, as illustrated in \cref{sec:tcrps_example}.
When $f$ is expensive to evaluate, a common approach is to use
active learning with Gaussian process surrogates.
A large variety of sampling criteria are available in the
literature~\citep[see, e.g.,][]{bryan2005activelearning,
bect2012sequential, chevalier2014fast, bogunovic2016}. %
To illustrate the benefits of \reGP for the estimation of an excursion
set, we choose the straddle heuristic of
\citet{bryan2005activelearning}, which consists in choosing~$x_{n+1}$
maximizing
\begin{equation}\label{eq:straddle}
  1.96 \sigma_n(x) - \left| \mu_n(x) - u \right|, \quad x \in \XX.
\end{equation}
This simple (but effective) criterion rewards locations where the
prediction is both close to~$u$ and uncertain.

As with the~EI and~UCB criteria,
the coordination of \reGP with the straddle
heuristic simply involves plugging the \reGP predictive distribution
into the criterion~\eqref{eq:straddle}.
Consequently, with observations~$(\xn, \, \zn)$
and using the framework described by \cref{algo:regp}, we are left
to specify a range of interest~$Q_n$
and a list $\RR \setminus Q_n = R_n^{(0)} \supset \dots \supset R_n^{(G - 1)}$
of relaxation range candidates. %
As in \cref{sec:tcrps_example},
symmetric ranges of interest~$Q_n = ( u - t_n^{(0)}, \, u + t_n^{(0)} )$ are considered.
For~$t_n^{(0)}$, the three heuristics proposed
in \cref{sec:bench_optim_methodo} are adapted by
considering~$\alpha$-quantiles of~$| u - z_i |$. %
Then, given~$t_n^{(0)}$,
we propose to use~$R_n^{(g)} = ( - \infty, \, u - t_n^{(g)}] \cup [u + t_n^{(g)}, \, + \infty)$
with~$t_n^{(g)}$ ranging logarithmically
from~$t_n^{(0)}$ to~$\max_i |f(x_i) - u|$.
As was done for optimization,
we use the convention that~$t_n^{(G - 1)} > \max_i |f(x_i) - u|$ so that
the $G$-th model is a vanilla GP.

As in \cref{sec:bench_optim_methodo},
the experiments are conducted using
Matern covariance functions with regularity~$\nu = 5/2$ and constant mean functions.  
We also use $G = 10$ and initial designs of size $n_0 = 10d$.
The criterion~\eqref{eq:straddle} is optimized using a subset simulation
algorithm.

Four test cases are considered for the straddle heuristic in this work.
First, the two functions~$c_6$ and~$c_6^7$ introduced in \cref{sec:tcrps_example},
with~$u = 0$.
Then, as in \cref{sec:bench_optim_methodo}, we also consider the Goldstein-Price function, 
as well as its log-version.
The vanilla Goldstein-Price function ranges from~$3$ to about~$10^6$.
To depart from optimization, the intermediate value~$u = 1000$ is chosen, which is
a spatial quantile with a level of about~$25.4\%$.
The corresponding value~$u = \log(1000)$ is also used for the log-version.
For each of these four test cases, we
ran~$n_{\mathrm{rep}} = 100$ (random) repetitions, each with a budget of
$n_{\mathrm{tot}} = 300$ evaluations.

As often done for the estimation of excursion sets, the performance of the
sequential strategies is measured using model predictions.
More specifically, writing~$\PP_n$ for a predictive distribution,
we use a Sobol' sequence $\tilde{x}_1, \dots, \tilde{x}_N$
to compute the estimate
\begin{equation}
\label{eq:levelset_metric}
\frac{1}{N} \sum_{i=1}^N
\PP_n \left( \xi ( \tilde{x}_i ) > u \right)
\one_{f ( \tilde{x}_i ) \leq u}
+
\PP_n \left( \xi ( \tilde{x}_i ) \leq u \right)
\one_{f ( \tilde{x}_i ) > u}
\end{equation}
of the (relative) expected volume of the symmetric difference
between~$\{ x \in \XX, \, f(x) \leq u\}$ and the random
set~$\{ x \in \XX, \, \xi(x) \leq u\}$.

\cref{fig:straddle_results} shows the evolution of quantiles of~\eqref{eq:levelset_metric}
across the~$n_{\mathrm{rep}}$ repetitions for the four test cases. %
Observe the large improvement brought by \reGP on the~$c_6$ test case.
Using GPs, the evaluation metric~\eqref{eq:levelset_metric} does not improve as~$n$ increases,
which is consistent with the observations of~\cite{feliot2017:_jogo}.
However, the metric drops using \reGP. Using the constant heuristic,
the metric eventually stagnates at a value substantially lower
than that obtained with vanilla GPs, whereas it continues to decrease
using the concentration and spatial heuristics.
\cref{fig:g10_type_problem_slack} illustrates the relaxation
obtained with~$n = 300$ observations after a run of the concentration heuristic.

The performances of all the methods are comparable on~$c_6^7$.
The straddle heuristic with vanilla GPs is more effective than on~$c_6$.
All \reGP instances are a little behind, with slightly higher medians but
overlapping colored areas (except at the very beginning).
Curiously, the concentration and spatial heuristics are less effective
on~$c_6^7$ than on~$c_6$ (while the level sets are the same, since $t \mapsto t^7$ is increasing).

The three \reGP instances give very similar results on the Goldstein-Price test case,
with a substantial improvement over vanilla GPs.
Finally, for the Log-Goldstein-Price test case, the \reGP instances begin
by giving very similar results to those obtained with GPs.
Then, from around $n = 150$, the performances of the concentration and spatial heuristics
improve to become substantially better.
Observe that this is also the case for a fraction of runs with the constant heuristic.

In conclusion, we observe that \reGP is a useful modeling technique for the estimation of excursion sets
with the straddle heuristic. It can give clear improvements in the best cases,
while giving results comparable to those obtained with GPs in the worst cases.

\begin{figure}
\begin{centering}
\input{figures/straddle/c6.pgf}
\input{figures/straddle/c67.pgf}
\input{figures/straddle/goldsteinprice-1000.pgf}
\input{figures/straddle/goldstein_price_log-6_90775.pgf}
\end{centering}
\caption{%
Evolution of quantiles of~\eqref{eq:levelset_metric}
against~$n$. The solid lines represent the medians and, for each $n$, the shaded areas are
deliminated by the~$10\%$ and~$90\%$ quantiles.
The red color stands for the straddle heuristic with vanilla GPs and
the blue, green, and black colors for straddle with reGP, using the constant,
concentration, and spatial heuristics respectively.}
\label{fig:straddle_results}
\end{figure}

\clearpage

\section{Extension to the case of noisy observations}\label{app:noise}

\subsection{Gaussian process regression and kernel ridge regression}

We now consider the case where observations are corrupted by additive
noise, a common setting when dealing with stochastic simulators
\citep{baker2020stochastic} or when optimizing validation loss in
machine learning tasks with mini-batch training
\citep{snoek-2012:_practical_bo}.

Let $\xn = (x_1, \ldots, x_n) \in \XX^n$ be the input locations and
let $\Zn = (Z_1, \ldots, Z_n)\tr \in \RR^n$ denote the noisy
observations, modeled as
$$
Z_i = f(x_i) + \epsilon_i,
$$
where $f \colon \XX \to \RR$ is the latent function of interest and
$\epsilon_i \sim \Ncal(0, \eta^2)$ are independent Gaussian noise
terms with common variance $\eta^2 > 0$.

We place a Gaussian process prior $\xi \sim \GP(\mu, k)$ on $f$, using
the same notations as in \cref{sec:basics}. The posterior distribution
of $\xi$ given the observations $\Zn$ remains Gaussian:
\begin{equation}\label{eq:posterior-distribution-noise}
  \xi \, | \, \Zn \sim \GP \left( \mu_n, \, k_n \right),
\end{equation}
where the posterior mean and covariance functions are
\begin{align*}
  \mu_n(x)
  &= \mu(x) + k(x, \xn) (K_n + \eta^2 I)^{-1} \left( \Zn - \mu(\xn) \right),
  \\
  k_n(x, y)
  &= k(x, y) - k(x, \xn) (K_n + \eta^2 I)^{-1} k(y, \xn)\tr.
\end{align*}
Here, $K_n$ is the $n \times n$ covariance matrix with entries $[K_n]_{ij} = k(x_i, x_j)$, and $I$ is the $n \times n$ identity matrix.

The posterior mean in the zero-mean case ($\mu = 0$) coincides with
the solution of a kernel ridge regression (KRR) problem. This
correspondence is a direct consequence of the representer theorem, as
formalized below.

\begin{proposition}{\citep{kw1970:_correspondance}}\label[proposition]{prop:kernel_ridge}
  The solution to the regularized least-squares problem
  \begin{equation*}
    \left\{\; \begin{aligned}
        \text{minimize}
        & \quad \ns{h}_{\HH(\XX)}^2 + \eta^{-2} \sum_{i=1}^n  \left(Z_i - h(x_i)\right)^2\\
        \text{subject to}
        & \quad h \in \HH(\XX)
      \end{aligned} \right.
  \end{equation*}
  is unique and given by
  $$
    s_{\Zn}^{(\eta)} = k(\cdot, \xn) (K_n + \eta^2 I)^{-1} \Zn.
  $$
\end{proposition}
Hence, when $\mu = 0$, the posterior mean $\mu_n$ coincides with the KRR solution $s_{\Zn}^{(\eta)}$.

\subsection{Relaxed Gaussian process regression}

As seen in \cref{prop:kernel_ridge}, the posterior mean of a Gaussian
process model does not interpolate noisy data. How can the
goal-oriented modeling framework introduced in
\cref{sec:slack} be extended to this setting?

A natural extension is to adapt \cref{def:rgpi} by optimizing the
likelihood with respect to noisy observations
falling in the relaxation range.

Assume that both the mean and the covariance functions belong to
parametric families, indexed by $\theta \in \Theta$. Fix the noise
variance $\eta^2 > 0$ (see \cref{app:noisy_ucb} for an estimator). Let
$\R = \bigcup_{j = 1}^J R_j$ be a union of disjoint closed intervals
and define the corresponding constraint set $C_{\R,n} \subset \RR^n$
as in \cref{sec:relaxed_interpolation}.

We define the relaxed GP predictive distribution as the posterior
distribution \eqref{eq:posterior-distribution-noise} of the GP
conditioned on $\Zn = \zn^\star$ and on the estimated parameter
$\hat{\theta}_n$, where
\begin{equation*}
  (\hat{\theta}_n, \zn^\star)
  \;=\; \argmin_{\theta \in \Theta, \, \underline{z} \in C_{\R,\,n}}
    \mathcal{L}(\theta; \underline{z}),
\end{equation*}
and $\mathcal{L}$ is the (negative) log-likelihood of the observations written as
\begin{equation*}
  \mathcal{L}(\theta; \underline{z}) 
  \propto \log \det(K_n + \eta^2 I)
  + (\underline{z} - \mu(\xn))\tr (K_n + \eta^2 I)^{-1} (\underline{z} - \mu(\xn))
  + \text{constant}.
\end{equation*}
Note that we recover \cref{def:rgpi} when $\eta^2 = 0$.

When the mean function is zero and the covariance function $k$ is
fixed, the resulting procedure can be formulated as a regularized regression problem:
\begin{proposition}\label[proposition]{prop:relaxed_kernel_ridge}
  Let $\HH(\XX)$ be the RKHS associated with $k$, and let $\eta^2 > 0$
  be fixed. Consider the relaxed regularized least-squares problem
  \begin{equation}\label{eq:relaxed_kernel_ridge}
    \left\{
    \begin{aligned}
      \text{minimize} \quad
      & \ns{h}_{\HH(\XX)}^2 + \eta^{-2} \sum_{i=1}^n e(Z_i, h(x_i); R), \\
      \text{subject to} \quad
      & h \in \HH(\XX),
    \end{aligned}
    \right.
  \end{equation}
  where the relaxed squared error $e(y_1, y_2; R)$ is defined for all $y_1, y_2 \in \RR$ by
  $$
    e(y_1, y_2; R) =
    \begin{cases}
      \inf_{r \in R_j} (y_2 - r)^2 & \text{if } y_1 \in R_j \text{ for some } j, \\
      (y_2 - y_1)^2 & \text{otherwise}.
    \end{cases}
  $$
  Then, the unique solution of \eqref{eq:relaxed_kernel_ridge} is
  $s_{\zn^\star}^{(\eta)}$, where $\zn^\star$ is the unique solution
  of
  \begin{equation}\label{eq:regp_r_fixed}
    \argmin_{\underline{z} \in C_{R,n}} \ \underline{z}\tr (K_n + \eta^2 I)^{-1} \underline{z}.
  \end{equation}
\end{proposition}


\begin{proof}
  Write~$\Delta = \eta^2 I$.
  For~$\underline{z} = (z_1, \dots, z_n)\tr \in \RR^n$, straightforward calculations
  (using the identity $I - K_n ( K_n + \Delta )^{-1} = \Delta \left( K_n + \Delta \right)^{-1}$) show that
  $$
      \ns{s_{\underline{z}}^{(\eta)}}_{\HH(\XX)}^2 +\eta^{-2} \sum_{i=1}^n \left( z_i -  s_{\underline{z}}^{(\eta)}(x_i) \right)^2
   =
      \underline{z}\tr 
      \left(K_n + \Delta \right)^{-1}  \underline{z}.
  $$
  The quadratic problem~\eqref{eq:regp_r_fixed} has a unique solution~$\zn^\star$ since the matrix is positive-definite
  and the feasible set is closed and convex. Write~$L^{\star}$ for the corresponding minimum value.

  For~$h \in \HH(\XX)$, the definition of~$e$ yields
  $$
  \ns{h}_{\HH(\XX)}^2 + \eta^{-2} \sum_{i=1}^n e \left(Z_i, \, h(x_i); \, R\right)
  =
  \inf_{\underline{z} = (z_1, \dots, z_n)\tr \in C_{R, n}} \ns{h}_{\HH(\XX)}^2 +\eta^{-2} \sum_{i=1}^n \left( z_i -  h(x_i) \right)^2.
  $$
  The infimum is reached by a~$\underline{z}(h) = (z_1(h), \dots, z_n(h))\tr \in C_{R, n}$.
  We then have
  \begin{align*}
    \ns{h}_{\HH(\XX)}^2 + \eta^{-2} \sum_{i=1}^n e \left(Z_i, \, h(x_i); \, R\right)
    & =
      \ns{h}_{\HH(\XX)}^2 +\eta^{-2} \sum_{i=1}^n \left( z_i(h) -  h(x_i) \right)^2\\
    & \geq
      \ns{s_{\underline{z}(h)}^{(\eta)}}_{\HH(\XX)}^2 +\eta^{-2} \sum_{i=1}^n \left( z_i(h) -  s_{\underline{z}(h)}^{(\eta)}(x_i) \right)^2\\
    & =
      \underline{z}(h)\tr 
      \left(K_n + \Delta \right)^{-1}  \underline{z}(h)\\
    & \geq L^{\star}.
  \end{align*}%
  If the previous inequalities are equalities,
  then~$h = s_{\zn^\star}^{(\eta)}$
  because of the unicity of the solution of~\eqref{eq:regp_r_fixed} and by \cref{prop:kernel_ridge}.

  Finally, writing~$\zn^\star = (z_1^{\star}, \dots, z_n^{\star})$, we have
  \begin{equation*}
    \ns{s_{\zn^\star}^{(\eta)}}_{\HH(\XX)}^2 + \eta^{-2} \sum_{i=1}^n e \left(Z_i, \, s_{\zn^\star}^{(\eta)}(x_i); \, R\right)
    \leq
      \ns{s_{\zn^\star}^{(\eta)}}_{\HH(\XX)}^2 + \eta^{-2} \sum_{i=1}^n \left(z_i^{\star} - s_{\zn^\star}^{(\eta)}(x_i) \right)^2
    =  L^{\star}.
  \end{equation*}%
\end{proof}%

This shows that, in the fixed-kernel case, the relaxed GP regression
procedure corresponds to a modified KRR problem: observations $Z_i$
outside the relaxation range $R$ are penalized by squared prediction
error, while observations inside $R_j$ are only penalized by the
squared distance from the prediction $h(x_i)$ to the interval $R_j$.

The resulting predictive distribution thus reflects a form of soft
constraint on the latent function: fidelity to the data is enforced
outside $R$, while inside $R$ the model is only constrained to remain
consistent with the relaxation range.

\subsection{Application to Bayesian optimization with the UCB criterion}\label{app:noisy_ucb}

We now illustrate the use of \reGP in Bayesian optimization with noisy
observations. We adopt the same UCB sampling strategy as described in
\cref{app:bench_ucb10}, using a fixed $\beta$ parameter. Importantly,
we do not inflate the posterior variance with the estimated noise
variance~$\eta^2$; the UCB criterion is applied to the predictive
distribution of the latent process~$\xi$.

Two test functions are considered: the Goldstein-Price function and
its logarithmic transformation. In both cases, Gaussian noise is added
to the evaluations. For the Goldstein-Price function, we use
$\eta = 100$ (standard deviation),
which is large relative to the local variations of the function near its minimum.
Specifically, while the function ranges from
$3$ to approximately $10^6$, the $25\%$ and $5\%$ spatial quantiles
are around~$1000$ and $100$, respectively. The noise level is thus
significant near the global minimum, whereas extreme values act as
outliers. This setting is favorable to the \reGP strategy, which
focuses on accurate modeling in target regions. For the
Log-Goldstein-Price function (value range: $1.10$ to~$13.83$), we set
$\eta = 3$.

In both cases, the noise variance~$\eta^2$ is assumed to be unknown
and must be estimated from the data. Jointly estimating~$\eta^2$, the
model parameters, and the relaxed values by maximizing the likelihood
is not advisable in the \reGP framework, as the relaxed values tend to
shrink toward the posterior mean function, potentially leading to a biased
underestimation of the noise level. Instead, we estimate~$\eta^2$
using standard GP maximum likelihood, restricted to observations
within the range of interest, and treat the result as a fixed value in
the relaxed procedure.

For \reGP, the relaxation threshold is selected from a predefined list
using the LOO-tCRPS criterion, as described in
\cref{sec:bench_optim_methodo}. The spatial heuristic is adapted by
setting the validation threshold~$t_n^{(0)}$ to a quantile of the
previous model's predictions at the previous
design~$\underline{x}_{n-1}$. Further adjustments are likely needed
for the concentration heuristic.

\begin{figure}
\begin{centering}
\input{figures/noisy-UCB10/multi_noisy.pgf}
\end{centering}
\caption{Bayesian optimization with noisy evaluations:
performance of the UCB strategy.
Left: Goldstein-Price function. Right: Log-Goldstein-Price function.
Each plot shows medians (solid lines)
and $10\%$/$90\%$ quantiles (shaded areas)
as functions of the number of evaluations~$n$.
Top: noisy evaluations $f(x_n) + \epsilon_n$.
Middle: oracle values $f(x_n^\star)$ at predicted minimizers $x_n^\star = \argmin \mu_n$.
Bottom: absolute estimation errors $\abs{f(x_n^\star) - \mu_n(x_n^\star)}$.}
\label{fig:ucb10_noisy}
\end{figure}

We run $n_{\mathrm{rep}} = 100$ independent repetitions with random
initial designs of size $n_0 = 10d$ and total budgets of
$n_{\mathrm{tot}} = 300$ evaluations. Results are reported in
\cref{fig:ucb10_noisy}. We define $x_n^\star = \argmin \mu_n$ as
the predicted minimizer at step~$n$, and use $f(x_n^\star)$ as
an oracle benchmark.

On the Goldstein-Price function, \reGP yields substantially lower
median noisy evaluations up to approximately $n = 150$. This is
confirmed by the superior oracle values $f(x_n^\star)$ obtained with
\reGP during the same phase. After that point, the difference between
noisy values and the global minimum becomes comparable to the noise
level, and the two approaches yield similar performance. Notably, the
estimation error $\abs{f(x_n^\star) - \mu_n(x_n^\star)}$ remains
consistently smaller with \reGP across all steps.

On the Log-Goldstein-Price function, both methods perform similarly
overall. However, standard GP models show higher $90\%$
quantiles for $f(x_n^\star)$ and $f(x_n) + \epsilon_n$,
indicating more variability in the predicted minima.

These results illustrate the robustness of \reGP in noisy optimization
tasks.

\clearpage

\bibliography{spetit-regp-paper}

\end{document}